\documentclass[12pt, reqno]{amsart}

\usepackage[margin=.9in, a4paper, foot=.3in]{geometry}

\usepackage{graphicx, color}

\definecolor{darkblue}{rgb}{0,0,.5}
\definecolor{darkred}{rgb}{.5,0,0}
\definecolor{darkgreen}{rgb}{0,0.5,0}

\usepackage{hyperref}
\hypersetup {
    pdfstartview=FitH,
    colorlinks,
    citecolor=darkgreen,
    linkcolor=darkred,
    urlcolor=darkblue,
    linktoc=page
}

\let\oldtocsection=\tocsection
\let\oldtocsubsection=\tocsubsection
\let\oldtocsubsubsection=\tocsubsubsection

\renewcommand{\tocsection}[2]{\hspace{0em}\oldtocsection{#1}{#2}}
\renewcommand{\tocsubsection}[2]{\hspace{1em}\oldtocsubsection{#1}{#2}}
\renewcommand{\tocsubsubsection}[2]{\hspace{2em}\oldtocsubsubsection{#1}{#2}}

\usepackage{cite}

\usepackage[noBBpl,slantedGreek]{mathpazo}
\usepackage[mathcal]{euscript}

\usepackage{amssymb}
\usepackage{bm, bbm}
\allowdisplaybreaks
\numberwithin{equation}{section}

\usepackage{mathtools}

\newcommand {\con}[2]{\langle #1, \, #2 \rangle}

\newcommand {\mhat}[3]{\hskip #2 \widehat{\hskip -#2 #1 \hskip -#3} \hskip #3}
\newcommand {\hDelta}{\mhat{\Delta}{.1em}{.05em}}
\newcommand {\hpi}{\mhat{\pi}{.1em}{.1em}}
\newcommand {\hQ}{\mhat{Q}{0em}{-.1em}}

\newcommand {\nover}[1]{\overset{\scriptscriptstyle (n)}{#1}{}}

\newcommand {\rme}{\mathrm e}

\newcommand {\ch}{\mathrm{ch}}
\renewcommand {\det}{\mathrm{det}}
\newcommand {\diag}{\mathrm{diag}}
\renewcommand {\dim}{\mathrm{dim}}
\newcommand {\End}{\mathrm{End}}
\renewcommand {\exp}{\mathrm{exp}}
\newcommand {\id}{\mathrm{id}}
\newcommand {\im}{\mathrm{im}}
\renewcommand {\ker}{\mathrm{ker}}
\renewcommand {\log}{\mathrm{log}}
\newcommand {\Osc}{\mathrm{Osc}}
\newcommand {\sgn}{\mathrm{sgn}}
\newcommand {\tr}{\mathrm{tr}}

\newcommand {\bbC}{\mathbb C}
\newcommand {\bbE}{\mathbb E}
\newcommand {\bbZ}{\mathbb Z}

\newcommand {\calC}{\mathcal C}
\newcommand {\calD}{\mathcal D}
\newcommand {\calF}{\mathcal F}
\newcommand {\calL}{\mathcal L}
\newcommand {\calO}{\mathcal O}
\newcommand {\calQ}{\mathcal Q}
\newcommand {\calR}{\mathcal R}
\newcommand {\calS}{\mathcal S}
\newcommand {\calT}{\mathcal T}
\newcommand {\calW}{\mathcal W}
\newcommand {\calX}{\mathcal X}
\newcommand {\calY}{\mathcal Y}

\newcommand {\gothh}{\mathfrak h}
\newcommand {\gothg}{\mathfrak g}
\newcommand {\gothk}{\mathfrak k}

\newcommand {\gllpo}{\mathfrak{gl}_{\, l + 1}}
\newcommand {\hgothh}{\widehat{\mathfrak h}}
\newcommand {\hlgothg}{\widehat{\mathcal L}(\gothg)}
\newcommand {\lgothg}{\mathcal L(\mathfrak g)}
\newcommand {\sllpo}{\mathfrak{sl}_{\, l + 1}}
\newcommand {\tgothh}{\widetilde{\mathfrak h}}
\newcommand {\tlgothg}{\widetilde{\mathcal L}(\mathfrak g)}
\newcommand {\uqgllpo}{\mathrm U_q(\mathfrak{gl}_{\, l + 1})}
\newcommand {\uqlbp}{\mathrm U_q(\mathcal L(\mathfrak b_+))}
\newcommand {\uqlbm}{\mathrm U_q(\mathcal L(\mathfrak b_-))}
\newcommand {\uqlg}{\mathrm U_q(\calL(\mathfrak g))}
\newcommand {\uqlsllpo}{\mathrm U_q(\mathcal L(\mathfrak{sl}_{\, l + 1}))}
\newcommand {\uqlslii}{\mathrm U_q(\mathcal L(\mathfrak{sl}_2))}

\newcommand {\ygln}{\mathrm Y(\mathfrak{gl}_n)}

\title[Quantum groups and functional relations for arbitrary rank]{Quantum groups and functional relations \\ for arbitrary rank}

\author[A. V. Razumov]{Alexander V. Razumov}
\address{Institute for High Energy Physics, NRC ``Kurchatov Institute", 142281 Protvino, Mos\-cow region, Russia}
\email{Alexander.Razumov@ihep.ru}

\begin{document}

\addtolength {\jot}{3pt}

\begin{abstract}
The quantum integrable systems associated with the quantum loop algebras $\uqlsllpo$ are considered. The factorized form of the transfer operators related to the infinite dimensional evaluation representations is found and the determinant form of the transfer operators related to the finite dimensional evaluation representations is obtained. The master $TQ$- and $TT$-relations are derived. The operatorial $T$- and $Q$-systems are found. The nested Bethe equations are obtained.
\end{abstract}

\maketitle

\tableofcontents

\section{Introduction}

Functional relations satisfied by commuting integrability objects are known to be a powerful tool for solving quantum integrable models.\footnote{For the terminology used we refer to the paper \cite{BooGoeKluNirRaz14a} and section \ref{s:urmio} of the present paper.} In this paper, using the quantum group approach, we construct and investigate functional relations for quantum integrable systems associated with the quantum loop algebras $\uqlsllpo$. The central object of the quantum group approach is the universal $R$-matrix being an element of the completed tensor product of two copies of the quantum loop algebra. The integrability objects are constructed by choosing representations for the factors of that tensor product. It is traditional to call the representation used for the first factor of this tensor product the auxiliary space, while treating the representation space of the second one as the quantum space. These conventions can also be interchanged. By choosing a representation of the quantum group in the auxiliary space, one fixes an integrability object, while subsequently fixing a representation in the quantum space one defines a physical model. The consistent application of the method based on the quantum group theory was initiated by Bazhanov, Lukyanov and Zamolodchikov \cite{BazLukZam96, BazLukZam97, BazLukZam99}. They considered the quantum version of KdV theory. Later on this method proved to be efficient for studying other quantum integrable models. Within the framework of this approach, $R$-operators \cite{KhoTol92, LevSoiStu93, ZhaGou94, BraGouZhaDel94, BraGouZha95, BooGoeKluNirRaz10, BooGoeKluNirRaz11}, monodromy operators and $L$-operators were constructed \cite{BazTsu08,  BooGoeKluNirRaz10, BooGoeKluNirRaz11, BooGoeKluNirRaz13, Raz13, BooGoeKluNirRaz14a}. The corresponding families of functional relations were found \cite{BazHibKho02, Koj08, BazTsu08, BooGoeKluNirRaz14a, BooGoeKluNirRaz14b, NirRaz16a}. Recently the quantum group approach was used to derive and investigate equations satisfied by the reduced density operators of the quantum chains related to an arbitrary loop algebra \cite{KluNirRaz20, Raz20}. 

In fact, the most important functional relation is the factorized representation of the transfer operator. All other relations appears to be its consequences.  Usually, the functional relations are obtained by using the appropriate fusion rules for representations of the quantum loop algebra \cite{Res83a, Res83, KulRes86, BazRes90, KluPea92, KunNakSuz94, KunNakSuz11}. In the papers \cite{BazLukZam97, BazLukZam99, BazHibKho02, NirRaz16a, BooGoeKluNirRaz14b} to prove the factorization relations a direct operator approach was used for $l = 1$ and $l = 2$. For the higher ranks, the computational difficulties that arise seem to be almost insurmountable. In this paper, we use a different approach based on the analysis of the $\ell$-weights of the representations. The effectiveness of the method was demonstrated in the paper \cite{Raz21a} for $l = 1$ and $l = 2$. The present paper is devoted to the case of general $l$.

In section 2 we define the quantum group $\uqgllpo$ and discuss its Verma modules. These modules are used later in section 4 to define the evaluation representations of the quantum loop algebra $\uqlsllpo$. In section 3 we define the quantum loop algebras $\uqlg$. We give two equivalent definitions, the first in terms of the Drinfeld--Jimbo generators, and the second is the second Drinfeld's realization. The need for two definitions is that the first is convenient for defining evaluation representations of $\uqlsllpo$, and the generators used in the second definition contain an infinite commutative subalgebra used to define $\ell$-weights of the quantum loop algebras representations. In the same section we describe the construction of the Poincar\'e--Birkhoff--Witt basis of $\uqlsllpo$ used in the present paper. In section 4 we describe the category $\calO$ of $\uqlg$-modules, introduce the concept of $\ell$-characters and define the Grothendieck ring of $\calO$. In the same section we define the evaluation representations of the quantum groups $\uqlsllpo$, and the $q$-oscillator representations of their Borel subalgebras $\uqlbp$. We use the $q$-oscillator representations to construct the $Q$-operators. In the paper \cite{FreHer15} the prefundamental representations, introduced in the paper \cite{HerJim12}, are used for this purpose.  We use the $q$-oscillator representations because of their explicit form. In section 5 we define various types of integrability objects and discuss their properties. Here we also obtain the explicit form of $L$-operators related to $q$-oscillator representations. The functional relations for $\uqlsllpo$ are studied in section 6. We obtain the factorization representation for transfer operators in terms of $Q$-operators. This allows us construct $TQ$-relations, $TT$-relations and give the operator form of $T$-systems and $Q$-systems.

It is worth noting here that the functional relations for the integrable systems associated with the Yangians $\ygln$ were found and investigated in the paper \cite{BazLukMenSta10, BazFraLukMenSta11}. The authors of these papers used fusion of the elementary $L$-operators with a subsequent appropriate factorization. This method was also used for the case of the quantum loop algebra $\uqlslii$ in the paper \cite{KhoTsu14}. It seems that the generalization to the higher ranks is very cumbersome. Therefore, we use a completely different approach.

In our paper $l$ is a fixed positive integer. We also fix the deformation parameter $\hbar$ in such a way that $q = \exp(\hbar)$ is not a root of unity and assume that
\begin{equation*}
q^\nu = \exp (\hbar \nu)
\end{equation*}
for any $\nu \in \bbC$. We define $q$-numbers by the equation
\begin{equation*}
[\nu]_q = \frac{q^\nu - q^{- \nu}}{q - q^{-1}}, \qquad \nu \in \bbC,
\end{equation*}
and use the notation
\begin{equation*}
\kappa_q = q - q^{-1}.
\end{equation*}
An algebra, if it is not a Lie algebra, is understood as a unital associative algebra. All algebras and vector spaces are assumed to be complex. By a tuple $(s_i)_{i \in I}$ we mean a mapping from a finite ordered set $I$ to some set of objects $S$. When a tuple has only one component or is used as a multi-index we omit the parentheses in the notation.

\section{\texorpdfstring{Quantum group $\uqgllpo$}{Quantum group Uq(gll+1)}} \label{s:2}

\subsection{Definition}

We start with a brief reminder of some basic facts on  the Lie algebras $\gllpo$ and $\sllpo$. The Lie algebra $\gllpo$ is formed by the square matrices of order $l + 1$. The standard basis of the standard Cartan subalgebra $\gothk$ of the general linear Lie algebra $\gllpo$ consists of the matrices $K_i$ defined as\footnote{We denote by $\bbE_{i j}$ the usual matrix units, so that $(\bbE_{i j})_{k l} = \delta_{i k} \delta_{j l}$.} 
\begin{equation*}
K_i = \bbE_{i i}, \qquad i = 1, \ldots, l + 1.
\end{equation*}
Let $(\epsilon_i)_{i = 1}^{l + 1}$ be the dual basis of $\gothk^*$. Below we often identify an element $\mu \in \gothk^*$ with the $(l + 1)$-tuple formed by the components $\mu_i = \con \mu {K_i}$ of $\mu$ with respect to this basis. There are $l$ simple roots
\begin{equation*}
\alpha_i = \epsilon_i - \epsilon_{i + 1}, \qquad i = 1, \ldots, l,
\end{equation*}
and the full system $\Delta^+$ of positive roots is formed by the roots
\begin{equation*}
\alpha_{i j} = \sum_{k = i}^{j - 1} \alpha_k = \epsilon_i - \epsilon_j, \qquad  1 \le i < j \le l + 1.
\end{equation*}
The system of negative roots is $\Delta_- = - \Delta_+$, and the full root system is $\Delta = \Delta_+ \cup \Delta_-$. The root
\begin{equation*}
\theta = \alpha_{1, \, l + 1} = \sum_{i = 1}^l \alpha_i
\end{equation*}
is the highest root of $\Delta$.

The special Lie algebra $\sllpo$ is a subalgebra of $\gllpo$ formed by the traceless matrices. The standard basis of the standard Cartan subalgebra $\gothh$ of $\sllpo$ is formed by the matrices
\begin{equation*}
H_i = K_i - K_{i + 1}, \qquad i = 1, \ldots l.
\end{equation*}
As the positive and negative roots one takes the restriction of the roots of $\gllpo$ to $\gothh$. We have
\begin{equation*}
\langle \alpha_i, \, H_j \rangle = a_{j i},
\end{equation*}
where
\begin{equation*}
a_{i j} = {} - \delta_{i, \, j + 1} + 2 \delta_{i j} - \delta_{i + 1, \, j}.
\end{equation*}
The matrix $A = (a_{i j})_{i, \, j = 1}^l$ is the Cartan matrix of the Lie algebra $\sllpo$.

We define the quantum group $\uqgllpo$ as an algebra generated by the elements
\begin{equation*}
E_i, \quad F_i, \quad i = 1, \ldots, l, \qquad q^X, \quad X \in \gothk,
\end{equation*}
satisfying the following defining relations
\begin{gather*}
q^0 = 1, \qquad q^{X_1} q^{X_2} = q^{X_1 + X_2}, \\
q^X E_i \, q^{-X} = q^{\langle \alpha_i, \, X \rangle} E_i, \qquad q^X F_i \, q^{-X} = q^{-\langle \alpha_i, \, X \rangle} F_i, \\
[E_i, \, F_j] = \delta_{ij} \, \frac{q^{K_i - K_{i+1}} - q^{- K_i + K_{i + 1}}}{q - q^{-1}}.
\end{gather*}
In addition, there are the Serre relations, whose explicit form is not used in the present paper.

\subsection{\texorpdfstring{Poincar\'e--Birkhoff--Witt basis}{Poincare-Birkhoff-Witt basis}} \label{s:pbwb}

The abelian group
\begin{equation*}
Q = \bigoplus_{i = 1}^l \bbZ \, \alpha_i
\end{equation*}
is called the root lattice of $\gllpo$. Assuming that
\begin{equation*}
E_i \in \uqgllpo_{\alpha_i}, \qquad F_i \in \uqgllpo_{- \alpha_i}, \qquad q^X \in \uqgllpo_0,
\end{equation*}
we endow the algebra $\uqgllpo$ with a $Q$-gradation. An element $x$ of $\uqgllpo$ is called a root vector corresponding to a root $\gamma$ of $\gllpo$ if $x \in \uqgllpo_\gamma$. In particular, $E_i$ and $F_i$ are root vectors corresponding to the roots $\alpha_i$ and $- \alpha_i$, respectively. Following Jimbo \cite{Jim86a}, we introduce the elements $E_{\gamma}$ and $F_{\gamma}$, corresponding to all roots $\gamma \in \Delta$, with the help of the relations
\begin{equation*}
E_{\alpha_{i, \, i + 1}} = E_i, \qquad E_{\alpha_{i j}} = E_{\alpha_{i, \, j - 1}} \, E_{\alpha_{j - 1, \, j}} - q \, E_{\alpha_{j - 1, \, j}} \, E_{\alpha_{i, \, j - 1}}, \quad j - i > 1,
\end{equation*} 
and
\begin{equation*} 
F_{\alpha_{i, \, i + 1}} = F_i, \qquad F_{\alpha_{i j}} = F_{\alpha_{j - 1, \, j}} \, F_{\alpha_{i, \, j - 1}} - q^{-1} F_{\alpha_{i, \, j - 1}} \, F_{\alpha_{j - 1, \, j}}, \quad j - i > 1.
\end{equation*}
It is clear that the elements $E_{\alpha_{i j}}$ and $F_{\alpha_{i j}}$ are root vectors corresponding to the roots $\alpha_{i j}$ and $- \alpha_{i j}$, respectively. They are linearly independent and together with the elements $q^X$, $X \in \gothk$ are called the Cartan--Weyl generators of $\uqgllpo$. One can demonstrate that the ordered monomials constructed from the Cartan--Weyl generators form a Poincar\'e--Birkhoff--Witt basis of the quantum group $\uqgllpo$. In this paper we choose the following ordering of monomials. First endow $\Delta_+$ with the colexicographical order. It means that $\alpha_{i j} \preccurlyeq \alpha_{m n}$ if $j < n$, or if $j = n$ and $i < m$. This is a normal ordering of roots in the sense of \cite{LezSav74, AshSmiTol79}, see also \cite{Tol89}. This normal ordering is also a normal ordering of the roots of $\sllpo$. Now we say that a monomial is ordered if it has the form
\begin{equation*}
F_{\alpha_{i_1 j_1}} \ldots F_{\alpha_{i_r j_r}} \, q^X \, E_{\alpha_{m_1 n_1}} \ldots E_{\alpha_{m_s n_s}},
\end{equation*}
where $\alpha_{i_1 j_1} \preccurlyeq \cdots \preccurlyeq \alpha_{i_r j_r}$, $\alpha_{m_1 n_1} \preccurlyeq \cdots \preccurlyeq \alpha_{m_s n_s}$ and $X$ is an arbitrary element of $\gothk$. It is demonstrated in the paper \cite{NirRaz17b} that  such monomials really form a basis of $\uqgllpo$, see also the paper \cite{Yam89} for the case of an alternative ordering.

\subsection{Verma modules} \label{s:vm}

We use the standard terminology of the representation theory. In particular, we say that a $\uqgllpo$-module $V$ is a weight module if
\begin{equation*}
V = \bigoplus_{\mu \in \gothk^*}  V_\mu,
\end{equation*}
where
\begin{equation*}
V_\mu = \{v \in V \mid q^X v = q^{\langle \mu, \, X \rangle} v \mbox{ for any } X \in \gothk \}.
\end{equation*}
The space $V_\mu$ is called the weight space of weight $\mu$, and a nonzero element of $V_\mu$ is called a weight vector of weight $\mu$. We say that $\mu \in \gothk^*$ is a weight of $V$ if $V_\mu \ne \{ 0 \}$.

A weight $\uqgllpo$-module $V$ is called a highest weight module of highest weight $\mu$ if there exists a weight vector $v \in V$ of weight $\mu$ such that
\begin{equation*}
E_i \, v = 0, \quad i = 1, \ldots, l, \quad \mbox{and} \quad V = \uqgllpo \, v.
\end{equation*}
The vector with the above properties is unique up to a scalar factor. We call it the highest weight vector of $V$.

Given $\mu \in \gothk^*$, denote by $\widetilde V^\mu$ the corresponding Verma $\uqgllpo$-module. This is a highest weight module of highest weight $\mu$. We denote by $\widetilde \pi^\mu$ the representation of $\uqgllpo$ corresponding to $\widetilde V^\mu$. Denote by $\bm m$ the $l (l + 1)/2$-tuple of non-negative integers $m_{i j}$, $1 \le i < j \le l + 1$, arranged in the colexicographical order of $(i, \, j)$. More explicitly,
\begin{equation*}
{\bm m} = (m_{12}, \, m_{13}, \, m_{23}, \, \ldots, \, m_{1j}, \, \ldots, \, m_{j-1, \, j}, \, \ldots, \,
m_{1, \, l + 1}, \, \ldots, \, m_{l, \, l + 1}).
\end{equation*}
The vectors
\begin{equation}
v_{\bm m} = F_{12}^{m_{12}} \, F_{13}^{m_{13}} \, F_{23}^{m_{23}} \, 
\ldots F_{1, \, j}^{m_{1, \, j}} \ldots F_{j - 1, \, j}^{m_{j - 1, \, j}} \ldots 
F_{1, \, l + 1}^{m_{1, \, l + 1}} \ldots F_{l, \, l + 1}^{m_{l, \, l + 1}} \, v_{\bm 0}, \label{vml}
\end{equation}
where for consistency we denote the highest weight vector by $v_{\bm 0}$, form a basis of $\widetilde V^\mu$. The explicit relations describing the action of the generators $E_i$, $F_i$ and $q^X$ of the quantum group $\uqgllpo$ on a general basis vector $v_{\bm m}$ are obtained in the paper \cite{NirRaz17b}.

Note that $\widetilde V^\mu$ is an infinite-dimensional $\uqgllpo$-module. However, if
\begin{equation*}
\mu_i - \mu_{i + 1} \in \bbZ_{\ge 0}
\end{equation*}
for all $i = 1, \ldots, l$, there is a maximal submodule of $\widetilde V^\mu$, such that the respective quotient module is simple and finite-dimensional. We denote this $\uqgllpo$-module and the corresponding representation by $V^\mu$ and $\pi^\mu$, respectively. Note that any finite-dimensional $\uqgllpo$-module can be constructed in this way.

The weights $\omega_i \in \gothk^*$, $i = 1, \ldots, l + 1$, defined as
\begin{equation*}
\omega_i = \sum_{k = 1}^i \epsilon_k = (\underbracket[.5pt]{1, \, \ldots, 1}_i, \, \underbracket[.5pt]{0, \, \ldots, \, 0}_{l + 1 - i})
\end{equation*}
correspond to finite-dimensional representations called fundamental. The restriction of $\omega_i$, $i = 1, \ldots, l$, to $\gothh$ are the fundamental weights of $\sllpo$ so that
\begin{equation*}
\langle \omega_i, \, H_j \rangle = \delta_{i j}
\end{equation*}
for $i, j = 1, \ldots, l$.

\section{\texorpdfstring{Quantum loop algebra $\uqlg$}{Quantum loop algebra Uq(L(g))}}

\subsection{\texorpdfstring{Definition in terms of Drinfeld--Jimbo generators}{Definition in terms of Drinfeld-Jimbo generators}} \label{ss:dtjmg}

Let $\gothg$ be a complex finite-dimen\-sional simple Lie algebra of rank $l$, $\gothh$ a Cartan subalgebra of $\gothg$, and $\Delta$ the root system of $\gothg$ relative to $\gothh$, see, for example, the books \cite{Ser01, Hum80}. Fix a system of simple roots $\alpha_i$, $i = 1, \ldots, l$. The corresponding coroots $h_i$ form a basis of $\gothh$, so that
\begin{equation*}
\gothh = \bigoplus_{i = 1}^l \bbC \, h_i.
\end{equation*}
The Cartan matrix $A = (a_{i j})_{i, \, j = 1}^l$ of $\gothg$ is given by the equation
\begin{equation*}
a_{i j} = \langle \alpha_j, \, h_i \rangle.
\end{equation*}
Note that the Cartan matrix $A$ is always symmetrizable. It means that there exists a diagonal matrix $D = \diag(d_1, \, \ldots, d_l)$ such that the matrix $D A$ is symmetric and $d_i$, $i = 1, \ldots, l$, are positive integers. It is evident that $D$ is defined up to a nonzero scalar factor. We fix its normalization assuming that the integers $d_i$ are relatively prime.

Following Kac \cite{Kac90}, we denote by $\lgothg$ the loop algebra of $\gothg$, by $\tlgothg$ its standard central extension by the one-dimensional center $\bbC \, c$, and by $\hlgothg$ the Lie algebra obtained from $\tlgothg$ by adding a natural derivation $d$. By definition,
\begin{equation*}
\hlgothg = \lgothg \oplus \bbC \, c \oplus \bbC \, d, 
\end{equation*}
and the Cartan subalgebra of $\hlgothg$ is the space
\begin{equation*}
\hgothh = \gothh \oplus \bbC \, c \oplus \bbC \, d.
\end{equation*}
The Lie algebra $\hlgothg$ is isomorphic to the affine (Kac--Moody) algebra associated with the extended Cartan matrix (Dynkin diagram) $A^{(1)}$ of $\gothg$, see, for example, the book \cite[p.~166]{OniVin90}. Here the basis coroots are $h_i$, $i = 1, \ldots, l$, and
\begin{equation*}
h_0 = c - \sum_{i = 1}^l a \char20 \hskip -.3em_i \, h_i,
\end{equation*}
The integers $a \char20 \hskip -.3em_i$, $i = 1, \ldots, l$, together with $ a \char20 \hskip -.3em_i = 1$ are the dual Kac labels of the Dynkin diagram associated with the Cartan matrix $A^{(1)}$. Thus, we have
\begin{equation*}
\hgothh = \bigoplus_{i = 0}^l \bbC \, h_i \oplus \bbC \, d.
\end{equation*}
To introduce the corresponding simple roots, we identify the space $\gothh^*$ with the subspace of $\widehat \gothh^*$ defined as
\begin{equation*}
\{\lambda \in \widehat \gothh^* \mid \langle \lambda, \, c \rangle = 0, 
\ \langle \lambda, \, d \rangle = 0 \},
\end{equation*}
and denote by $\delta$ the element of $\hgothh^*$ defined by the equations
\begin{equation*}
\langle \delta, \, h_i \rangle = 0, \quad i = 0, 1, \ldots, l, \qquad \langle \delta, \, d \rangle = 1.
\end{equation*}
Then the simple roots are $\alpha_i$, $i = 1, \ldots, l$, and
\begin{equation*}
\alpha_0 = \delta - \theta,
\end{equation*}
where
\begin{equation*}
\theta = \sum_{i = 1}^l a_i \alpha_i
\end{equation*}
is the highest root of $\Delta$. The integers $a_i$, $i = 1, \ldots, l$, together with $ a_0 = 1$ are the Kac labels of the Dynkin diagram associated with the Cartan matrix $A^{(1)}$. One can demonstrate that the equation
\begin{equation*}
a_{i j} = \langle \alpha_j, \, h_i \rangle, \qquad i, j = 0, 1, \ldots, l,
\end{equation*}
gives the entries of the Cartan matrix $A^{(1)}$. Complementing the numbers $d_i$, $i = 1, \ldots, l$, with a suitable number $d_0$, one can demonstrate that the matrix $A^{(1)}$ is symmetrizable. Note that for $\gothg = \sllpo$, $d_i = 1$ for all $i = 0, 1, \ldots, l$.

The system of positive roots of the affine algebra $\hlgothg$ is
\begin{multline*}
\widehat \Delta_+ = \{\gamma + m \delta \mid  \gamma \in \Delta_+, \ m \in \bbZ_{\ge 0} \} \\ \cup \{m \delta \mid m \in \bbZ_{>0} \} \cup \{(\delta - \gamma) + m \delta \mid  \gamma \in \Delta_+, \ m \in \bbZ_{\ge 0}\},
\end{multline*}
where $\Delta_+$ is the system of positive roots of the Lie algebra $\gothg$. The system of negative roots $\widehat \Delta_-$ of $\hlgothg$ is $\widehat \Delta_- = - \widehat \Delta_+$, and the full system of roots is
\begin{equation*}
\widehat \Delta = \widehat \Delta_+ \sqcup \widehat \Delta_- 
= \{ \gamma + m \delta \mid \gamma \in \Delta, \ m \in \bbZ \} \cup \{m \delta \mid m \in \bbZ \setminus \{0\} \}.
\end{equation*}
The roots $m \delta$ are imaginary, and all other roots are real, see the book \cite{Kac90} for definition.

It is convenient for our purposes to denote
\begin{equation*}
\tgothh = \gothh \oplus \bbC \, c = \bigoplus_{i = 0}^l \bbC \, h_i.
\end{equation*}
It is easy to show that for any $\lambda \in \gothh^*$ there is a unique element $\widetilde \lambda \in \widetilde \gothh^*$ such that
\begin{equation*}
\langle \widetilde \lambda, \, c \rangle = 0, \qquad \langle \widetilde \lambda, \, x \rangle = \langle \lambda, \, x \rangle, \quad x \in \gothh.
\end{equation*}

We define the quantum group $\uqlg$ as an algebra generated by the elements $e_i$, $f_i$, $i = 0, 1, \ldots, l$, and $q^x$, $x \in \tgothh$, satisfying the relations
\begin{gather*}
q^{\nu \, c} = 1, \quad \nu \in \bbC, \qquad q^{x_1} q^{x_2} = q^{x_1 + x_2}, \\
q^x e_i \, q^{-x} = q^{\langle \alpha_i, \, x \rangle} e_i, \qquad q^x f_i \, q^{-x} = q^{- \langle \alpha_i, \, x \rangle} f_i, \\
[e_i, \, f_j] = \delta_{ij} \, \frac{q_i^{h_i} - q_i^{-h_i}}{q^{}_i - q_i^{-1}}.
\end{gather*}
for all $i = 0, 1, \ldots, l$. Here, for each $i = 0, 1, \ldots, l$ we set
\begin{equation*}
q_i = q^{d_i}. 
\end{equation*}
There are also the Serre relations, whose explicit form is not used in the present paper.

The quantum loop algebra $\uqlg$ is a Hopf algebra. The comultiplication $\Delta$, the antipode $S$, and the counit $\varepsilon$ are given by the relations
\begin{gather*}
\Delta(q^x) = q^x \otimes q^x, \qquad \Delta(e^{}_i) = e^{}_i \otimes 1 + q_i^{h_i} \otimes e^{}_i, \qquad \Delta(f^{}_i) = f^{}_i \otimes q_i^{- h_i} + 1 \otimes f^{}_i, \\
S(q^x) = q^{- x}, \qquad S(e^{}_i) = - q_i^{- h_i} e^{}_i, \qquad S(f^{}_i) = - f^{}_i \, q_i^{h_i}, \\
\varepsilon(q^x) = 1, \qquad \varepsilon(e^{}_i) = 0, \qquad \varepsilon(f^{}_i) = 0.
\end{gather*}
We do not use these relations in the present paper. They are given only to fix the conventions used.

To distinguish from the tensor product of mappings, we denote the tensor product of any two representations of $\uqlg$, say $\varphi$ and $\psi$, as
\begin{equation}
\varphi \otimes_\Delta \psi = (\varphi \otimes \psi) \circ \Delta \label{fodf}
\end{equation}
and, similarly, the tensor product of the corresponding $\uqlg$-modules $V$ and $W$ as $V \otimes_\Delta W$. More generally, given two homomorphisms $\varphi$ and $\psi$ from $\uqlg$ to algebras $A$ and $B$, respectively, equation (\ref{fodf}) defines a homomorphism of $\uqlg$ into $A \otimes B$.

The abelian group
\begin{gather*}
\hQ = \bigoplus_{i = 0}^l \bbZ \alpha_i
\end{gather*}
is called the root lattice of $\hlgothg$. Assuming that
\begin{gather*}
e_i \in \uqlg_{\alpha_i}, \qquad f_i \in \uqlg_{- \alpha_i}, \qquad q^x \in \uqlg_0
\end{gather*}
for any $i = 0, 1, \ldots, l$ and $x \in \tgothh$, we endow the algebra $\uqlg$ with a $\hQ$-gradation. An element $x$ of $\uqlg$ is called a root vector corresponding to a root $\gamma \in \hDelta$ if $x \in \uqlg_\gamma$. One can construct linearly independent root vectors corresponding to all roots from~$\hDelta$, see, for example, the papers \cite{TolKho92, KhoTol92, KhoTol93, KhoTol94} or the papers \cite{Bec94a, Dam98} for a different approach.  We use the approach by Khoroshkin and Tolstoy \cite{TolKho92, KhoTol92, KhoTol93, KhoTol94}. The linearly independent root vectors together with the elements $q^x$, $x \in \tgothh$ are called the Cartan--Weyl generators of $\uqlg$. If some ordering of roots is chosen, then appropriately ordered monomials constructed from the Cartan--Weyl generators form a Poincar\'e--Birkhoff--Witt basis of $\uqlg$. In fact, in applications to the theory of quantum integrable systems one uses the so-called normal orderings. An example of such an ordering for $\gothg = \sllpo$ is described in section \ref{ss:pbwb}. We denote the root vector corresponding to a real positive root $\gamma \in \hDelta$ by $e_\gamma$, and the root vector corresponding to a real negative root $\gamma \in \hDelta$ by $f_{-\gamma}$. In fact, the root vectors corresponding to the imaginary roots $m \delta$ and $- m \delta$, $m > 0$, are indexed by the simple roots of $\gothg$ and denoted as $e'_{m \delta; \, \alpha_i}$ and $f'_{m \delta; \, \alpha_i}$. The prime in the notation is explained by the fact that one also uses another set of root vectors corresponding to imaginary roots. They are introduced by the functional equations
\begin{equation}
- \kappa_q \, e_{\delta; \, \gamma}(u) = \log(1 - \kappa_q \, e'_{\delta; \, \gamma}(u)), \qquad \kappa_q \, f_{\delta; \,  \gamma}(u^{-1}) = \log(1 + \kappa_q \, f'_{\delta; \, \gamma}(u^{-1})), \label{efdg}
\end{equation}
where the generating functions
\begin{align*}
& e'_{\delta; \, \gamma}(u) = \sum_{n = 1}^\infty e'_{n \delta; \, \gamma} \, u^n \in \uqlg[[u]], && f'_{\delta, \, \gamma}(u^{-1}) = \sum_{n = 1}^\infty f'_{n \delta; \, \gamma} \, u^{- n} \in \uqlg[[u^{-1}]], \\
& e_{\delta; \, \gamma}(u) 
= \sum_{n = 1}^\infty e_{n \delta; \, \gamma} \, u^n \in \uqlg[[u]], && f_{\delta; \, \gamma}(u^{-1}) = \sum_{n = 1}^\infty f_{n \delta; \, \gamma} \, u^{- n} \in \uqlg[[u^{-1}]]
\end{align*}
are defined as formal power series.

\subsection{Spectral parameters}

In applications to the theory of quantum integrable systems, one usually considers families of representations of a quantum loop algebra and its subalgebras, parameterized by a complex parameter called a spectral parameter. We introduce a spectral parameter in the following way. Assume that a quantum loop algebra $\uqlg$ is $\bbZ$-graded,
\begin{equation*}
\uqlg = \bigoplus_{m \in \bbZ} \uqlg_m, \qquad \uqlg_m \, \uqlg_n \subset \uqlg_{m + n},
\end{equation*}
so that any element $x \in \uqlg$ can be uniquely represented as
\begin{equation*}
x = \sum_{m \in \bbZ} x_m, \qquad x_m \in \uqlg_m.
\end{equation*}
Given $\zeta \in \bbC^\times$, we define the grading automorphism $\Gamma_\zeta$ by the equation
\begin{equation*}
\Gamma_\zeta(x) = \sum_{m \in \bbZ} \zeta^m x_m.
\end{equation*}
Now, for any representation $\varphi$ of $\uqlg$ we define the family $\varphi_\zeta$ of representations as
\begin{equation*}
\varphi_\zeta = \varphi \circ \Gamma_\zeta.
\end{equation*}
If $V$ is the $\uqlg$-module corresponding to the representation $\varphi$, we denote by $V_\zeta$ the $\uqlg$-module corresponding to the representation $\varphi_\zeta$.

In the present paper we endow $\uqlg$ with a $\bbZ$-gradation assuming that
\begin{equation*}
q^h \in \uqlg_0, \qquad e_i \in \uqlg_{s_i}, \qquad f_i \in \uqlg_{-s_i},
\end{equation*}
where $s_i$ are arbitrary integers. We denote
\begin{equation*}
s = \sum_{i = 0}^l a_i s_i,
\end{equation*}
where $a_i$ are the Kac labels of the Dynkin diagram associated with the extended Cartan matrix $A^{(1)}$.

\subsection{Drinfeld's second realization} \label{s:2dr}

The quantum loop algebra $\uqlg$ can be realized in a different way \cite{Dri87, Dri88, Bec94b} as an algebra with generators $\xi^\pm_{i, \, m}$ with $i = 1, \ldots, l$ and $m \in \bbZ$, $q^x$ with $x \in \gothh$, and $\chi_{i, \, m}$ with $i = 1, \ldots, l$ and $m \in \bbZ \setminus \{0\}$. They satisfy the defining relations
\begin{gather*}
q^0 = 1, \qquad q^{x_1} q^{x_2} = q^{x_1 + x_2}, \\
[q^x, \, \chi_{j, \, m}] = 0, \qquad [\chi^{\mathstrut}_{i, \, m}, \, \chi^{\mathstrut}_{j, \, n}] = 0, \\
q^x \xi^\pm_{i, \, m} q^{- x} = q^{\pm \langle \alpha_i, \, x \rangle} \xi^\pm_{i, \, m}, \qquad [\chi^{\mathstrut}_{i, \, m}, \, \xi^\pm_{j, n}] = \pm \frac{1}{m} [m \, a_{i j}]_{q^i} \, \xi^\pm_{j, \, m + n}, \\
\xi^\pm_{i, \, m + 1} \xi^\pm_{j, \, n} - q^{\pm a_{i j}}_i \, \xi^\pm_{j, \, n} \, \xi^\pm_{i, \, m + 1} = q^{\pm a_{i j}}_i \, \xi^\pm_{i, \, m} \, \xi^\pm_{j, \, n + 1} - \xi^\pm_{j, \, n + 1} \xi^\pm_{i, \, m}, \\
[\xi^+_{i, \, m}, \, \xi^-_{j, \, n}] = \delta_{i j} \, \frac{q^{h_i}_i \phi^+_{i, \, m + n} - q^{-h_i}_i \phi^-_{i, \, m + n}}{q^{}_i - q^{-1}_i},
\end{gather*}
and the Serre relations whose explicit form is not important for our consideration. The quantities $\phi^\pm_{i, \, \pm m}$, $i = 1, \ldots, l$, $m \in \bbZ$, are defined by the equation
\begin{equation}
1 + \sum_{m = 1}^\infty \phi^\pm_{i, \, \pm m} u^{\pm m} = \exp \left( \pm \kappa_q \sum_{m = 1}^\infty \chi_{i, \, \pm m} u^{\pm m} \right) \label{phipm}
\end{equation}
and by the conditions
\begin{gather*}
\phi^+_{i, \, 0} = 1, \qquad \phi^-_{i, \, 0} = 1, \\
\phi^+_{i, \, m} = 0, \quad m < 0, \qquad \phi^-_{i, \, m} = 0, \quad m > 0.
\end{gather*}
Stress that we use the definition of $\phi^\pm_{i, \, m}$ slightly different from the commonly used.

The generators of Drinfeld's second realization are related to the Cartan--Weyl generators in the following way \cite{KhoTol93, KhoTol94}. The generators $q^x$ of the quantum loop algebra in the Drinfeld--Jimbo's and Drinfeld's second realizations are the same, except that in the first case $x \in \widetilde \gothh$, while in the second case $x \in \gothh \subset \widetilde \gothh$. For the generators $\xi^\pm_{i, \, m}$ and $\chi_{i, \, m}$ of the Drinfeld's second realization we have
\begin{align}
& \xi^+_{i, \, m} = o_i^m e_{\alpha_i + m \delta}, \quad m \ge 0, && \xi^+_{i, \, m} = - o_i^m q^{-h_i}_i f_{(\delta - \alpha_i) - (m + 1)\delta}, \quad  m < 0, \label{ksipin} \\
& \xi^-_{i, \, m} = - o_i^m e_{(\delta - \alpha_i) + (m - 1) \delta} \, q^{h_i}_i, \quad m > 0, && \xi^-_{i, \, m} = o_i^m f_{\alpha_i - m \delta}, \quad  m \le 0, \label{ksimin} \\
& \chi_{i, \, m} = - o_i^m e_{m \delta; \, \alpha_i}, \quad  m > 0, && \chi_{i, \, m} = - o_i^m f_{- n\delta; \, \alpha_i}, \quad m < 0, \label{chiin}
\end{align}
where for each $i = 1, \ldots, l$ the number $o_i$ is either $+1$ or $-1$, such that $o_i = - o_j$ 
whenever $a_{i j} < 0$. It follows from (\ref{efdg}), (\ref{phipm}) and (\ref{chiin}) that
\begin{equation*}
\phi^+_{i, \, m} = - \kappa_q \, o_i^m e'_{m \delta; \, \alpha_i}, \qquad \phi^-_{i, \, m} = \kappa_q \, o_i^m f'_{- m \delta; \, \alpha_i}.
\end{equation*}
Defining the generating functions $\phi^+_i(u)$ and $\phi^-_i(u)$ as
\begin{equation*}
\phi^+_i(u) = 1 + \sum_{m = 1}^\infty \phi^+_{i, \, m} u^m, \qquad \phi^-_i(u^{-1}) = 1 + \sum_{m = 1}^\infty \phi^-_{i, \, -m} u^{- m},
\end{equation*}
we also obtain
\begin{equation}
\phi^+_i(u) = 1 - \kappa_q \, e'_{\delta; \, \alpha_i}(o_i u), \qquad \phi^-_i(u^{-1}) = 1 + \kappa_q \, f'_{\delta; \, \alpha_i}(o_i u^{-1}). \label{phiiu}
\end{equation}

\subsection{\texorpdfstring{Poincar\'e--Birkhoff--Witt basis for $\uqlsllpo$}{Poincar\'e--Birkhoff--Witt basis for Uq(L(sll+1))}} \label{ss:pbwb}

In this article, we often give general definitions for general $\gothg$, however, concrete results are obtained for the case of the loop algebra $\uqlsllpo$. In this case $d_i =1$,  $a_i = 1$, and $a \char20 \hskip -.3em_i = 1$, for all $0 \le i \le l$, and we assume that $o_i = (-1)^i$.\footnote{We use slightly different correspondence between root vectors and the generators of the Drinfeld's second realization and the choice of $o_i$ for $\uqlsllpo$ from those used in the papers \cite{BooGoeKluNirRaz16, BooGoeKluNirRaz17b}. In fact, despite the changes made, the expressions for the $\ell$-weights from \cite{BooGoeKluNirRaz16, BooGoeKluNirRaz17b} remain valid in our case.}

To define an appropriate ordering of positive roots of $\uqlsllpo$, we start with the normal ordering of the positive roots of $\sllpo$ defined in subsection \ref{s:pbwb}. Then we assume that $\alpha + m \delta \preccurlyeq \beta + n \delta$, with $\alpha, \beta \in \Delta_+$ and $m, n \in \bbZ_{\ge 0}$, if $\alpha \preccurlyeq \beta$, or $\alpha = \beta$ and $m < n$. Further, $(\delta - \alpha) + m \delta \preccurlyeq (\delta - \beta) + n \delta$, with $\alpha, \beta \in \Delta_+$, if $\alpha \preccurlyeq \beta$, or $\alpha = \beta$ and $m > n$. Finally, we assume that the relation
\begin{equation*}
\alpha + k \delta \preccurlyeq m \delta \preccurlyeq (\delta - \beta) + n \delta \label{akd}
\end{equation*}
is valid. As the result we come to a normal ordering of positive roots of $\uqlsllpo$. For more details see the paper \cite{NirRaz19}.

Following \cite{TolKho92, KhoTol92, KhoTol93, KhoTol94}, we define the root vectors inductively . In the case under consideration the procedure looks as follows. We start with the root vectors corresponding to the roots $\alpha_i$ and $- \alpha_i$ for $1 \le i \le l$, which we identify with the generators $e_i$ and $f_i$ so that
\begin{equation*}
e_{\alpha_i} = e_{\alpha_{i, \, i + 1}} = e_i, \qquad f_{\alpha_i} = f_{\alpha_{i, \, i + 1}} = f_i.
\end{equation*}
The next step is to construct root vectors $e_{\alpha_{i j}}$ and $f_{\alpha_{i j}}$ for all roots $\alpha_{i j} \in \Delta_+$. We assume that
\begin{equation*}
e_{\alpha_{i j}} = e_{\alpha_{i, \, j - 1}} e_{\alpha_{j - 1, \, j}} - q \, e_{\alpha_{j - 1, \, j}} e_{\alpha_{i, \, j - 1}}, \qquad f_{\alpha_{i j}} = f_{\alpha_{j - 1, \, j}} f_{\alpha_{i, \, j - 1}} - q^{-1} f_{\alpha_{i, \, j - 1}} f_{\alpha_{j - 1, \, j}}
\end{equation*}
for $j - i > 1$. Further, taking into account that
\begin{equation*}
\alpha_0 = \delta - \theta = \delta - \alpha_{1, \, l + 1},
\end{equation*}
we put
\begin{equation*}
e_{\delta - \alpha_{1, \, l + 1}} = e_0, \qquad f_{\delta - \alpha_{1, \, l + 1}} = f_0,
\end{equation*}
and define
\begin{align}
& e_{\delta - \alpha_{i, \, l + 1}} = e_{\alpha_{i - 1, \, i}} e_{\delta - \alpha_{i - 1, \, l + 1}} - q \, e_{\delta - \alpha_{i - 1, \, l + 1}} e_{\alpha_{i - 1, \, i}}, \label{edmaa} \\
& f_{\delta - \alpha_{i, l + 1}} = f_{\delta - \alpha_{i - 1, \, l + 1}} f_{\alpha_{i - 1, \, i}}  - q^{-1} f_{\alpha_{i - 1, \, i}} f_{\delta - \alpha_{i - 1, \, l + 1}}
\end{align}
for $i > 1$, and then
\begin{align}
& e_{\delta - \alpha_{i j}} = e_{\alpha_{j, \, j + 1}} e_{\delta - \alpha_{i, \, j + 1}} - q \, e_{\delta - \alpha_{i, \, j + 1}} e_{\alpha_{j, \, j + 1}}, \label{edmab} \\
& f_{\delta - \alpha_{i j}} = f_{\delta - \alpha_{i, \, j + 1}} f_{\alpha_{j, \, j + 1}}  - q^{-1} f_{\alpha_{j, \, j + 1}} f_{\delta - \alpha_{i, \, j + 1}}
\end{align}
for $j < l + 1$. The root vectors corresponding to the roots $\delta$ and $-\delta$ are indexed by the elements of $\Delta_+$ and defined by the relations
\begin{equation*}
e'_{\delta; \, \alpha_{i j}} = e_{\alpha_{i j}} e_{\delta - \alpha_{i j}} - q^2  e_{\delta - \alpha_{i j}} e_{\alpha_{i j}}, \qquad f'_{\delta; \, \alpha_{i j}} = f_{\delta - \alpha_{i j}} f_{\alpha_{i j}} - q^{- 2}  f_{\alpha_{i j}} f_{\delta - \alpha_{i j}}.
\end{equation*}
The remaining definitions are
\begin{align}
e_{\alpha_{i j} + n \delta} & = [2]_q^{-1} \big( e_{\alpha_{i j} + (n - 1)\delta} \, e'_{\delta, \, \alpha_{i j}} - e'_{\delta, \, \alpha_{i j}} e_{\alpha_{i j} + (n - 1)\delta} \big), \label{cwby1} \\
e_{(\delta - \alpha_{i j}) + n \delta} & = [2]_q^{-1} \big( e'_{\delta, \, \alpha_{i j}} e_{(\delta - \alpha_{i j}) + (n - 1)\delta} - e_{(\delta - \alpha_{i j}) + (n - 1)\delta} \, e'_{\delta, \, \alpha_{i j}} \big), \label{cwby3} \\
e'_{n \delta; \, \alpha_{i j}} & = e_{\alpha_{i j} + (n - 1)\delta} e_{\delta - \alpha_{i j}} - q^2  e_{\delta - \alpha_{i j}} e_{\alpha_{i j} + (n - 1)\delta}, \label{cwby5} \\
f_{\alpha_{i j} + n \delta} & = [2]_q^{-1} \big( f'_{\delta, \, \alpha_{i j}} f_{\alpha_{i j} + (n - 1)\delta} -  f_{\alpha_{i j} + (n - 1)\delta} \, f'_{\delta, \, \alpha_{i j}} \big), \label{cwby2} \\
f_{(\delta - \alpha_{i j}) + n \delta} & = [2]_q^{-1} \big( f_{(\delta - \alpha_{i j}) + (n - 1)\delta} f'_{\delta, \, \alpha_{i j}} -  f'_{\delta, \, \alpha_{i j}} f_{(\delta - \alpha_{i j}) + (n - 1)\delta} \big), \label{cwby4 } \\
f'_{n \delta; \, \alpha_{i j}} & = f_{\delta - \alpha_{i j}} f_{\alpha_{i j} + (n - 1)\delta} - q^{- 2}  f_{\alpha_{i j} + (n - 1)\delta} f_{\delta - \alpha_{i j}}. \label{cwby6}
\end{align}
It can be seen that we label the root vectors corresponding to the imaginary roots by all roots of $\sllpo$ and not only by the simple ones. This is, of course, redundant for building a Poincar\'e--Birkhoff--Witt basis, but useful as a technical tool.

\section{\texorpdfstring{Highest $\ell$-weight representations}{Highest l-weight representations}}

\subsection{\texorpdfstring{$\ell$-weights of $\uqlg$-modules}{l-weights of UqL(g)-modules}} \label{ss:lw}

A $\uqlg$-module $V$ is called a weight module if
\begin{equation}
V = \bigoplus_{\lambda \in \gothh^*}  V_\lambda, \label{vvl}
\end{equation}
where
\begin{equation*}
V_\lambda = \{v \in V \mid q^x v = q^{\langle \widetilde \lambda, \, x} \rangle v \mbox{ for any } x \in \tgothh \}.
\end{equation*}
The meaning of the element $\widetilde \lambda \in \tgothh^*$ for $\lambda \in \gothh^*$ is explained in section \ref{ss:dtjmg}. The space $V_\lambda$ is called the weight space of weight $\lambda$, and a nonzero element of $V_\lambda$ is called a weight vector of weight $\lambda$. We say that $\lambda \in \gothh^*$ is a weight of $V$ if $V_\lambda \ne \{ 0 \}$.

A $\uqlg$-module $V$ is said to be in category $\calO$ if
\begin{itemize}
\item[(i)] $V$ is a weight module all of whose weight spaces are finite-dimensional;
\item[(ii)] there exists a finite number of elements $\mu_1, \ldots, \mu_s \in \gothh^*$ such that every weight of $V$ belongs to the set $\bigcup_{i = 1}^s D(\mu_i)$ , where $D(\mu) = \{\lambda \in \gothh^* \mid \lambda \leq \mu \}$ with $\leq$ being the usual partial order in $\gothh^*$, see, for example, the book \cite{Hum80}.
\end{itemize}

Let a $\uqlg$-module $V$ be in category $\calO$. The algebra $\uqlg$ contains an infinite-dimensional commutative subalgebra generated by the elements $\phi^\pm_{i, \, \pm m}$, $i = 1, \ldots, l$, $m \in \bbZ_{>0}$ and $q^h$, $h \in \widetilde \gothh$. We can refine the weight decomposition (\ref{vvl}) in the following way. Let $\lambda$ be a weight of $V$. By definition, the space $V_\lambda$ is finite-dimensional. The restriction of the action of the elements $\phi^\pm_{i, \, \pm m}$ to $V_\lambda$ constitutes a countable set of pairwise commuting linear operators on $V_\lambda$. Hence, there is a basis of $V_\lambda$ which consists of eigenvectors and generalized eigenvectors of all those operators, see, for example, the book \cite{Lax07}. This leads to the following definitions.

An $\ell$-weight is a triple
\begin{equation*}
\bm \Lambda = (\lambda, \,  \bm \Lambda^+ \! , \, \bm \Lambda^-),
\end{equation*}
where $\lambda \in \gothh^*$, $\bm \Lambda^+ = (\Lambda^+_i(u))^l_{i = 1}$ and $\bm \Lambda^- = (\Lambda^-_i(u^{-1}))^l_{i = 1}$ are $l$-tuples of formal power series
\begin{equation*}
\Lambda^+_i(u) = 1 + \sum_{m \in \bbZ_{> 0}} \Lambda^+_{i, \, m} u^m \in \bbC[[u]], \qquad \Lambda^-_i(u^{-1}) = 1 + \sum_{m \in \bbZ_{> 0}} \Lambda^-_{i, \, - m} u^{- n} \in \bbC[[u^{-1}]].
\end{equation*}
We denote the set of $\ell$-weights by $\gothh^*_\ell$ .

Define a surjective homomorphism $\varpi \colon \gothh^*_\ell \to \gothh^*$ by the relation
\begin{equation*}
\varpi(\bm \Lambda) = \lambda
\end{equation*}
if $\bm \Lambda = (\lambda, \, \bm \Lambda^+ \! , \, \bm \Lambda^-)$. Now we have
\begin{equation*}
V_\lambda = \bigoplus_{\varpi(\bm \Lambda) = \lambda} V_{\bm \Lambda},
\end{equation*}
where $V_{\bm \Lambda}$ is the subspace of $V_\lambda$ such that for any $v \in V_{\bm \Lambda}$ there is $p \in \bbZ_{> 0}$ such that
\begin{equation*}
(\phi^+_{i, \, m} - \Lambda^+_{i, \, m})^p v = 0, \qquad (\phi^-_{i, \, - m} - \Lambda^-_{i, \, - m})^p v = 0
\end{equation*}
for all $i = 1, \ldots, l$ and $m \in \bbZ_{> 0}$. The space $V_{\bm \Lambda}$ is called the $\ell$-weight space of $\ell$-weight $\bm \Lambda$. We say that $\bm \Lambda$ is an $\ell$-weight of $V$ if $V_{\bm \Lambda} \ne \{0\}$. A nonzero element $v \in V_{\bm \Lambda}$ such that
\begin{equation*}
\phi^+_{i, \, m} v = \Lambda^+_{i, \, m} v, \qquad \phi^-_{i, \, - m} v = \Lambda^-_{i, \, - m} v
\end{equation*}
for any $i = 1, \ldots, l$ and $m \in \bbZ_{> 0}$ is said to be an $\ell$-weight vector of $\ell$-weight $\bm \Lambda$. Every nontrivial $\ell$-weight space contains an $\ell$-weight vector.

For any two $\ell$-weights $\bm \Lambda = (\lambda, \, \bm \Lambda^+ \! , \, \bm \Lambda^-)$ and $\bm \Xi = (\xi, \, \bm \Xi^+ \! , \, \bm \Xi^-)$ we define the $\ell$-weight $\bm \Lambda \, \bm \Xi$ as
\begin{equation}
\bm \Lambda \, \bm \Xi = (\lambda + \xi, \, (\bm \Lambda \, \bm \Xi)^+, \, (\bm \Lambda \, \bm \Xi)^-), \label{lm}
\end{equation}
where
\begin{equation*}
(\bm \Lambda \, \bm \Xi)^+ = (\Lambda^+_i(u) \Xi^+_i(u))_{i = 1}^l, \qquad (\bm \Lambda \, \bm \Xi)^- = (\Lambda^-_i(u^{-1}) \Xi^-_i(u^{-1}))_{i = 1}^l.
\end{equation*}
The product (\ref{lm}) is an associative operation with respect to which $\gothh^*_\ell$ is an abelian group. Here
\begin{equation*}
\bm \Lambda^{-1} = (- \lambda, (\bm \Lambda^+)^{-1}, \, (\bm \Lambda^-)^{-1}),
\end{equation*}
where
\begin{equation*}
(\bm \Lambda^+)^{-1} = (\Lambda^+_i(u)^{-1})^l_{i = 1}, \qquad (\bm \Lambda^-)^{-1} = (\Lambda^-_i(u)^{-1})^l_{i = 1}.
\end{equation*}
and the role of the unity is played by the $\ell$-weight $(0, \, (\underbracket[.5pt]{1, \, \ldots, \, 1}_l), \, (\underbracket[.5pt]{1, \, \ldots, \, 1}_l))$. Note that by definition each $\Lambda_i^+(u)$ and each $\Lambda_i^-(u^{-1})$ is an invertible power series.

A $\uqlg$-module $V$ in category $\calO$ is called a highest $\ell$-weight module of highest $\ell$-weight $\bm \Lambda$ if there exists an $\ell$-weight vector $v \in V$ of $\ell$-weight $\bm \Lambda$ such that
\begin{equation*}
e_i \, v = 0, \quad i = 1, \ldots, l, \quad \mbox{and} \quad V = \uqlg \, v.
\end{equation*}
The vector with the above properties is unique up to a scalar factor. We call it the  highest $\ell$-weight vector of $V$.

Let $V$ and $W$ be highest $\ell$-weight $\uqlg$-modules in category $\calO$ of highest $\ell$-weights $\bm \Lambda$ and $\bm \Xi$, respectively. The submodule of $V \otimes_\Delta W$ generated by the tensor product of the highest $\ell$-weight vectors is a highest $\ell$-weight module of highest $\ell$-weight $\bm \Lambda \, \bm \Xi$.

\subsection{\texorpdfstring{Evaluation representations}{Evaluation representations}} \label{s:lwer}

To construct representations of the quantum loop algebra $\uqlsllpo$, it is common to use the Jimbo's homomorphism $\varepsilon$ from $\uqlsllpo$ to the quantum group $\uqgllpo$ defined by the equations \cite{Jim86a}
\begin{align*}
& \varepsilon(q^{\nu h_0}) = q^{\nu (K_{l+1} - K_1)}, &&  \varepsilon(q^{\nu h_i}) = q^{\nu (K_{i} - K_{i+1})}, \\
& \varepsilon(e_0) = F_{1, \, l + 1} \, q^{K_1 + K_{l+1}}, && \varepsilon(e_i) = E_{i, \, i + 1}, \\
& \varepsilon(f_0) = E_{1, \, l + 1} \, q^{-K_1 - K_{l+1}}, && \varepsilon(f_i) = F_{i, \, i + 1},  
\end{align*}
where $i = 1, \ldots, l$. If $\pi$ is a representation of $\uqgllpo$, then $\pi \circ \varepsilon$ is a representation of $\uqlsllpo$ called an evaluation representation. Given $\mu \in \gothk^*$, we denote the representation $\widetilde \pi^\mu \circ \varepsilon$, where the representation  $\widetilde \pi^\mu$ is described in section \ref{s:vm}, by $\widetilde \varphi^\mu$. Since the representation space of $\widetilde \pi^\mu$ and $\widetilde \varphi^\mu$ is the same, slightly abusing notation, we denote both corresponding modules as $\widetilde V^\mu$. The same convention is used for the respective finite-dimensional counterparts, if any. It is common to call $\widetilde V^\mu$ an evaluation module. It is a highest weight $\uqlsllpo$-module in category $\calO$ of highest weight being the restriction of $\mu$ to $\gothh$.

For any $\mu$ in $\gothh^*$ the module $\widetilde V^\mu$ is a highest $\ell$-weight module in category $\calO$ with highest $\ell$-weight vector $v_{\bm 0}^\mu$. The $\ell$-weight spaces and the corresponding $\ell$-weights for $l = 1$ and $l = 2$ was found in the paper \cite{BooGoeKluNirRaz16}. It appears that in this case all $\ell$-weight spaces are one-dimensional and, therefore, generated by $\ell$-weight vectors.  Although the $\ell$-weight vectors do not coincide with the basic vectors $v^\mu_{\bm m}$, defined by equation (\ref{vml}), they can also be labeled by the $(l + 1)l/2$-tuple $\bm m$. It is natural to assume that this is the case for an arbitrary $l$, and we use for the $\ell$-weights the notation
\begin{equation*}
\bm \Lambda^\mu_{\bm m}(\zeta) = (\lambda^\mu_{\bm m}, \, \bm \Lambda_{\bm m}^{\mu +}(\zeta), \, \bm \Lambda_{\bm m}^{\mu -}(\zeta)).
\end{equation*}
It was found in the paper \cite{NirRaz17b} that for an arbitrary $l$,
\begin{equation}
\lambda^\mu_{\bm m} = \sum_{i = 1}^l \big( \mu_i - \mu_{i + 1} + \sum_{k = 1}^{i - 1} (m_{k, \, i} - m_{k, \, i + 1}) - 2 m_{i, \, i+1} - \sum_{k = i + 2}^{l + 1} (m_{i \, k} - m_{i + 1, \, k}) \big) \omega_i. \label{psimum}
\end{equation}
In particular, for the highest weight we have
\begin{equation}
\lambda^\mu_{\bm 0} = \sum_{i = 1}^l (\mu_i - \mu_{i + 1}) \omega_i. \label{slhw}
\end{equation}
Using formulas of the paper \cite{BooGoeKluNirRaz16}, one can demonstrate that for the components of $\bm \Lambda_{\bm 0}^{\mu +}(\zeta)$ and $\bm \Lambda_{\bm 0}^{\mu -}(\zeta)$, $i = 1, \ldots, l$, corresponding to the highest $\ell$-weight vector $v_{\bm 0}^\mu$ of $(\widetilde V^\mu)_\zeta$ one has the expression
\begin{equation}
\Lambda_{\bm 0, \, i}^{\mu +}(\zeta, u) = \frac{1 - q^{2 \mu_{i + 1} - i + 1} \zeta^s u} {1 - q^{2 \mu_i - i + 1} \zeta^s u},  \qquad \Lambda_{\bm 0, \, i}^{\mu -}(\zeta, u) = \frac{1 - q^{- 2 \mu_{i + 1} + i - 1} \zeta^{-s} u^{-1}} {1 - q^{-2 \mu_i + i - 1} \zeta^{-s} u^{-1}}, \label{slhlw}
\end{equation}
see appendix \ref{a:hwfr}.

Below we use the $(l + 1)$-dimensional representation $\varphi^{\omega_1}_\zeta$ on the vector space $\bbC^{l + 1}$. For this representation we have
\begin{align}
& \varphi^{\omega_1}_\zeta(q^{\nu h_0}) = q^{-\nu} \bbE_{1 1} + q^\nu \bbE_{l + 1, \, l + 1} + \sum_{m = 2}^l \bbE_{m m}, \label{fooa} \\
& \varphi^{\omega_1}_\zeta(q^{\nu h_i}) = q^\nu \bbE_{i i} +  q^{-\nu} \bbE_{i + 1, \, i + 1} + \sum_{\substack{m = 1 \\  m \ne i, \, i + 1}}^{l + 1} \bbE_{m m}, \qquad i = 1, \ldots l,
\end{align}
and, further,
\begin{align}
& \varphi^{\omega_1}_\zeta(e_0) = \zeta^{s_0} q \, \bbE_{l + 1, \, 1}, &&  \varphi^{\omega_1}_\zeta(e_i) = \zeta^{s_i} \bbE_{i, \, i + 1}, && i = 1, \ldots, l, \\
& \varphi^{\omega_1}_\zeta(f_0) = \zeta^{- s_0} q^{-1} \bbE_{1, \, l + 1}, && \varphi^{\omega_1}_\zeta(f_i) = \zeta^{-s_i} \bbE_{i + 1, \, i}, && i = 1, \ldots, l. \label{food}
\end{align}
Let $v_m$, $m = 1, \ldots, l + 1$, be the elements of the standard basis of $\bbC^{l + 1}$. It is clear that
\begin{equation*}
v_m = f_{m -1} \ldots f_1 v_1
\end{equation*}
for any $m = 2, \ldots, l + 1$. We see that the vector $v_m$ is a weight vector of weight $\lambda_m$ being the restriction of the element $\mu_m \in \gothk^*$ defined as
\begin{equation*}
\mu_m = \omega_1 - \sum_{i = 1}^{m - 1} \alpha_i 
\end{equation*}
to the Cartan subalgebra $\gothh$ of $\sllpo$. Explicitly, we have
\begin{equation}
\lambda_1 = \omega_1, \quad \lambda_2 = \omega_2 - \omega_1, \quad \ldots, \quad \lambda_l = \omega_l - \omega_{l - 1}, \quad \lambda_{l + 1} = - \omega_l. \label{lmoms}
\end{equation}

\subsection{\texorpdfstring{$\ell$-weights of $\uqlbp$-modules}{l-weights of UqL(b+)-modules}}

There are two standard Borel subalgebras of the quantum loop algebra $\uqlg$. In terms of the Drinfeld--Jumbo generators they are defined as follows. The Borel subalgebra $\uqlbp$ is defined as the subalgebra of $\uqlg$ generated by $e^{}_i$ with $0 \le i \le l$ and $q^x$ with $x \in \tgothh$, and the Borel subalgebra $\uqlbm$ is the subalgebra generated by $f^{}_i$ with $0 \le i \le l$ and $q^x$ with $x \in \tgothh$. It is clear that these subalgebras are Hopf subalgebras of $\uqlg$. The description of $\uqlbp$ and $\uqlbm$ in terms of the Drinfeld generators is more intricate.

The category $\calO$ for $\uqlbp$ or $\uqlbm$ is defined literally by the same words as it is defined in section \ref{ss:lw} for the case of $\uqlg$. By definition, any module $V$ in category $\calO$ allows for the weight decomposition
\begin{equation}
V = \bigoplus_{\lambda \in \gothh^*}  V_\lambda, \label{wwp}
\end{equation}
which can be refined again by considering $\ell$-weights.

Consider the Borel subalgebra $\uqlbp$. It does not contain the elements $\phi^-_{i, \, - m}$, $i = 1, \ldots, l$, $m \in \bbZ_{>0}$, and its infinite-dimensional commutative subalgebra is generated only by the elements $\phi^+_{i, \, m}$ $i = 1, \ldots, l$, $m \in \bbZ_{>0}$, and $q^h$, $h \in \widetilde \gothh$. Respectively, now an $\ell$-weight $\bm \Lambda$ is a pair
\begin{equation}
\bm \Lambda = (\lambda, \, \bm \Lambda^+), \label{llp}
\end{equation}
where $\lambda \in \gothh^*$ and $\bm \Lambda^+$ is an $l$-tuple $ \bm \Lambda^+ = (\Lambda^+_i(u))_{i = 1}^l$ of formal series
\begin{equation*}
\Lambda^+(u) = 1 + \sum_{n = 1}^\infty \Lambda^+_{i, \, n} \, u^n \in \bbC[[u]].
\end{equation*}
We denote the set of $\ell$-weights (\ref{llp}) by $\gothh^{* +}_\ell$.

For any two $\ell$-weights $\bm \Lambda = (\psi, \, \bm \Lambda^+)$ and $\bm \Xi = (\xi, \, \bm \Xi^+)$ we define the $\ell$-weight $\bm \Lambda \, \bm \Xi$ as
\begin{equation}
\bm \Lambda \, \bm \Xi = (\lambda + \xi, \, (\bm \Lambda \, \bm \Xi)^+), \label{pk}
\end{equation}
where
\begin{equation*}
(\bm \Lambda \, \bm \Xi)^+ = (\Lambda^+_i(u) \Xi^+_i(u))_{i = 1}^l.
\end{equation*}
The product (\ref{pk}) is an associative operation with respect to which $\gothh^{* +}_\ell$ is an abelian group. Here
\begin{equation*}
\bm \Lambda^{-1} = (- \lambda, (\bm \Lambda^+)^{-1}),
\end{equation*}
where
\begin{equation*}
(\bm \Lambda^+)^{-1} = (\Lambda^+_i(u)^{-1})_{i = 1}^l,
\end{equation*}
and the role of the unity is played by the $\ell$-weight $(0, \, (\underbracket[.5pt]{1, \, \ldots, \, 1}_l))$. Note that each $\Lambda_i^+(u)$ is an invertible power series.

Define a surjective homomorphism $\varpi \colon \gothh^{* +}_\ell \to \gothh^*$ by the relation
\begin{equation*}
\varpi(\bm \Lambda) = \lambda
\end{equation*}
if $\bm \Lambda = (\lambda, \, \bm \Lambda^+)$. Now for any $V_\lambda$ entering the decomposition (\ref{wwp}) we have
\begin{equation*}
V_\lambda = \bigoplus_{\varpi(\bm \Lambda) = \lambda} V_{\bm \Lambda},
\end{equation*}
where $V_{\bm \Lambda}$ is the subspace of $V_\lambda$ such that for any $v$ in $V_{\bm \Lambda}$ there is $p \in \bbZ_{> 0}$ such that
\begin{equation*}
(\phi^+_{i, \, m} - \Lambda^+_{i, \, m})^p w = 0
\end{equation*}
for all $i = 1, \ldots, l$ and $m \in \bbZ_{> 0}$. The space $V_{\bm \Lambda}$ is called the $\ell$-weight space of $\ell$-weight $\bm \Lambda$. We say that $\bm \Lambda$ is an $\ell$-weight of $V$ if $V_{\bm \Lambda} \ne \{0\}$. A nonzero element $v \in V_{\bm \Lambda}$ such that
\begin{equation*}
\phi^+_{i, \, m} v = \Lambda^+_{i, \, m} v
\end{equation*}
for all $i = 1, \ldots, l$ and $m \in \bbZ_{> 0}$ is said to be an $\ell$-weight vector of $\ell$-weight $\bm \Lambda$. Every nontrivial $\ell$-weight space contains an $\ell$-weight vector.

A $\uqlbp$-module $V$ in is called a highest $\ell$-weight module of highest $\ell$-weight $\bm \Lambda$ if there exists an $\ell$-weight vector $v \in V$ of $\ell$-weight $\bm \Lambda$ such that
\begin{equation*}
e_i \, v = 0, \quad i = 1, \ldots, l, \quad \mbox{and} \quad V = \uqlbp \, v.
\end{equation*}
The vector with the above properties is unique up to a scalar factor. We call it the  highest $\ell$-weight vector of $V$. Let $V$ and $W$ be highest $\ell$-weight $\uqlbp$-modules in category $\calO$ of highest $\ell$-weights $\bm \Lambda$ and $\bm \Xi$ respectively. The submodule of $V \otimes_\Delta W$ generated by the tensor product of the highest $\ell$-weight vectors is a highest $\ell$-weight module of highest $\ell$-weight $\bm \Lambda \, \bm \Xi$.

For any $\uqlbp$-module $V$ in category $\calO$ and an element $\delta \in \gothh^*$, we define the shifted $\uqlbp$-module $V[\delta]$ shifting the action of the generators $q^h$. Namely, if $\varphi$ is the representation of $\uqlbp$ corresponding to the module $V$ and $\varphi[\xi]$ is the representation corresponding to the module $V[\xi]$, then
\begin{equation}
\varphi[\delta](e_i) = \varphi(e_i), \quad i = 1, \ldots, l, \qquad \varphi[\delta](q^x) = q^{\langle \widetilde \delta, 
\, x \rangle} \varphi(q^x), \quad x \in \tgothh. \label{phixi}
\end{equation}
For an arbitrary $\uqlg$-module $V$ we define the shifted module $V[\delta]$, $\delta \in \gothh^*$, as a $\uqlbp$-module obtained first by restricting $V$ to $\uqlbp$ and then shifting the obtained $\uqlbp$-module. It is clear that the module $V[\delta]$ is in category $\calO$. If $\bm \Lambda = (\lambda, \, \bm \Lambda^+)$ is an $\ell$-weight of the $\uqlbp$-module $V$, then $\bm \Lambda[\delta] = (\lambda + \delta, \, \bm \Lambda^+)$ is an $\ell$-weight of the module $V[\delta]$. 

\subsection{Oscillator representations}

One can get representations of $\uqlbp$ by restricting to it the representations of $\uqlsllpo$. However, the most interesting for theory of integrable systems are the representations which cannot be obtained by such a procedure. In the present paper we use a class of such representations called the $q$-oscillator representations. We define the $q$-oscillator representations in two steps. First, we introduce a homomorphism of $\uqlbp$ into a tensor power of the $q$-oscillator algebra $\Osc_q$, then for each factor of the tensor product we choose a certain representation of $\Osc_q$ and come to the required representations.

The $q$-oscillator algebra $\Osc_q$ is an algebra defined by the generators $b^\dagger$, $b$, $q^{\nu N}$, $\nu \in \bbC$, and the relations
\begin{gather*}
q^0 = 1, \qquad q^{\nu_1 N} q^{\nu_2 N} = q^{(\nu_1 + \nu_2)N}, \\
q^{\nu N} b^\dagger q^{-\nu N} = q^\nu b^\dagger, \qquad q^{\nu N} b q^{-\nu N} = q^{-\nu} b, \\
b^\dagger b = \frac{q^N - q^{-N}}{q - q^{-1}}, \qquad b b^\dagger = \frac{q q^N - q^{-1} q^{-N}}{q - q^{-1}}.
\end{gather*}
We are particularly interested in the two basic representations of $\Osc_q$. First, let $W^{\scriptscriptstyle +}$ be the free vector space with a basis $(w_n)_{n \in \bbZ_{ \ge 0}}$. One can show that the relations
\begin{gather*}
q^{\nu N} w_n = q^{\nu n} w_n,  \\[.3em]
b^\dagger w_n = w_{n + 1}, \qquad b \, w_n = [n]_q w_{n - 1},
\end{gather*}
where we assume that $w_{-1} = 0$, endow $W^{\scriptscriptstyle +}$ with the structure of an $\Osc_q$-module. We denote the corresponding representation of the algebra $\Osc_q$ by $\chi^{\scriptscriptstyle +}$. Further, let $W^{\scriptscriptstyle -}$ be again a free vector space with a basis $(w_n)_{n \in \bbZ_{ \ge 0}}$. The relations
\begin{gather*}
q^{\nu N} w_n = q^{- \nu (n + 1)} v_n,  \\[.3em]
b \, w_n = w_{n + 1}, \qquad b^\dagger w_n = - [n]_q w_{n - 1},
\end{gather*}
where we again assume that $w_{-1} = 0$, endow $W^{\scriptscriptstyle -}$ with the structure of an $\Osc_q$-module. We denote the corresponding representation of $\Osc_q$ by $\chi^{\scriptscriptstyle -}$.

Denoting
\begin{equation*}
\tr_{\chi^+} = \tr_{\End(W^+)} \circ \chi^+, \qquad \tr_{\chi^-} = \tr_{\End(W^-)} \circ \chi^-,
\end{equation*}
we see that
\begin{equation}
\tr_{\chi^+} ((b^\dagger)^k q^{\nu N}) = 0, \qquad \tr_{\chi^+} (b^k q^{\nu N}) = 0 \label{trbp}
\end{equation}
for $k \in \bbZ_{>0}$ and $\nu \in \bbC$, and that
\begin{equation}
\tr_{\chi^+}(q^{\nu N}) = (1 - q^\nu)^{-1} \label{trbm}
\end{equation}
for $|q^\nu| < 1$. For $|q^\nu| > 1$ we define the trace $\tr_{\chi^+}$ by analytic continuation. Since the monomials $(b^\dagger)^k q^{\nu N}$, $b^k q^{\nu N}$ and $q^{\nu N}$ for $k \in \bbZ_{>0}$ and $\nu \in \bbC$ form a basis of $\Osc_q$, the above relations are enough to determine the trace of any element of $\Osc_q$. It appears that
\begin{equation}
\tr_{\chi^-} = - \tr_{\chi^+}. \label{trtr}
\end{equation}

Consider the tensor product of $l$ copies of the $q$-oscillator algebra, and denote
\begin{gather*}
b_i = \underbracket[.5pt]{1 \otimes \cdots \otimes 1}_{i - 1} {} \otimes b \otimes \underbracket[.5pt]{1 \otimes \cdots \otimes 1}_{l - i} \, , \qquad
b_i^\dagger =  \underbracket[.5pt]{1 \otimes \cdots \otimes 1}_{i - 1} {} \otimes b^\dagger \otimes \underbracket[.5pt]{1 \otimes \cdots \otimes 1}_{l - i} \, , \\
q^{\nu N_i} =  \underbracket[.5pt]{1 \otimes \cdots \otimes 1}_{i - 1} {} \otimes q^{\nu N} \otimes \underbracket[.5pt]{1 \otimes \cdots \otimes 1}_{l - i} \, .
\end{gather*}
In the paper \cite{NirRaz17b} a homomorphism $o$ of $\uqlbp$ into $(\Osc_q)^{\otimes \, l}$ described by the equations
\begin{align}
& o(q^{\nu h_0}) = q^{\nu (N_1 + \sum_{k = 1}^l N_k)}, 
&& o(e_0) = b^\dagger_1 \, q^{\sum_{k = 2}^l N_k}, 
\label{roh0e0}\\[.3em]
& o(q^{\nu h_i}) = q^{\nu (N_{i + 1} - N_{i})}, 
&& o(e_i) = - b^{}_i \, b^{\mathstrut \dagger}_{i + 1} \, q^{N_i - N_{i + 1} - 1}, 
\label{rohiei}\\[.3em]
& o(q^{\nu h_l}) = q^{- \nu (\sum_{k = 1}^l N_k + N_l)},
&& o(e_l) = - \kappa_q^{-1} \, b^{}_l \, q^{N_l},
\label{rohlel}
\end{align}
where $i = 1, \ldots, l$, was obtained by some limiting procedure starting from the evaluation representations $\widetilde \varphi^\lambda$ of $\uqlsllpo$. Define the representation
\begin{equation*}
\theta = (\underbracket[.5pt]{\, \chi^{\scriptscriptstyle{+}} \otimes \cdots \otimes \chi^{\scriptscriptstyle{+}}}_l) \circ o.
\end{equation*}
The basis of this representation is formed by the vectors
\begin{equation*}
w_{\bm n} = b^{\dagger \, n_1}_1 \ldots b^{\dagger \, n_l}_l \, w^{}_{\bm 0},
\end{equation*}
where $n_i \in \bbZ_{\ge 0}$ for all $1 \le i \le l$, and we use the notation $\bm n = (n_i)_{i = 1}^l$. Here the vector $w_{\bm 0}$ is the vacuum vector, satisfying the equations
\begin{equation*}
b_i^{} \, w_{\bm 0} = 0, \qquad 1 \le i \le l.
\end{equation*}
The representation $\theta$ is in category $\calO$. It is a highest $\ell$-weight representation with highest $\ell$-weight vector $w_{\bm 0}$.

There is an automorphism of $\uqlsllpo$ defined by the equations
\begin{equation}
\sigma(q^{\nu h_i}) = q^{\nu h_{i + 1}}, \qquad \sigma(e_i) = e_{i + 1}, \qquad \sigma(f_i) = f_{i + 1}, \qquad 0 \le i \le l, \label{sqsesf}
\end{equation}
where it is assumed that $q^{\nu h_{l + 1}} = q^{\nu h_0}$, $e_{l + 1} = e_0$ and $f_{l + 1} = f_0$. One can restrict $\sigma$ to an automorphism of $\uqlbp$. It is useful to have in mind that $\sigma^{l + 1} = 1$. We define a collection of homomorphisms from $\uqlbp$ into $\Osc_q^{\otimes l}$ as
\begin{equation}
o_a = o \circ \sigma^{-a},
\label{roaovroa}
\end{equation}
and a family of representations
\begin{equation*}
\theta_a = \chi_a \circ o_a, \qquad a = 1, \ldots, l + 1,
\end{equation*}
where
\begin{equation}
\chi_a = \underbracket[.5pt]{\, \chi^- \otimes \cdots \otimes \chi^-}_{l - a + 1} \otimes \underbracket[.5pt]{\, \chi^+ \otimes \cdots \otimes \chi^+}_{a - 1}. \label{chia}
\end{equation}
Note that $\theta = \theta_{l + 1}$. The corresponding basis vectors are 
\begin{equation*}
w_{a, \, \bm n} = b_1^{n_1} \ldots b_{l - a + 1}^{n_{l - a + 1}} \, b^{\dagger \, n_{l - a + 2}}_{l - a + 2} \ldots b^{\dagger \, n_l}_{l} \, w_{a, \, \bm 0}.
\end{equation*}
The vacuum vector $w_{a, \, \bm 0}$ satisfies the equations
\begin{equation*}
b_i^\dagger w_{a, \, \bm 0} = 0, \quad 1 \le i \le l - a + 1, \qquad b_i^{} \,w_{a, \, \bm 0} = 0, \quad l - a + 2 \le i \le l.
\end{equation*}
All the representations $\theta_a$ are highest $\ell$-weight representations in category $\calO$. The vectors $w_{a, \, \bm n}$ are the $\ell$-weight vectors, and $w_{a, \, \bm 0}$ is the highest $\ell$-weight vector of the representation $\theta_a$. We denote the $\uqlbp$-module corresponding to the representation $\theta_a$ by $W_a$.

The explicit form of the $\ell$-weights $\bm \Psi_{a, \, \bm n}(\zeta) = (\psi_{a, \, \bm n}, \, \bm \Psi^+_{a, \, \bm n}(\zeta))$ for the modules $(W_a)_\zeta$ was found in the paper~\cite{BooGoeKluNirRaz17b}. For the future usage, we collect below the relevant expressions.

\subsection{\texorpdfstring{$\ell$-weights of the oscillator representations}{l-weights of the oscillator representations}} \label{s:lwor}

In fact, below we need the $\ell$-weights for the modules $(W_a)_\zeta$.

\subsubsection{\texorpdfstring{$a = 1$}{a = 1}}

\begin{align*}
& \psi_{1, \, \bm n} = - \Big( 2 n_1 + \sum_{j = 2}^l n_j + l + 1 \Big) \, \omega_1 - \sum_{i = 2}^l (n_i - n_{i - 1}) \, \omega_i, \\
& \Psi^+_{1, \, {\bm n}, \, 1}(\zeta, u) = \frac{(1 - q^{- 2 \sum_{j = 2}^l n_j - l + 2} \, \zeta^s \, u)} {(1 - q^{- 2 \sum_{j = 1}^l n_j - l} \, \zeta^s \, u) (1 - q^{- 2 \sum_{j = 1}^l n_j - l + 2} \, \zeta^s \, u)}, \\
& \Psi^+_{1, \, {\bm n}, \, i}(\zeta, u) = \frac{1 - q^{ - 2\sum_{j = i - 1}^l n_j - l + i - 1} \, \zeta^s \, )} {1 - q^{ - 2 \sum_{j = i}^l n_j - l + i - 1} \, \zeta^s \, u} \\
& \hspace{15em} \times \frac{1 - q^{ - 2 \sum_{j = i + 1}^l n_j - l + i + 1} \, \zeta^s \, u} {1 - q^{ - 2 \sum_{j = i}^l n_j - l + i + 1} \, \zeta^s \, u}, \quad i = 2, \ldots, l.
\end{align*}

\subsubsection{\texorpdfstring{$a = 2, \ldots, l$}{a = 2, ..., l}}

\begin{align*}
& \psi_{a, \, \bm n} = \sum_{i = 1}^{a - 2} (n_{l + i - a + 2} - n_{l + i - a + 1}) \, \omega_i - \sum_{i = a + 1}^l (n_{i - a + 1} - n_{i - a}) \, \omega_i \\
& \hspace{5em} {} + \Big( \sum_{j = 1}^{l - a + 1} n_j - \sum_{j = l - a + 2}^{l - 1} n_j - 2 n_l + l - a + 1 \Big) \, \omega_{a - 1} \\
& \hspace{10em} {} - \Big( 2 n_1 + \sum_{j = 2}^{l - a + 1} n_j - \sum_{j = l - a + 2}^l n_j + l - a + 2 \Big) \, \omega_a, \\
& \Psi^+_{a, \, {\bm n}, \, i}(\zeta, u) = 1, \quad i = 2, \ldots, a - 2, \qquad \Psi^+_{a, \, {\bm n}, \, a - 1}(u) = 1 - q^{ - 2 \sum_{j = 1}^{l - a + 1} n_j - l + a} \, \zeta^s \, u, \\
& \Psi^+_{a, \, {\bm n}, \, a}(\zeta, u) = \frac{\big( 1 - q^{ - 2 \sum_{j = 2}^{l - a + 1} n_j - l + a + 1} \, \zeta^s \, u \big) }{\big( 1 - q^{- 2 \sum_{j = 1}^{l - a + 1} n_j - l + a - 1} \, \zeta^s \, u \big) \, \big( 1 - q^{- 2 \sum_{j = 1}^{l - a + 1} n_j - l + a + 1} \, \zeta^s \, u \big)}, \\
& \Psi^+_{a, \, {\bm n}, \, i}(\zeta, u) = \frac{1 - q^{ - 2 \sum_{j = i - a}^{l - a + 1} n_j - l + i - 1} \, \zeta^s \, u} {1 - q^{ - 2\sum_{j = i - a + 1}^{l - a + 1} n_j - l + i - 1} \, \zeta^s \, u} \\
& \hspace{13em} \times \frac{1 - q^{ - 2 \sum_{j = i - a + 2}^{l - a + 1} n_j - l + i + 1} \, \zeta^s \, u} {1 - q^{ - 2 \sum_{j = i - a + 1}^{l - a + 1} n_j - l + i + 1} \, \zeta^s \, u}, \quad i = a + 1, \ldots, l.
\end{align*}

\subsubsection{\texorpdfstring{$a = l + 1$}{a = l + 1}}

\begin{align*}
& \psi_{l + 1, \, \bm n} = \sum_{i = 1}^{l - 1} (n_{i + 1} - n_i) \, \omega_i - \Big( \sum_{j = 1}^{l - 1} n_j + 2 n_l \Big) \, \omega_l, \\
& \Psi^+_{l + 1, \, {\bm n}, \, i}(\zeta, u) = 1, \quad i = 1, \ldots, l - 1, \qquad \Psi^+_{l + 1, \, {\bm n}, \, l, \, l + 1}(\zeta, u) = (1 - q \, \zeta^s \, u).
\end{align*}

\subsection{\texorpdfstring{$q$-characters and Grothendieck ring}{q-characters and Grothendieck ring}}

Let $V$ be a $\uqlg$-module in category $\calO$. We define the character of $V$ as a formal sum
\begin{equation*}
\ch(V) = \sum_{\lambda \in \gothh^*} \dim \, V_\lambda \, [\lambda].
\end{equation*}
By the definition of the category $\calO$,  $\dim V_\lambda = 0$ for $\lambda$ outside the union of a finite number of sets of the form $D(\mu)$, $\mu \in \gothh^*$. For any two $\uqlg$-modules $V$ and $U$ in category $\calO$ we have
\begin{equation*}
\ch(V \oplus U) = \ch(V) + \ch(U).
\end{equation*}
More generally, if $\uqlg$-modules $V$, $W$ and $U$ in category $\calO$ can be included in a short exact sequence
\begin{equation}
0 \rightarrow V \rightarrow W \rightarrow U \rightarrow 0, \label{vwu}
\end{equation}
then
\begin{equation*}
\ch(W) = \ch(V) + \ch(U).
\end{equation*}
It can be also shown that
\begin{equation*}
\ch(V \otimes_\Delta U) = \ch(V) \, \ch(U)
\end{equation*}
for any $\uqlg$-modules $V$ and $U$ in category $\calO$. Here, to multiply characters we assume that
\begin{equation*}
[\lambda] \, [\mu] = [\lambda + \mu]
\end{equation*}
for any $\lambda, \mu \in \gothh$.

The Grothendieck group of the category $\calO$ of $\uqlg$-modules is defined as the quotient of the free abelian group on the set of all isomorphism classes of objects in $\calO$ modulo the relations
\begin{equation*}
\langle V \rangle = \langle U \rangle + \langle W \rangle
\end{equation*}
if the objects $V$, $W$ and $U$ can be included in a short exact sequence (\ref{vwu}). Here for any object $V$, $\langle V \rangle$ denotes the isomorphism class of $V$. Defining 
\begin{equation*}
\langle V \rangle \langle W \rangle = \langle V \otimes_\Delta W \rangle,
\end{equation*}
we come to the Grothendieck ring of $\calO$. It is a commutative unital ring, for which the role of the unit is played by the trivial $\uqlg$-module. We see that the character can be considered as a mapping from the Grothendieck ring of $\calO$. However, it is not injective, and, therefore, does not uniquely distinguish between its elements. This role is played by the $q$-character.

We define the $q$-character of a $\uqlg$-module $V$ in category $\calO$ as a formal sum
\begin{equation*}
\ch_q(V) = \sum_{\bm \Lambda \in \gothh^*_\ell} \dim (V_{\bm \Lambda}) \, [\bm \Lambda].
\end{equation*}
One can easily demonstrate that
\begin{equation*}
\varpi(\ch_q(V)) = \ch(V).
\end{equation*}
Here we assume that
\begin{equation*}
\varpi([\bm \Lambda]) = [\varpi(\bm \Lambda)],
\end{equation*}
and extend this rule by linearity. The $q$-character has the same properties as the usual character. Namely, if $\uqlg$-modules $V$, $W$ and $U$ in category $\calO$ can be included in a short exact sequence (\ref{vwu}), then
\begin{equation*}
\ch_q(W) = \ch_q(V) + \ch_q(U),
\end{equation*}
and for any two $\uqlg$-modules $V$ and $U$ in category $\calO$, we have
\begin{equation}
\ch_q(V \otimes_\Delta W) = \ch_q(V) \, \ch_q(W), \label{cvtu}
\end{equation}
see the paper \cite{FreRes99}. To define the product of $q$-characters, we assume that
\begin{equation*}
[\bm \Lambda][\bm \Xi] = [\bm \Lambda \bm \Xi]
\end{equation*}
for any $\bm \Lambda, \bm \Xi \in \gothh^*_\ell$. Thus, $q$-character can also be considered as a mapping from the Grothendieck ring of $\calO$, and now one can demonstrate that this mapping is injective. In other words, different elements of the Grothendieck ring of $\calO$ have different $q$-characters.

It follows from equation (\ref{cvtu}) that
\begin{equation*}
\ch_q(V \otimes_\Delta W) = \ch_q(V) \ch_q(W) = \ch_q(W) \ch_q(V) = \ch_q(W \otimes_\Delta V).
\end{equation*}
It means that the $\uqlg$-modules $V \otimes_\Delta W$ and $W \otimes_\Delta V$ belong to the same equivalence class in the Grothendieck ring of $\calO$.

All above can be naturally extended to the categories $\calO$ of the modules over $\uqlbp$ and $\uqlbm$.

\section{\texorpdfstring{Universal $R$-matrix and integrability objects}{Universal R-matrix and integrability objects}} \label{s:urmio}

\subsection{\texorpdfstring{Quantum group as a $\bbC[[\hbar]]$-algebra}{Quantum group as a C[[h]]-algebra}}

\subsubsection{\texorpdfstring{Universal $R$-matrix}{Universal R-matrix}}

As any Hopf algebra the quantum loop algebra $\uqlg$ has another comultiplication called the opposite comultiplication. It is defined by the equation
\begin{equation*}
\Delta' = \Pi \circ \Delta,
\end{equation*}
where
\begin{equation*}
\Pi(x \otimes y) = y \otimes x
\end{equation*}
for all $x, y \in \uqlg$.

There are several definitions of quantum groups. In particular, $\hbar$ can be treated not only as a complex number \cite{Jim86a, EtiFreKir98, Dam98}, but also as an indeterminate. In this case the quantum group $\uqlg$ is a $\bbC[[\hbar]]$-algebra \cite{Dri87, TolKho92, KhoTol92, KhoTol93, KhoTol94}. Let us assume temporally that it is the case. Herewith $\uqlg$ is a quasitriangular Hopf algebra. It means that there exists an element $\calR$ of the completed tensor product $\uqlg \mathop{\widehat \otimes} \uqlg$, called the universal $R$-matrix, such that
\begin{equation}
\Delta'(x) = \calR \, \Delta(x) \, \calR^{-1} \label{urmd}
\end{equation}
for all $x \in \uqlg$, and\footnote{For an explanation of the notation see any book on quantum groups, or the paper \cite{NirRaz19}.}
\begin{equation}
(\Delta \otimes \id) (\calR) = \calR^{(1 3)} \calR^{(2 3)}, \qquad (\id \otimes \Delta) (\calR) = \calR^{(1 3)} \calR^{(1 2)}. \label{urm}
\end{equation}
The latter equations are considered as equalities in the completed tensor product of three copies of $\uqlg$. In fact, it follows from the explicit expression for the universal $R$-matrix \cite{TolKho92, KhoTol92, KhoTol93, KhoTol94} that it is an element of a completed tensor product of two Borel subalgebras $\uqlbp$ and $\uqlbm$.

\subsubsection{Integrability objects}

Let $\varphi$ be a homomorphism from $\uqlbp$ to an algebra $A$, and $\psi$ a homomorphism from $\uqlbm$ to another algebra $B$. The corresponding integrability object $\bm X_{\varphi | \psi}$ is defined by the equation
\begin{equation}
\rho_{\varphi | \psi} \bm X_{\varphi | \psi} = [(\varphi \otimes \psi)(\calR)], \label{xfpa}
\end{equation}
where $\rho_{\varphi | \psi}$ is a scalar normalization factor. It is evident that this object is an element of $A \otimes B$. Certainly, as $\varphi$ and $\psi$ one can use the restrictions to $\uqlbp$ and $\uqlbm$ of homomorphisms from $\uqlg$. However, this is not always the case.

It follows from equation (\ref{urmd}) that
\begin{equation}
(\varphi \otimes \psi)(\Pi(\Delta(x))) = \bm X_{\varphi | \psi} (\varphi \otimes \psi)(\Delta(x)) (\bm X_{\varphi | \psi})^{-1}, \label{urmx}
\end{equation}
and equation (\ref{urm}) gives
\begin{equation}
\bm X\strut_{\varphi_1 \otimes_\Delta \varphi_2 | \psi} = \bm X_{\varphi_1 | \psi}^{(1 3)}\bm  X_{\varphi_2 | \psi}^{(2 3)}, \qquad \bm X\strut_{\varphi | \psi_1 \otimes_\Delta \psi_2} = \bm X_{\varphi | \psi_2}^{(1 3)} \bm X_{\varphi | \psi_1}^{(1 2)}, \label{xfff}
\end{equation}
where in the first equation $\varphi_1$, $\varphi_2$ are homomorphisms from $\uqlbp$ to algebras $A_1$, $A_2$, and $\psi$ is a homomorphism from $\uqlbm$ to an algebra $B$, while in the second one $\varphi$ is a homomorphism  from $\uqlbp$ to an algebra $A$, and $\psi_1$, $\psi_2$ are homomorphisms from $\uqlbm$ to algebras $B_1$, $B_2$. In fact, hereinafter, we assume that
\begin{equation*}
\rho_{\varphi_1 \otimes_\Delta \varphi_2 | \psi} = \rho_{\varphi_1 | \psi} \, \rho_{\varphi_2 | \psi}, \qquad \rho_{\varphi_1 | \psi_1 \otimes_\Delta \psi_2} = \rho_{\varphi | \psi_1} \, \rho_{\varphi | \psi_2}.
\end{equation*}

Each integrability object $\bm X_{\varphi | \psi} \in A \otimes B$ is accompanied by an integrability object of a different type, defined as follows. Let $\chi$ be a representation of the algebra $A$ on a vector space $V$. Define a trace on $A$ by the equation
\begin{equation}
\tr_\chi = \tr_{\End(V)} \circ \chi, \label{traa}
\end{equation}
where $\tr_{\End(V)}$ is the usual trace on the endomorphism algebra of the vector space $V$. When $\varphi$ is already a representation of $\uqlbp$ on a vector space $V$ we use as $\chi$ the identity mapping $\id_V$ and omit $\chi$ in the notation. The integrability object in question is
\begin{equation*}
\bm Y_{\varphi | \psi}^\chi = (\tr_\chi \mathop{\otimes} \id_B)(\bm X_{\varphi | \psi} (\varphi(q^t) \otimes 1_B)),
\end{equation*}
where
\begin{equation}
q^t = q^{\sum_{i = 1}^l t_i h_i}, \label{tqfh}
\end{equation}
and $t_i$ are complex parameters. The element $q^t \in \uqlbp \cap \uqlbm$ is a group like element\footnote{An element $a$ of a Hopf algebra is called group like if $\Delta(x) = x \otimes x$. Note that any group like element is invertible.} called a twisting element. It is necessary for convergence of the trace in the case when $\varphi$ is an infinite dimensional representation. We see that $\bm Y_{\varphi | \psi}^\chi$ is an element of the algebra $B$.

\subsubsection{Universal integrability objects}

For any homomorphism $\varphi$ from $\uqlbp$ to an algebra $A$, it is productive to define universal integrability object
\begin{equation}
\calX_\varphi = (\varphi \otimes \id_{\uqlbm})(\calR), \label{cxf}
\end{equation}
being an element of $A \otimes \uqlbm$. It is clear that for any homomorphism $\psi$ from $\uqlbm$ the integrability object $\bm X_{\varphi | \psi}$ is related to $\calX_\varphi$ by the equation
\begin{equation}
(\id_A \mathop{\otimes} \psi) (\calX_\varphi) = \rho_{\varphi | \psi} \bm X_{\varphi | \psi}. \label{xfpcxf}
\end{equation}

Similarly, each universal integrability object $\calX_\varphi$ is accompanied by universal integrability object
\begin{equation}
\calY_\varphi^\chi = (\tr_\chi \mathop{\otimes} \id_{\uqlg})(\calX_\varphi (\varphi(q^t) \otimes 1_{\uqlbm})), \label{cyf}
\end{equation}
being an element of $\uqlbm$. By definition, for any homomorphism $\psi$ from $\uqlbm$ to an algebra $B$ we have
\begin{equation*}
\psi(\calY_\varphi^\chi) = \rho_{\varphi | \psi}\bm Y_{\varphi| \psi}^\chi.
\end{equation*}
Furthermore, let $\varphi_1$, $\ldots$, $\varphi_m$ be homomorphisms from $\uqlbp$ to algebras $A_1$, $\ldots$, $A_m$, and $\chi_1$, $\ldots$, $\chi_m$ are representations of $A_1$, $\ldots$, $A_m$, then
\begin{equation}
\psi(\calY_{\varphi_1}^{\chi_1} \ldots \calY_{\varphi_m}^{\chi_m}) = \psi(\calY_{\varphi_1}^{\chi_1}) \ldots \psi(\calY_{\varphi_m}^{\chi_m}) = \rho_{\varphi_1 | \psi} \ldots \rho_{\varphi_m | \psi} \bm Y_{\varphi_1 | \psi}^{\chi_1} \ldots \bm Y_{\varphi_m | \psi}^{\chi_m}. \label{yfpcyf}
\end{equation}

\subsection{\texorpdfstring{Quantum group as a $\bbC$-algebra}{Quantum group as a C-algebra}}

\subsubsection{Integrability objects}

The expression for the universal $R$-matrix of a quantum loop algebra $\uqlg$ considered as a $\bbC[[\hbar]]$-algebra can be constructed using the procedure proposed by Khoroshkin and Tolstoy \cite{TolKho92, KhoTol93, KhoTol94}. However, in this paper, we define the quantum loop algebra $\uqlg$ as a $\bbC$-algebra. In fact, one can use the expression for the universal $R$-matrix from the papers \cite{TolKho92, KhoTol93, KhoTol94} to construct the integrability objects also in this case, having in mind that $\uqlg$ is quasitriangular only in some restricted sense, see the paper \cite{Tan92}, the book \cite[p.~327]{ChaPre94}, and the discussion below.

We restrict ourselves to the following case. Let $\varphi$ be a homomorphism from $\uqlbp$ to an associative algebra $A$ and $\psi$ a representation of $\uqlbm$ on a vector space $U$. Define an integrability object $\bm X_{\varphi | \psi}$ as an element of $A \otimes \End(U)$ by the equation
\begin{equation}
\rho_{\varphi | \psi} \bm X_{\varphi | \psi} = (\varphi \otimes \psi)(\calR_{\prec \delta} \, \calR_{\sim \delta} \, \calR_{\succ \delta}) \bm K_{\varphi | \psi}. \label{xfpb}
\end{equation}
Here $\calR_{\prec \delta}$, $\calR_{\sim \delta}$ and $\calR_{\succ \delta}$ are elements of $\uqlbp \mathop{\widehat \otimes} \uqlbm$, $\bm K_{\varphi | \psi}$ is an element of $A \otimes \End(U)$, and $\rho_{\varphi | \psi}$ a scalar normalization factor.

The element $\bm K_{\varphi | \psi} \in A \otimes \End(U)$ is given by the equation
\begin{equation}
\bm K_{\varphi | \psi} = \sum_{\lambda \in \gothh^*} \varphi \big( q^{- \sum_{i, \, j = 1}^l h^{}_i c^{}_{i j} \, d_j^{-1} \langle \lambda, \, h^{}_j \rangle} \big) \otimes \Pi^\psi_\lambda,
\label{kfpb}
\end{equation}
where $\Pi_\lambda \in \End(U)$ is the projector on the component $U^\psi_\lambda$ of the weight decomposition
\begin{equation*}
U = \bigoplus_{\lambda \in \gothh^*}  U^\psi_\lambda
\end{equation*}
with respect to the representation $\psi$, and $c_{i j}$ are the matrix entries of the matrix $C$ inverse to the Cartan matrix $A$ of the Lie algebra $\gothg$.

It can be shown that with an appropriate choice of $\calR_{\prec \delta}$, $\calR_{\sim \delta}$ and $\calR_{\succ \delta}$ the integrability objects defined by equation (\ref{xfpb}) satisfy equations~(\ref{urmx}) and (\ref{xfff}), see the paper \cite{Tan92} and next section.

The $Y$ type companion of $\bm X_{\varphi | \psi}$ is defined by the equation
\begin{equation}
\bm Y_{\varphi | \psi}^\chi = (\tr_\chi \mathop{\otimes} \id_{\End(U)})(\bm X_{\varphi | \psi} (\varphi(q^t) \otimes 1_{\End(U)})). \label{byfp}
\end{equation}
It is not difficult to demonstrate that
\begin{equation}
\rho_{\varphi | \psi} \bm Y^\chi_{\varphi | \psi} = \rho_{\chi \circ \varphi | \psi} \bm Y_{\chi \circ \varphi | \psi}. \label{rfpy}
\end{equation}
It is clear that the integrability object $\bm Y_{\varphi | \psi}^\chi$ depend on $\chi$ and $\varphi$ only through the equivalence class in the Grothendieck ring of the category $\calO$ to which the representation $\chi \circ \varphi$ belongs.

\subsubsection{\texorpdfstring{The case $\gothg = \sllpo$}{The case g = sll+1}}

We give explicit expressions for the elements $\calR_{\prec \delta}$ , $\calR_{\sim \delta} \, \calR_{\succ \delta}$ and $\bm K_{\varphi | \psi}$ only for the most important for this paper case $\gothg = \sllpo$. The element $\calR_{\prec \delta}$ is the product over the set of roots $\alpha_{i j} + n \delta$, $1 \le i < j \le l + 1$, $n \in \bbZ_{\ge 0}$, of the $q$-exponentials
\begin{equation*}
\calR_{\alpha_{i j} + n \delta} = \exp_{q^2} \left( - \kappa_q \, e_{\alpha_{i j} + n \delta} \otimes f_{\alpha_{i j} + n \delta} \right).
\end{equation*}
Hereinafter, the $q$-exponential is defined as
\begin{equation*}
\exp_q(x) = \sum_{n = 0}^\infty q^{- n (n - 1) / 4} \frac{x^n}{[n]_q^{1/2}!}. 
\end{equation*}
The order of the factors in $\calR_{\prec \delta}$ coincides with the normal order of the roots $\alpha_{i j} + n \delta$, described in section \ref{ss:pbwb}.

The element $\calR_{\sim \delta}$ is defined as
\begin{equation*}
\calR_{\sim \delta} = \exp \Big( - \kappa_q \sum_{n \in \bbZ_{>0}} \, \sum_{i, j
= 1}^l u_{n, \, i j} \, e_{n \delta; \, \alpha_i} \otimes f_{n \delta; \, \alpha_j} \Big),
\end{equation*}
where for each $n \in \bbZ_{> 0}$ the quantities $u_{n, \, i j}$ are given by the equations
\begin{align*}
u_{n, \, i j} = (-1)^{n (i + j)} \frac{n}{[n]_q} \frac{[i]_{q^n}[l - j + 1]_{q^n}}{[l + 1]_{q^n}}, \quad i \le j, \\
u_{n, \, i j} = (-1)^{n (i + j)} \frac{n}{[n]_q} \frac{[l - i + 1]_{q^n} [j]_{q^n}}{[l + 1]_{q^n}}, \quad i > j.
\end{align*}

The definition of the element $\calR_{\succ \delta}$ is similar to the definition of the element $\calR_{\prec \delta}$. It is the product over the set of roots $(\delta - \alpha_{i j}) + n \delta$ of the $q$-exponentials
\begin{equation*}
\calR_{(\delta - \alpha_{i j}) + n \delta} = \exp_{q^2} \left( - \kappa_q \, e_{(\delta - \alpha_{i j}) + n \delta}^{} \otimes f_{(\delta - \alpha_{i j}) + n \delta}^{}
\right). \label{rdmgm}
\end{equation*}
The order of the factors in $\calR_{\succ \delta}$ coincides with the normal order of the roots $(\delta - \alpha_{i j}) + n \delta$, described in section \ref{ss:pbwb}.

The matrix $A$ is tridiagonal, and the matrix elements of the matrix $C$ entering the expression (\ref{kfpb}) for $\bm K_{\varphi | \psi}$ can be found using the results of the paper \cite{Usm94}. We obtain
\begin{equation}
c_{i j} = \frac{i (l - j + 1)}{l + 1}, \quad i \le j, \qquad c_{i j} = \frac{(l - i + 1) j}{l + 1}, \quad i > j. \label{cij}
\end{equation}

\subsubsection{Universal integrability objects}

When $\uqlg$ is defined as a $\bbC[[\hbar]]$-algebra, we use equations (\ref{cxf}) and (\ref{cyf}) to define universal integrability objects. This is not possible if $\uqlg$ is defined as a $\bbC$-algebra. Here the universal integrability objects are defined only as formal objects with specific rules of use. If $\varphi$ is a homomorphism from $\uqlbp$ to an algebra $A$, the universal integrability object $\calX_{\varphi}$ behaves as an element of $A \otimes \uqlbm$. To obtain usual integrability object we use the rule  
\begin{equation}
(\id_A \mathop{\otimes} \psi) (\calX_\varphi) = \rho_{\varphi | \psi} \bm X_{\varphi | \psi}. \label{uxto}
\end{equation}
The universal object $\calY^\chi_{\varphi}$, where $\varphi$ is a homomorphism from $\uqlbp$ to an algebra $A$ and $\chi$ a representation of $A$, behaves as an element of $\uqlbm$. It obeys the following rule. Let $\varphi_1, \ldots, \varphi_m$ be homomorphisms from $\uqlbp$ to algebras $A_1, \ldots, A_m$, and $\chi_1, \ldots, \chi_m$ representations of  $A_1, \ldots, A_m$, respectively, then
\begin{equation}
\psi(\calY_{\varphi_1}^{\chi_1} \ldots \calY_{\varphi_m}^{\chi_m}) = \rho_{\varphi_1 | \psi} \ldots \rho_{\varphi_m | \psi} \bm Y_{\varphi_1 | \psi}^{\chi_1} \ldots \bm Y_{\varphi_m | \psi}^{\chi_m} \label{uyto}
\end{equation}
for any representation $\psi$ of $\uqlbm$. Formally, the two above relations coincide with equations (\ref{xfpcxf}) and (\ref{yfpcyf}). However, (\ref{xfpcxf}) and (\ref{yfpcyf}) are consequences of the definitions, while (\ref{uxto}) and (\ref{uyto}) is a part of the definitions. Using (\ref{rfpy}), we obtain
\begin{equation*}
\psi(\calY_\varphi^\chi) = \rho_{\varphi | \psi} \bm Y_{\varphi | \psi}^\chi = \rho_{\chi \circ \varphi | \psi} \bm Y_{\chi \circ \varphi | \psi} = \psi(\calY_{\chi \circ \varphi}). 
\end{equation*}
and come to the following equation for formal objects
\begin{equation*}
\calY_\varphi^\chi = \calY_{\chi \circ \varphi}.
\end{equation*}
It is clear that the integrability object $\calY^\chi_\varphi$ depend on $\chi$ and $\varphi$ only through the equivalence class in the Grothendieck ring of the category $\calO$ to which the representation $\chi \circ \varphi$ belongs.

\subsubsection{Commutativity of integrability objects}

Let $\varphi_1$ and $\varphi_2$ be homomorphisms from the Borel subalgebra $\uqlbp$ to algebras $A_1$ and $A_2$, $\chi_1$ and $\chi_2$ representations of the algebras $A_1$ and $A_2$, and $\psi$ a representation of $\uqlbm$ on a vector space $U$. The definition (\ref{byfp}) of $\bm Y^\chi_{\varphi | \psi}$ implies
\begin{equation*}
\bm Y^{\chi_1 \otimes \chi_2}_{\varphi_1 \otimes_\Delta \varphi_2 | \psi} = (\tr_{\chi_1 \otimes \chi_2} \otimes \id_{\End(U)})(\bm X_{\varphi_1 \otimes_\Delta \varphi_2 | \psi} ((\varphi_1 \otimes_\Delta \varphi_2)(q^t)) \otimes 1_{\End(U)}).
\end{equation*}
Using the equations
\begin{equation*}
\tr_{\chi_1 \otimes \chi_2} = \tr_{\chi_1} \otimes \tr_{\chi_2}
\end{equation*}
and
\begin{equation*}
(\varphi_1 \otimes_\Delta \varphi_2)(q^t) = \varphi_1(q^t) \otimes \varphi_2(q^t),
\end{equation*}
and taking into account the first equation of (\ref{xfff}), we obtain
\begin{equation*}
\bm Y^{\chi_1 \otimes \chi_2}_{\varphi_1 \otimes_\Delta \varphi_2 | \psi} = (\tr_{\chi_1} \otimes \tr_{\chi_2} \otimes \id_{\End(U)})(\bm X^{(1 3)} _{\varphi_1 | \psi} \bm X^{(2 3)} _{\varphi_2 | \psi} ((\varphi_1(q^t) \otimes \varphi_2(q^t)) \otimes 1_{\End(U)}).
\end{equation*}
Hence, we see that
\begin{equation*}
\bm Y^{\chi_1 \otimes \chi_2}_{\varphi_1 \otimes_\Delta \varphi_2 | \psi} = \bm Y^{\chi_1}_{\varphi_1 | \psi} \bm Y^{\chi_2}_{\varphi_2 | \psi},
\end{equation*}
or, equivalently,
\begin{equation*}
\bm Y_{(\chi_1 \circ \varphi_1) \otimes_\Delta (\chi_2 \circ \varphi_2) | \psi} = \bm Y_{\chi_1 \circ \varphi_1 | \psi} \bm Y_{\chi_2 \circ \varphi_2 | \psi}.
\end{equation*}
In terms of universal integrability objects we have
\begin{equation*}
\calY^{\chi_1 \otimes \chi_2}_{\varphi_1 \otimes_\Delta \varphi_2 | \psi} = \calY^{\chi_1}_{\varphi_1 | \psi} \calY^{\chi_2}_{\varphi_2 | \psi},
\end{equation*}
or, equivalently,
\begin{equation}
\calY_{(\chi_1 \circ \varphi_1) \otimes_\Delta (\chi_2 \circ \varphi_2) | \psi} = \calY_{\chi_1 \circ \varphi_1 | \psi} \calY_{\chi_2 \circ \varphi_2 | \psi}.  \label{yffp}
\end{equation}

Assume that the representations $(\chi_1 \circ \varphi_1)$ and $(\chi_2 \circ \varphi_2)$ and $(\chi_2 \circ \varphi_2) \otimes_\Delta (\chi_2 \circ \varphi_1)$ are in category $\calO$. Since the homomorphisms $(\chi_1 \circ \varphi_1) \otimes_\Delta (\chi_2 \circ \varphi_2)$ and $(\chi_2 \circ \varphi_2) \otimes_\Delta (\chi_2 \circ \varphi_1)$ belongs to the same equivalence class in the Grothendieck ring of $\calO$, we have
\begin{equation*}
\bm Y_{(\chi_1 \circ \varphi_1) \otimes_\Delta (\chi_2 \circ \varphi_2) | \psi} = \bm Y_{(\chi_2 \circ \varphi_2) \otimes_\Delta (\chi_1 \circ \varphi_1) | \psi},
\end{equation*}
and, therefore,
\begin{equation}
\bm Y_{\chi_1 \circ \varphi_1 | \psi} \bm Y_{\chi_2 \circ \varphi_2 | \psi} = \bm Y_{\chi_2 \circ \varphi_2 | \psi} \bm Y_{\chi_1 \circ \varphi_1 | \psi}, \label{yyyy}
\end{equation}
or, in terms of universal integrability objects,
\begin{equation}
\calY_{\chi_1 \circ \varphi_1} \calY_{\chi_2 \circ \varphi_2} = \calY_{\chi_2 \circ \varphi_2} \calY_{\chi_1 \circ \varphi_1}. \label{yyyyu}
\end{equation}

\subsubsection{Commutation with group like elements}

Let $\varphi$ be a homomorphism from $\uqlbp$ to an algebras $A$, and $\psi$ a representation of $\uqlbm$ on a vector space $U$. If $x \in \uqlbp \cap \uqlbm$ is a group like element, then using equation (\ref{urmx}), we obtain
\begin{equation*}
(\varphi(x) \otimes \psi(x)) \bm X_{\varphi | \psi} =  \bm X_{\varphi | \psi} (\varphi(x) \otimes \psi(x)).
\end{equation*}
Assuming that $a$ commutes with the twisting element $q^t$, we come to the equation
\begin{multline*}
(1_A \otimes \psi(x))\Big[ \bm X_{\varphi | \psi} (\varphi(q^t) \otimes 1_{\End(U)})\Big] \\=  (\varphi(x)^{-1} \otimes 1_{\End(U)}) \Big[ \bm X_{\varphi | \psi} (\varphi(q^t) \otimes 1_{\End(U)}) \Big] (\varphi(x) \otimes 1_{\End(U)}) (1_A \otimes \psi(x)).
\end{multline*}
Applying to both sides of this equation the mapping $\tr_\chi$, where $\chi$ is a representation of $A$, we come to the equation
\begin{equation}
\bm Y^\chi_{\varphi | \psi} \psi(x) = \psi(x) \, \bm Y^\chi_{\varphi | \psi}. \label{ypa}
\end{equation}
For the universal integrability object $\calY^\chi_\varphi$ we  have
\begin{equation*}
\calY^\chi_\varphi \, x = x \, \calY^\chi_\varphi
\end{equation*}
for any group like element $a$ commuting with the twisting element $q^t$.

\subsubsection{\texorpdfstring{Behavior under action of automorphism $\sigma$}{Behavior under action of automorphism sigma}}

Return again to the case $\gothg = \sllpo$, and find the connection between the integrability objects $\bm X_{\varphi \circ \sigma | \psi \circ \sigma}$ and $\bm X_{\varphi | \psi}$, where $\sigma$ is the automorphism of $\uqlsllpo$ defined by equation (\ref{sqsesf}). Adapting the reasoning given in the paper \cite{KhoTol92} to our situation, we can demonstrate that
\begin{equation}
((\varphi \circ \sigma) \otimes (\psi \circ \sigma))(\calR_{\prec \delta} \, \calR_{\sim \delta} \, \calR_{\succ \delta}) = (\varphi \otimes \psi)(\calR_{\prec \delta} \, \calR_{\sim \delta} \, \calR_{\succ \delta}). \label{fsps}
\end{equation}
Further, we have
\begin{equation}
\bm K_{\varphi \circ \sigma | \psi \circ \sigma} = \sum_{\lambda \in \gothh^*} \varphi \circ \sigma \big( q^{- \sum_{i, \, j = 1}^l h_i c_{i j} \, \langle \lambda, \, h_j \rangle} \big) \otimes \Pi^{\psi \circ \sigma}_\lambda.  \label{kfs}
\end{equation}
Using the fact that $q^{\nu \sum_{k = 0}^l h_k} = 1$, we find 
\begin{equation*}
\sigma(q^{\nu h_i}) = q^{\nu \sum_{k = 1}^l S_{i k} h_k}, \qquad i = 1, \ldots, l,
\end{equation*}
where $S_{i k}$ are the matrix entries of the square matrix $S$ of size $l$ defined as
\begin{equation*}
S = \sum_{k = 1}^{l - 1} \bbE_{k, \, k + 1} - \sum_{k = 1}^l \bbE_{l, \, k}.
\end{equation*}
One can get convinced that $S^t C = C \, S^{-1}$, and obtain
\begin{equation}
\sigma \big( q^{- \sum_{i, \, j = 1}^l h_i c_{i j} \, \langle \lambda, \, h_j \rangle} \big) = q^{- \sum_{i, \, j = 1}^l h_i c_{i j} \, \langle \lambda^\sigma, \, h_j \rangle}, \label{smo}
\end{equation}
where
\begin{equation*}
\lambda^\sigma = \sum_{i = 1}^l \Big( \sum_{j = 1}^l  (S^{-1})_{i j} \lambda_j \Big) \omega_i
\end{equation*}
for $\lambda = \sum_{i = 1}^l \lambda_i \, \omega_i$. It is not difficult to demonstrate also that
\begin{equation}
\Pi^{\varphi \circ \sigma}_\lambda = \Pi^\varphi_{\lambda^\sigma}. \label{pfs}
\end{equation}
Substituting (\ref{smo}) and (\ref{pfs}) into (\ref{kfs}), we see that $\bm K_{\varphi \circ \sigma | \psi \circ \sigma} = \bm K_{\varphi | \psi}$, and, taking into account equation (\ref{fsps}), obtain $\bm X_{\varphi \circ \sigma | \psi \circ \sigma} = \bm X_{\varphi | \psi}$. Here we set $\rho_{\varphi \circ \sigma | \psi \circ \sigma} = \rho_{\varphi | \psi}$. It is evident that after a suitable adjustment of the normalizations, we obtain
\begin{equation}
\bm X_{\varphi \circ \sigma^a | \psi \circ \sigma^a} = \bm X_{\varphi | \psi}. \label{xfsfs}
\end{equation}
for any $a \in \bbZ$.

\subsubsection{Behavior under shift of representation}

Consider the behavior of an integrability objects $X_{\varphi | \psi}$ and $\calX_{\varphi | \psi}$ under a shift of the homomorphism $\varphi$, see equation (\ref{phixi}). First observe that
\begin{equation*}
(\varphi[\xi] \otimes \psi)(\calR_{\prec \delta} \, \calR_{\sim \delta} \, \calR_{\succ \delta}) = (\varphi \otimes \psi)(\calR_{\prec \delta} \, \calR_{\sim \delta} \, \calR_{\succ \delta}).
\end{equation*}
Then, using equation (\ref{phixi}), we obtain
\begin{multline*}
\bm K_{\varphi[\xi] | \psi} = \sum_{\lambda \in \gothh^*} \varphi[\xi] \big( q^{- \sum_{i, \, j = 1}^l h^{}_i c^{}_{i j} d_j^{-1} \, \langle \lambda, \, h^{}_j \rangle} \big) \otimes \Pi^\psi_\lambda \\*
= \sum_{\lambda \in \gothh^*} \varphi \big( q^{- \sum_{i, \, j = 1}^l h^{}_i c^{}_{i j} d_j^{-1} \, \langle \lambda, \, h^{}_j \rangle} \big) \otimes q^{- \sum_{i, \, j = 1}^l \langle \xi, \, h^{}_i \rangle c^{}_{i j} d_j^{-1} \, \langle \lambda, \, h^{}_j \rangle} \Pi^\psi_\lambda \\*
= \sum_{\lambda \in \gothh^*} \varphi \big( q^{- \sum_{i, \, j = 1}^l h^{}_i c^{}_{i j} d_j^{-1} \, \langle \lambda, \, h^{}_j \rangle} \big) \otimes \psi \big( q^{- \sum_{i, \, j = 1}^l \langle \xi, \, h^{}_i \rangle c^{}_{i j} d_j^{-1} \, h^{}_j} \big) \Pi^\psi_\lambda.
\end{multline*}
Since
\begin{equation*}
\psi(q^x) \Pi^\psi_\lambda = \Pi^\psi_\lambda \psi(q^x) 
\end{equation*}
for any $x \in \tgothh$, we come to the equation
\begin{equation*}
\bm K_{\varphi[\xi] | \psi} = K_{\varphi | \psi} \big( 1_A \otimes \psi(q^{- \sum_{i, \, j = 1}^l \langle \xi, \, h^{}_i \rangle c^{}_{i j} d_j^{-1} \, h^{}_j} \big)
\end{equation*}
and determine that
\begin{equation*}
\bm X_{\varphi[\xi] | \psi} = \bm X_{\varphi | \psi} \, \big( 1_A \otimes \psi(q^{- \sum_{i, \, j = 1}^l \langle \xi, \, h^{}_i \rangle c^{}_{i j} d_j^{-1} \, h_j}) \big), \quad  \calX_{\varphi[\xi]} = \calX_\varphi \, \big( 1_A \otimes q^{- \sum_{i, \, j = 1}^l \langle \xi, \, h^{}_i \rangle c^{}_{i j} d_j^{-1} \, h_j} \big).
\end{equation*}
For the integrability objects $Y_{\varphi | \psi}$ and $\calY_\varphi$ we obtain the equations
\begin{equation}
\bm Y^\chi_{\varphi[\xi] | \psi} = \bm Y^\chi_{\varphi | \psi} \, \psi \big( q^{- \sum_{i, \, j = 1}^l \langle \xi, \, h^{}_i \rangle c^{}_{i j} d_j^{-1} \, h'_j} \big), \qquad \calY^\chi_{\varphi[\xi]} = \calY^\chi_\varphi \, q^{-  \sum_{i, \, j = 1}^l \langle \xi, \, h^{}_i \rangle c^{}_{i j} d_j^{-1} \, h'_j}, \label{yfsp}
\end{equation}
where
\begin{equation*}
h'_i = h_i - t_i.
\end{equation*}

\subsubsection{Adding spectral parameters}

To define integrability objects depending on spectral parameters, we use as a homomorphism $\varphi$ the tensor product of $m$ homomorphisms $\varphi_{\zeta_1}$, $\ldots$, $\varphi_{\zeta_m}$, and as $\psi$ the $n$th tensor power of a representation $\psi$.\footnote{One can also use as $\psi$ the tensor product $\psi_{\eta_1} \otimes_\Delta \cdots \otimes_\Delta \psi_{\eta_n}$. However, we do not consider such a generalization in this paper.} The corresponding universal integrability objects are denoted as
\begin{equation*}
\calX_\varphi(\zeta_1, \, \ldots, \, \zeta_m) =  \calX_{\varphi_{\zeta_1} \otimes_\Delta \cdots \otimes_\Delta \varphi_{\zeta_m}}, \qquad \calY^\chi_\varphi(\zeta_1, \, \ldots, \, \zeta_m) =  \calY^\chi_{\varphi_{\zeta_1} \otimes_\Delta \cdots \otimes_\Delta \varphi_{\zeta_m}}
\end{equation*}
while for usual integrability objects we use the notation 
\begin{equation}
\nover{\bm X}_{\varphi | \psi} (\zeta_1, \, \ldots, \, \zeta_m) 
= \bm X_{\varphi_{\zeta_1} \otimes_\Delta \cdots \otimes_\Delta \varphi_{\zeta_m} | \underbracket[.5pt]{\scriptstyle \psi \otimes_\Delta \cdots \otimes_\Delta \psi}_n} \label{xfpz}
\end{equation}
and
\begin{equation}
\nover{\bm Y}^\chi_{\varphi | \psi} (\zeta_1, \, \ldots, \, \zeta_m) 
=  \bm Y^{\chi_1 \otimes \cdots \chi_m}_{\varphi_{\zeta_1} \otimes_\Delta \cdots \otimes_\Delta \varphi_{\zeta_m} | \underbracket[.5pt]{\scriptstyle \psi \otimes_\Delta \cdots \otimes_\Delta \psi}_n} \, . \label{yfpz}
\end{equation}
When $n = 1$ we usually write just $\bm X_{\varphi | \psi} (\zeta_1, \, \ldots, \, \zeta_m)$ and $\bm Y^\chi_{\varphi | \psi} (\zeta_1, \, \ldots, \, \zeta_m)$. In accordance with our conventions, we assume that
\begin{equation}
\nover{\rho}_{\varphi | \psi}(\zeta_1, \, \ldots, \, \zeta_m) = \rho_{\varphi_{\zeta_1} \otimes_\Delta \cdots \otimes_\Delta \varphi_{\zeta_m} | \underbracket[.5pt]{\scriptstyle \psi \otimes_\Delta \cdots \otimes_\Delta \psi}_n} = \prod_{i = 1}^m (\rho_{\varphi_{\zeta_i} | \psi})^n = \prod_{i = 1}^m \rho_{\varphi | \psi}(\zeta_i)^n, \label{rfp}
\end{equation}
see equations (\ref{xfff}). It is worth to note that from the point of view of quantum integrable systems the integrability objects $\nover{\bm X}_{\varphi | \psi} (\zeta_1, \, \ldots, \, \zeta_m)$ and $\nover{\bm Y}_{\varphi | \psi} (\zeta_1, \, \ldots, \, \zeta_m)$ correspond to the quantum chains of length $n$.

If $m = n = 1$, an integrability object $\bm X_{\varphi | \psi}(\zeta) = \overset{\scriptscriptstyle (1)}{\bm X}{}_{\varphi | \psi}(\zeta)$ defined by equation (\ref{xfpz}) is said to be a basic integrability object. With the help of equations (\ref{xfff}), all other integrability objects of type $X$ can be expressed through the basic ones. 

All integrability objects that we use in this paper, are constructed as described above. However, depending on the role they play in the integration procedure, they are given different names. Below we describe the main classes of integrability objects. It is worth to have in mind that the proposed classification is rather conditional, although it is widespread.

The most famous integrability objects are the $R$ -operators. They form a special class of integrability objects of type $X$ used to permute integrability objects of type $Y$. However, there is a more general method for demonstrating commutativity of integrability objects of type $Y$, described in the previous section, and we do not define and use $R$-operators in the present paper.

\subsection{Monodromy operators and transfer operators}

When $\varphi$ is a homomorphism from the quantum loop algebra $\uqlg$ to an algebra $A$, $\psi$ a representation of $\uqlbm$ on a vector space $U$, the integrability object $\nover{\bm X}_{\varphi | \psi}(\zeta)$ is called a monodromy operator and denoted as $\nover{\bm M}_{\varphi | \psi}(\zeta)$.

The type $Y$ companion of a monodromy operator $\nover{\bm M}_{\varphi | \psi}(\zeta)$ is called a transfer operator and denoted $\nover{\bm T}^\chi_{\varphi | \psi}(\zeta)$. Explicitly we have
\begin{equation*}
\nover{\bm T}^\chi_{\varphi | \psi}(\zeta) = \big( \tr_\chi \otimes \id_{\End(W^{\otimes n})} \big) \big( \nover{\bm M}_{\varphi | \psi}(\zeta) (\varphi_\zeta(q^t) \otimes 1_{\End(W^{\otimes n})}) \big),
\end{equation*}
where $q^t$ is a twisting element, which we define by equation (\ref{tqfh}), $\chi$ is a representation of the algebra $A$, and the trace $\tr_\chi$ is given by equation (\ref{traa}).

Let $\varphi_1$ and $\varphi_2$ be homomorphisms from $\uqlg$ to algebras $A_1$ and $A_2$, respectively, $\psi$ a representation of $\uqlbm$, and $\chi_1$ and $\chi_2$ representations of $A_1$ and $A_2$. It follows from equation (\ref{yyyy}) that
\begin{equation*}
\nover{\bm T}^{\chi_1}_{\varphi_1 | \psi}(\zeta_1) \, \nover{\bm T}^{\chi_2}_{\varphi_2 | \psi}(\zeta_2) = \nover{\bm T}^{\chi_2}_{\varphi_2 | \psi}(\zeta_2) \, \nover{\bm T}^{\chi_1}_{\varphi_1 | \psi}(\zeta_1),
\end{equation*}
for any $\zeta_1, \zeta_2 \in \bbC^\times$. Similar commutativity takes place for the universal transfer operators,
\begin{equation*}
\calT^{\chi_1}_{\varphi_1}(\zeta_1) \calT^{\chi_2}_{\varphi_2}(\zeta_2) = \calT^{\chi_2}_{\varphi_2}(\zeta_2) \calT^{\chi_2}_{\varphi_2}(\zeta_2).
\end{equation*}
Emphasize that it is important for commutativity that the twist element is group like.

In this paper we construct the monodromy operators using as $\varphi$ the evaluation representations $\widetilde \varphi^\lambda$ and $\varphi^\lambda$ defined in section \ref{s:lwer}, and as $\psi$ the finite-dimensional evaluation representation $\varphi^{\omega_1}$. Here the following notation is used
\begin{equation*}
\nover{\bm T}^\lambda(\zeta) = \nover{\bm T}{}_{\widetilde \varphi^\lambda | \varphi^{\omega_1}}(\zeta), \qquad \nover{\bm T}{}^\lambda(\zeta) = \nover{\bm T}{}_{\varphi^\lambda | \varphi^{\omega_1}}(\zeta).
\end{equation*}
Similarly, for the corresponding universal transfer operators we use the notation
\begin{equation*}
\widetilde \calT^\lambda(\zeta) = \calT_{\widetilde \varphi^\lambda | \varphi^{\omega_1}}(\zeta), \qquad \calT^\lambda(\zeta) = \calT_{\varphi^\lambda | \varphi^{\omega_1}}(\zeta).
\end{equation*}

One can demonstrate that the transfer operators for the evaluation representations of $\uqlbp$ depend on $\zeta$ via $\zeta^s$, see the papers \cite{NirRaz16a, BooGoeKluNirRaz14b} for $l = 1$ and $l = 2$.

\subsection{\texorpdfstring{$L$-operators and $Q$-operators}{L-operators and Q-operators}}

\subsubsection{Definition}

Now let $\varphi$ be a homomorphism from $\uqlbp$ to an algebra $A$ which cannot be extended to a homomorphism of $\uqlg$, and $\psi$ a representation of the Borel subalgebra $\uqlbm$ on a vector space $U$. In this case the integrability object $\nover{\bm X}_{\varphi | \psi}(\zeta)$ defined by equation (\ref{xfpz}) is called an $L$-operator and denoted $\nover{\bm L}_{\varphi | \psi}(\zeta)$.

The companion of an $L$-operator $\underset{n}{\bm L}{}_{\varphi | \psi}(\zeta)$ of type $Y$ is called a $Q$-operator and is denoted $\nover{\bm Q}^\chi_{\varphi | \psi}(\zeta)$. Explicitly, we have
\begin{equation*}
\nover{\bm Q}^\chi_{\varphi | \psi}(\zeta) = \big( \tr_\chi \otimes \id_{\End(U^{\otimes n})} \big) \big( \underset{n}{\bm L}{}_{\varphi | \psi}(\zeta) (\varphi_\zeta(q^t) \otimes 1_{\End(U^{\otimes n})}) \big),
\end{equation*}
where $q^t$ is a twisting element, which we define by equation (\ref{tqfh}), $\chi$ is a representation of the algebra $A$ and the trace $\tr_\chi$ is given by equation (\ref{traa}).

Let $\varphi_1$ and $\varphi_2$ be homomorphisms from $\uqlbp$ to algebras $A_1$ and $A_2$ which cannot be extended to homomorphisms of $\uqlg$, $\psi$ a representation of $\uqlg$, and $\chi_1$ and $\chi_2$ representations of $A_1$ and $A_2$. It follows from equation (\ref{yyyy}) that
\begin{equation*}
\nover{\bm Q}^{\chi_1}_{\varphi_1 | \psi}(\zeta_1) \, \nover{\bm Q}^{\chi_2}_{\varphi_2 | \psi}(\zeta_2) = \nover{\bm Q}^{\chi_2}_{\varphi_2 | \psi}(\zeta_2) \, \nover{\bm Q}^{\chi_1}_{\varphi_1 | \psi}(\zeta_1)
\end{equation*}
for any $\zeta_1, \zeta_2 \in \bbC^\times$, and similarly for the universal $Q$-operators,
\begin{equation*}
\calQ^{\chi_1}_{\varphi_1}(\zeta_1) \calQ^{\chi_2}_{\varphi_2}(\zeta_2) = \calQ^{\chi_2}_{\varphi_2}(\zeta_2) \calQ^{\chi_2}_{\varphi_2}(\zeta_2).
\end{equation*}

One can demonstrate that the $Q$-operators for the oscillator representations of the Borel subalgebra $\uqlbp$ depend on $\zeta$ via $\zeta^s$, see the papers \cite{NirRaz16a, BooGoeKluNirRaz14b} for $l = 1$ and $l = 2$.

\subsubsection{Case of oscillator representations}

Let us find the basic $L$-operator for the case when $\varphi$ is the homomorphism $o$ defined by equations (\ref{roh0e0})--(\ref{rohlel}), and $\psi$ the representation $\varphi^{\omega_1}$ described by equations (\ref{fooa})--(\ref{food}). We denote this $L$-operator simply as $\bm L(\zeta)$. We have
\begin{equation*}
\rho_{o | \varphi^{\omega_1}} \bm L(\zeta) = (o \otimes \varphi^{\omega_1})(\calR_{\prec \delta} \calR_{\sim \delta} \calR_{\prec \delta}) \bm K_{o^{}_\zeta | \varphi^{\omega_1}}.
\end{equation*}
First find the expression for the factor $\bm K_{o | \varphi^{\omega_1}}$. To this end we use equation (\ref{kfpb}). Recall that the representation $\varphi^{\omega_1}$ is $(l + 1)$-dimensional, the basic vectors $v_m$ are the weight vectors of weights $\lambda_m$ given by equation (\ref{lmoms}). It is clear that the projector on the weight vector $v_m$ is $\bbE_{m m}$. Thus, we have
\begin{equation*}
\bm K_{o^{}_\zeta | \varphi^{\omega_1}} = \sum_{m = 1}^{l + 1} o_\zeta \big(q^{- \sum_{i, j = 1}^l h_i c_{i j} \langle \lambda_m, \, h_j \rangle} \big) \otimes \bbE_{m m}.
\end{equation*}
Using equation (\ref{cij}), we obtain
\begin{equation*}
q^{- \sum_{i, j = 1}^l h_i c_{i j} \langle \lambda_m, \, h_j \rangle} = q^{\frac{1}{l + 1} (\sum_{i = 1}^{m - 1} i h_i  - \sum_{i = m}^l (l - i + 1) h_i)} = q^{\frac{1}{l + 1} \sum_{i = 1}^l i h_i  - \sum_{i = m}^l h_i}.
\end{equation*}
Now, taking into account (\ref{rohiei}) and (\ref{rohlel}), we see that
\begin{equation*}
o_\zeta \big(q^{- \sum_{i, j = 1}^l h_i c_{i j} \langle \lambda_m, \, h_j \rangle} \big) = q^{N_m}, \quad m = 1, \ldots, l, \qquad o_\zeta \big( q^{- \sum_{i, j = 1}^l h_i c_{i j} \langle \lambda_{l + 1}, \, h_j \rangle} \big) = q^{- \sum_{i = 1}^l N_i}.
\end{equation*}
Thus, we have
\begin{equation*}
\bm K_{o^{}_\zeta | \varphi^{\omega_1}} = \sum_{m = 1}^l q^{N_m} \otimes \bbE_{m m} + q^{- N_{1, \, l + 1}} \otimes \bbE_{l + 1, \, l + 1}.
\end{equation*}
Here and below we use the notation $N_{i j} = \sum_{k = i}^{j - 1} N_k$.

It follows from equations (\ref{cwby1})--(\ref{cwby5}) that
\begin{align*}
& o_\zeta(e_{\alpha_{i j} + n \delta}) = - \zeta^{s_{i j}} \, \delta_{n 0} \, q^{- i + j - 2} b^{}_i b^\dagger_j q^{N_{i j} - N_j}, && 1 \le i < j \le l, \\
& o_\zeta(e_{\alpha_{i, \, l + 1} + n \delta}) = - \zeta^{s_{i, \, l + 1}} \delta_{n 0} \, \kappa_q^{-1} q^{- i + l} b_i \, q^{N_{i, \, l + 1}}, && 1 \le i \le l, \\
& o_\zeta(e_{(\delta - \alpha_{i j}) + n \delta}) = - \zeta^{(s - s_{i j}) + n \delta} \delta_{n 0} \, \delta_{l + 1, \, j} \, (-1)^i b^\dagger_i q^{N_{i + 1, \, l + 1}}, && 1 \le i < j \le l + 1, \\
& o_\zeta(e'_{n \delta; \, \alpha_{i j}}) = \zeta^{n s} \delta_{n 1} \delta_{l + 1, \, j} \, \kappa_q^{-1} (-1)^i q^{- i + l + 1} q^{2 N_{i + 1, \, l + 1}}, && 1 \le i < j \le l + 1.
\end{align*}
Using the first equation of (\ref{efdg}), we obtain
\begin{equation*}
o_\zeta(e_{n \delta; \, \alpha_{i j}}) = \zeta^{n s} \delta_{l + 1, \, j} \, \kappa_q^{-1} (-1)^{i n} q^{ - (i - l - 1) n} q^{2 n N_{i + 1, \, l + 1}} 
\end{equation*}
for all $1 \le i < j \le l + 1$. In turn, equations (\ref{cwby2})--(\ref{cwby6}) give
\begin{align}
& \varphi^{\omega_1}(f_{\alpha_{i j} + n \delta}) = (-1)^{i n} q^{- (i + 1) n} \bbE_{j i}, \label{fij} \\
& \varphi^{\omega_1}(f_{(\delta - \alpha_{i j}) + n \delta}) = - (-1)^{i (n + 1)} q^{- (i + 1) n - i} \bbE_{i j}, \\
& \varphi^{\omega_1}(f'_{n \delta, \, \alpha_{i j}}) = - (-1)^{i n} q^{- (i + 1) n + 1}(\bbE_{i i} - q^{- 2} \bbE_{j j}), \label{fzeff}
\end{align}
and the second equation of (\ref{efdg}) implies
\begin{equation*}
\varphi^{\omega_1}(f_{n \delta, \, \alpha_{i j}}) = - (-1)^{i n} q^{- i n} \frac{[n]_q}{n} (\bbE_{i i} - q^{- 2 n} \bbE_{j j}).
\end{equation*}
Now we find the expression
\begin{multline*}
(o_\zeta \otimes \varphi^{\omega_1}) \big( - \kappa_q \sum_{n = 1}^\infty \sum_{i, \, j = 1}^l u_{n, \, i j} \, e_{n \delta, \, \alpha_i} \otimes f_{n \delta, \, \alpha_j} \big) \\*
= F_{l + 1}(\zeta^s ) \big( 1_{\Osc^{\otimes l}} \otimes \mathbbm 1 \big) + \log (1 - q^{-l} \zeta^s) \big( 1_{\Osc^{\otimes l}} \otimes \bbE_{l + 1, \, l + 1} \big)
\end{multline*}
and obtain
\begin{equation*}
(o_\zeta \otimes \varphi^{\omega_1})(\calR_{\sim \delta}) = \rme^{F_{l + 1}(\zeta^s)} \big( 1_{\Osc^{\otimes l}} \otimes (\sum_{k = 1}^l \bbE_{k k} - q^{-l} \zeta^s \bbE_{l + 1, \, l + 1}) \big).
\end{equation*}
Here the transcendental function $F_{l + 1}(\zeta)$ is defined as
\begin{equation}
F_{l + 1}(\zeta) = \sum_{n = 1}^\infty \frac{1}{[l + 1]_{q^n}} \frac{\zeta^n}{n}. \label{fmz}
\end{equation}
It is not difficult to determine that
\begin{multline*}
(o_\zeta \otimes \varphi^{\omega_1})(\calR_{\prec \delta}) = \big( 1_{\Osc^{\otimes l}} \otimes \mathbbm 1 \big) \\*
+ \sum_{\substack{i, \, j = 1 \\
 i < j}}^l \zeta^{s_{i j}} q^{-i + j + 2} b^{}_i b^\dagger_j q^{N_{i j} - N_j} \otimes \bbE_{j i} + \sum_{i = 1}^l \zeta^{s_{i, \, l + 1}} q^{- i + l} b^{}_i q^{N_{i, \, l + 1}} \otimes \bbE_{l + 1, \, i}
\end{multline*}
and
\begin{equation*}
(o_\zeta \otimes \varphi^{\omega_1})(\calR_{\succ \delta}) = \big( 1_{\Osc^{\otimes l}} \otimes \mathbbm 1 \big) - \kappa_q \sum_{i = 1}^l \zeta^{s - s_{i j}} q^{- i} b^\dagger_i q^{N_{i + 1, \, l +1}} \otimes \bbE_{i, \, l + 1}.
\end{equation*}
Here we use the identities
\begin{gather*}
(\bbE_{i j} \otimes \bbE_{j i})^k = 0
\end{gather*}
for all $1 \le i < j \le l + 1$ and $k > 1$, and
\begin{gather*}
(\bbE_{i j} \otimes \bbE_{j i})(\bbE_{k m} \otimes \bbE_{m k}) = 0
\end{gather*}
for all $1 \le i < j \le l + 1$ and $1 \le k < m \le l + 1$. 

Assuming that the normalization factor is given by the equation  
\begin{equation}
\rho_{o | \varphi^{\omega_1}}(\zeta) = \rme^{F_{l + 1}(\zeta^s)}, \label{rof}
\end{equation}
we finally obtain\footnote{There are some inaccuracies in the corresponding formulas of the paper \cite{NirRaz17a}.}
\begin{multline}
\bm L(\zeta) = \sum_{\substack{i, \, j = 1 \\ j < i}}^l \zeta^{s_{j i}} \kappa_q b^{}_j b^\dagger_i q^{N_j + N_{j i} - N_i - j + i - 2} \otimes \bbE_{i j} \\*
+ \sum_{j = 1}^l \zeta^{s_{j, \, l + 1}} b_j q^{N_j + N_{j, l + 1} - j + l} \otimes \bbE_{l + 1, \, j} - \sum_{i = 1}^l \zeta^{s - s_{i, \, l + 1}} \kappa_q b^\dagger_i q^{N_{1 i} - N_i + i - 2} \otimes \bbE_{i, \, l + 1} \\+ \sum_{i = 1}^l q^{N_i} \otimes \bbE_{i i} + (q^{- N_{1, l + 1}} - \zeta^s q^{N_{1, l + 1} + l}) \otimes \bbE_{l + 1, \, l + 1}. \label{lzeta}
\end{multline}

Using the set of the homomorphisms $o_a$, given by equation (\ref{roaovroa}), we define a family of $L$-operators
\begin{equation*}
\bm L_a(\zeta) = \bm L_{o_a | \varphi^{\omega_1}}(\zeta).
\end{equation*}
By definition, $\bm L_{a + l + 1}(\zeta) = \bm L_a$, so that there are only $l + 1$ distinct $L$-operators of this type.

It is easy to demonstrate that for any $x \in \uqlsllpo$,
\begin{equation}
(\varphi \circ \sigma^{-1})_\zeta(x) = \left (\varphi_\zeta \circ \sigma^{-1})(x) \right|_{s \to \sigma(s)}, \label{fszx}
\end{equation}
where $s \to \sigma(s)$ stands for the set of substitutions
\begin{equation*}
s_0 \to s_1, \quad s_1 \to s_2, \quad \ldots, \quad s_l \to s_0.
\end{equation*}
Using equations (\ref{xfsfs}) and (\ref{fszx}), we obtain
\begin{equation} 
\bm L_{\varphi \circ \sigma^{-1} | \psi \circ \sigma^{-1}}(\zeta) = \left. \bm L_{\varphi | \psi}(\zeta) \right|_{s \to \sigma(s)}. \label{xfsps}
\end{equation}
Taking this equation into account, we come to
\begin{equation*}
\bm L_{o_a | \varphi_{}^{\omega_1}}(\zeta) = \bm L_{o_{a - 1} \circ \sigma^{-1} | \varphi^{\omega_1} \circ \sigma \circ \sigma^{-1}}(\zeta) =  \left. \bm L_{o_{a - 1} | \varphi^{\omega_1} \circ \sigma}(\zeta) \right|_{s \to \sigma(s)}.
\end{equation*}
One can verify that for any $x \in \uqlsllpo$
\begin{equation*}
(\varphi^{\omega_1} \circ \sigma)(x) = O \, \varphi^{\omega_1}(x) \, O^{-1},
\end{equation*}
where $O$ is the square matrix of size $l + 1$ defined as
\begin{equation*}
O = q^{-1} \bbE_{1, \, l + 1} + \sum_{i = 1}^l \bbE_{i + 1, \, i}.
\end{equation*}
Hence, we have
\begin{equation}
\bm L_a(\zeta) = \left. (1_{\Osc^{\otimes l}} \otimes O) \bm L_{a - 1}(\zeta) (1_{\Osc^{\otimes l}} \otimes O^{-1}) \right|_{s \to \sigma(s)}. \label{lazeta}
\end{equation}

To define the corresponding $Q$-operators we use the trace $\tr_{\chi_a}$, where $\chi_a$ is the representation of $\Osc^{\otimes l}$ defined by equation (\ref{chia}), and use the notation
\begin{equation*}
\calQ'_a(\zeta) = \calQ^{\chi_a}_{o_a}(\zeta) = \calQ_{\theta_a}(\zeta)
\end{equation*}
for the universal $Q$-operators, and
\begin{equation*}
\nover{\bm Q}'_a(\zeta) = \nover{\bm Q}_{o_a | \varphi^{\omega_1}}^{\chi_a}(\zeta)= \nover{\bm Q}_{\theta_a | \varphi^{\omega_1}}(\zeta)
\end{equation*}
for the usual $Q$-operators. We use a prime to indicate that we redefine these operators below, see equation (\ref{qqp}). 

\section{Functional relations}

\subsection{Factorization of transfer operators}

In this section we generalize the consideration given in the paper \cite{Raz21a} for $l = 1$ and $l = 2$ to the case of general $l$. Below we treat the $\uqlsllpo$-modules $\widetilde V^\mu_\zeta$ and $V^\mu_\zeta$ as $\uqlbp$-modules.

Let us consider the following tensor product of $l + 1$ oscillator $\uqlbp$-modules
\begin{equation*}
W^\mu(\zeta) = (W_1)_{\zeta^\mu_1} \otimes_\Delta  \cdots \otimes_\Delta (W_{l + 1})_{\zeta^\mu_{l + 1}},
\end{equation*}
where
\begin{equation*}
\zeta^\mu_a = q^{2 (\mu_a  + \rho_a) / s} \zeta.
\end{equation*}
with
\begin{equation*}
\rho_a = l/2 - a + 1.
\end{equation*}
It is worth to note that
\begin{equation}
\rho = \sum_{a = 1}^{l + 1} \rho_a \, \epsilon_a \label{rho}
\end{equation}
is the half sum of positive roots of $\gllpo$.

Using the formulas of section \ref{s:lwor}, equations (\ref{slhw}) and (\ref{slhlw}), we see that the product $\bm \Psi_{1, \, \bm 0}(\zeta_1^\mu) \ldots \bm \Psi_{l + 1, \, \bm 0}(\zeta_{l + 1}^\mu)$ of the highest $\ell$-weights of the factors, forming $W^\mu(\zeta)$, coincides with the highest $\ell$-weight $\bm \Lambda^\mu_{\bm 0}[\delta](\zeta)$ of the shifted evaluation $\uqlsllpo$-module $(\widetilde V^\mu[\delta])_\zeta$ with
\begin{equation}
\delta = \sum_{i = 1}^l(- 2 - \mu_i + \mu_{i + 1}) \, \omega_i. \label{delta}
\end{equation}
This means that the submodule of $W^\mu(\zeta)$ generated by the tensor product of the highest $\ell$-weight vectors of the oscillator factors is isomorphic to $(\widetilde V^\mu[\delta])_\zeta$. Hence, the $q$-character of $W(\zeta)$ should contain a summand coinciding with the $q$-character of $(\widetilde V^\mu[\delta])_\zeta$. Let us show that the entire $q$-character of $W(\zeta)$ is the sum of such $q$-characters.

Introduce an independent labeling for the oscillators, using for it the tuple $\bm n = (n_{a, \, i})$, where $a = 1, \ldots, l + 1$ and $i = 1, \ldots, l$. Denote the $\ell$-weights of the module $W^\mu(\zeta)$ as $\bm \Xi^\mu_{\bm n}(\zeta) = (\xi^\mu_{\bm n}, \, \bm \Xi^{\mu +}_{\bm n}(\zeta))$. Using expressions from section \ref{s:lwor}, we obtain that
\begin{multline}
\xi^\mu_{\bm n} = \sum_{i = 1}^l \Big[ - 2 + \sum_{k = 1}^{i - 1} (n_{k, \, i - k} - n_{k, \, i - k + 1} + n_{i, \, k - i + l + 1} - n_{i + 1, \, k - i + l}) - 2 n_{i, \, 1} - 2 n_{i + 1, \, l} \\
- \sum_{k = i + 2}^{l + 1} (n_{i, \, k - i} - n_{i + 1, \, k - i - 1} + n_{k, \, i - k + l + 1} - n_{k, \, i - k + l + 2}) \Big] \omega_i \label{ximu}
\end{multline}
and $\Xi^{\mu +}_{\bm n}(\zeta) = (\Xi^{\mu +}_{\bm n, \, i}(\zeta, \, u))_{i = 1, \ldots, l}$, where
\begin{multline}
\Xi^{\mu +}_{\bm n, \, i}(\zeta, \, u) = \frac{\prod_{a = 1}^{i - 1} (1 - q^{2 \mu_a - 2 \sum_{j = i - a}^{l - a + 1} n_{a j} + i - 2 a + 1} \zeta^s u) }{\prod_{a = 1}^i (1 - q^{2 \mu_a - 2 \sum_{j = i - a + 1}^{l - a + 1} n_{a j} + i - 2 a + 1} \zeta^s u)} \\
\times \frac{\prod_{a = 1}^{i + 1} (1 - q^{2 \mu_a - 2 \sum_{j = i - a + 2}^{l - a + 1} n_{a j} + i - 2 a + 3} \zeta^s u)}{\prod_{a = 1}^i (1 - q^{2 \mu_a - 2 \sum_{j = i - a + 1}^{l - a + 1} n_{a j} + i - 2 a + 3} \zeta^s u)}. \label{ximup}
\end{multline}
One can easily get convinced that $\bm \Xi^{\mu +}_{\bm n}(\zeta)$ does not depend on $n_{a, \, i}$ with $a + i > l + 1$. For $a + i \le l + 1$ denote
\begin{equation}
 m_{a, \, i} = n_{a, \, i - a} \label{naimai}
\end{equation}
and rewrite (\ref{ximu}) as
\begin{multline*}
\xi^\mu_{\bm n} = \sum_{i = 1}^l \Big[ \mu_i - \mu_{i + 1} + \sum_{k = 1}^{i - 1} (m_{k, \, i} - m_{k, \, i + 1}) - 2 m_{i, \, i + 1} - \sum_{k = i + 2}^{l + 1} (m_{i, \, k} - m_{i + 1, \, k}) \Big] \omega_i \\
+ \sum_{i = 1}^l \Big[ - 2 - \mu_i + \mu_{i + 1} + \sum_{k = 1}^{i - 1} (n_{i, \, k - i + l + 1} - n_{i + 1, \, k - i + l}) \\
- 2 n_{i + 1, \, l} - \sum_{k = i + 2}^{l + 1} (n_{k, \, i - k + l + 1} - n_{k, \, i - k + l + 2}) \Big] \omega_i.
\end{multline*}
Denote also the subtuple of $\bm n$ formed by $n_{a, \, i}$ with $a + i > l + 1$ by $\bm n'$. Comparing the above equation with (\ref{psimum}), we see that the weights $\xi^\mu_{\bm n}$ for $\bm n' = \bm 0$, after the identification (\ref{naimai}), coincide with the weights of the module $\widetilde V^\mu_\zeta[\delta](\zeta)$, where $\delta$ is defined by equation (\ref{delta}). It is natural to assume that $\bm \Xi^{\mu +}_{\bm n}(\zeta)$ for $\bm n' = \bm 0$, after the identification (\ref{naimai}), coincides with the component $\bm \Lambda^{\mu +}_{\bm m}[\delta](\zeta)$ of the $\ell$-weight $\bm \Lambda^\mu_{\bm m}[\delta](\zeta)$ of the module $\widetilde V^\mu_\zeta[\delta]$. Stress that $\bm \Lambda^\mu_{\bm m}[\delta](\zeta) = \bm \Lambda^\mu_{\bm m}(\zeta)$.

 More generally, it is natural to assume that the $\ell$-weight $\bm \Xi^\mu_{\bm n}(\zeta)$ for any fixed $\bm n'$, after the identification (\ref{naimai}), coincides with the $\ell$-weight $\bm \Lambda^\mu_{\bm m}[\delta_{\bm n'}](\zeta)$ of the $\uqlsllpo$-module $\widetilde V^\mu_\zeta[\delta_{\bm n'}]$, where
\begin{multline*}
\delta_{\bm n'} = \sum_{i = 1}^l \big[ - 2 - \mu_i + \mu_{i + 1} + \sum_{k = 1}^{i - 1}(n_{i, \, k - i + l + 1} - n_{i + 1, \, k - i + l}) \\
{} - 2 n_{i + 1, \, l} - \sum_{k = i + 2}^{l + 1} (n_{k, \, i - k + l + 1} - n_{k, \, i - k + l + 2}) \big] \omega_i.
\end{multline*}
This very plausible assumption is supported at least by the fact that it is true for $l = 1$ and $l = 2$, and we will assume that it is true for arbitrary $l$. Similarly as above, this results in the relation
\begin{equation*}
\chi_q(W^\mu(\zeta)) = \sum_{\bm n'} \chi_q(\widetilde V^\mu_\zeta [\delta_{\bm n'}]),
\end{equation*}
satisfied by $q$-characters, which is equivalent to the relation in the Grothendieck ring
\begin{equation*}
\big \langle \, W^\mu(\zeta) \, \big \rangle = \sum_{\bm n'} \big \langle \, \widetilde V^\mu_\zeta [\delta_{\bm n'}] \, \big \rangle = \big \langle \bigoplus_{\bm n'} \widetilde V^\mu_\zeta [\delta_{\bm n'}] \, \big \rangle.
\end{equation*}
Here the summation over $\bm n'$ means the summation over the components of $\bm n'$ from $0$ to $\infty$. It follows that
\begin{equation*}
\calY_{W^\mu(\zeta)} = \sum_{\bm n'} \calY_{ \widetilde V^\mu_\zeta [\delta_{\bm n'}]}.
\end{equation*}
Having in mind that
\begin{equation*}
\calY_{W^\mu(\zeta)} = \calY_{\theta_1}(\zeta^\mu_1) \ldots \calY_{\theta_{l + 1}}(\zeta^\mu_{l + 1}) = \calQ'_1(\zeta^\mu_1) \ldots \calQ'_{l + 1}(\zeta^\mu_{l + 1}),
\end{equation*}
see equation (\ref{yffp}), and taking into account equation (\ref{yfsp}), we come to the equation
\begin{equation*}
 \calQ'_1(\zeta^\mu_1) \ldots \calQ'_{l + 1}(\zeta^\mu_{l + 1}) = q^{- \sum_{i, \, j = 1}^l \langle \delta_{\bm n'}, h^{}_i \rangle c^{}_{i j} \, h'_j} \, \widetilde \calT^\mu(\zeta).
\end{equation*}
By a direct calculation we obtain
\begin{multline*}
- \sum_{i, \, j = 1}^l \langle \delta_{\bm n'}, h^{}_i \rangle c^{}_{i j} \, h'_j \\*
=  {} - \frac{1}{l + 1} \sum_{i = 1}^{l + 1} \big( \sum_{j = 1}^{i - 1} j \, h'_j - \sum_{j = i}^l (l - j + 1) h'_j \big) \mu_i + \quad \sum_{\mathclap{1 \le i < j \le l + 1}} \quad h'_{i j} + \quad \sum_{\mathclap{1 \le i < j \le l + 1}} \quad h'_{i j} n_{j, \, i - j + l + 1},
\end{multline*}
where
\begin{equation*}
h'_{i j} = \sum_{k = i}^{j - 1} h'_k.
\end{equation*}
Thus, we have the following factorization relation
\begin{multline}
\bigg( \prod_{i = 1}^{l + 1} q^{- \frac{1}{l + 1} \big( \sum_{j = 1}^{i - 1} j \, h'_j - \sum_{j = i}^l (l - j + 1) h'_j \big) \mu_i}  \quad \prod_{\mathclap{1 \le i < j \le l + 1}} \quad  q^{h'_{i j}} (1 - q^{h'_{i j}})^{-1} \bigg) \, \widetilde \calT^\mu(\zeta) \\
{} = \calQ'^{(1)}(\zeta_1^\mu) \cdots \calQ'^{(l + 1)}(\zeta_{l + 1}^\mu). \label{tmuqq}
\end{multline}
Now introduce the notation
\begin{equation}
\calD_a = \frac{s}{2(l + 1)} \Big[ \sum_{j = 1}^{a - 1} j \, h'_j - \sum_{j = a}^l (l - j + 1) \, h'_j \Big] \label{cda}
\end{equation}
and define new universal $Q$-operators
\begin{equation}
\calQ_a(\zeta) = \zeta^{\calD_a} \calQ'_a(\zeta). \label{qqp}
\end{equation}
In terms of these operators the factorization relation takes the form
\begin{equation}
\calC^{}_l \, \widetilde \calT^\lambda(\zeta) = \calQ^{}_1(\zeta_1^\mu) \cdots \calQ^{}_{l + 1}(\zeta_{l + 1}^\mu), \label{tqq}
\end{equation}
where
\begin{equation*}
\calC_l = \quad \prod_{\mathclap{1 \le i < j \le l + 1}} \quad q^{h'_{i j}/2}(1 - q^{h'_{i j}})^{-1}.
\end{equation*}
The advantage of this formula over (\ref{tmuqq}) is that the factor $\calC_l$ does not depend on $\mu$, and this is necessary to obtain the determinant representation of the universal transfer operator $\calT^\lambda(\zeta)$.

\subsection{Determinant representation}

The Weyl group $\calW$ of the root system of $\gllpo$ is generated by the reflections $r_i: (\gllpo)^* \to (\gllpo)^*$, $i = 1, \ldots, l$, defined by the equation
\begin{equation*}
r_i(\mu) = \mu - \langle \mu, \, H_i\rangle \, \alpha_i.
\end{equation*}
The minimal number of generators $r_i$ necessary to represent an element $w \in \calW$ is said to be the length of $w$ and is denoted by $l(w)$. It is assumed that the identity element has the length equal to $0$.

It is easy to demonstrate that the reflection $r_i$ transposes the components $\mu_i$ and $\mu_{i + 1}$ of $\mu$, and the whole Weyl group $\calW$ can be identified with the symmetric group $\mathrm S_{l + 1}$. Here $(-1)^{l(w)}$ is evidently the sign of the permutation corresponding to the element $w \in \calW$.  The order of $\calW$ is equal to $l!$, and the length of the elements of $\calW$ runs from $0$ to $l (l + 1) / 2$.

Consider now a finite-dimensional $\uqlbp$-module $V^\mu$, see for the definition section~\ref{s:vm}. Introduce the following direct sums of infinite-dimensional  $\uqgllpo$-modules:
\begin{equation*}
\widehat V_k = \bigoplus_{\substack{w \in \calW \\ l(w) = k}} \widetilde V^{w \cdot \mu},
\end{equation*}
where $k = 0, 1, \ldots, l (l + 1)/2$, and $w \cdot \mu$ means the affine action of $w$ defined as
\begin{equation*}
w \cdot \mu = w(\mu + \rho) - \rho
\end{equation*}
with $\rho$ given by equation (\ref{rho}). The quantum version of the Bernstein--Gelfand--Gelfand resolution for the quantum group $\uqgllpo$ is the following exact sequence of $\uqgllpo$-modules and $\uqgllpo$-homomorphisms:
\begin{equation*}
0 \longrightarrow \widehat V_{l(l + 1)/2} \overset{f_{l(l + 1)/2}}{\longrightarrow} \cdots \longrightarrow \widehat V_2 \overset{f_2}{\longrightarrow} \widehat V_1 \overset{f_1}{\longrightarrow} \widehat V_0 \overset{f_0}{\longrightarrow} \widehat V_{-1} \longrightarrow 0,
\end{equation*}
where $\widehat V_{-1} = V^\mu$, see, for example, the papers \cite{Ros91, Mal92, HecKol07}.

Let $\hpi_k$ be the representation of $\uqgllpo$ corresponding to the $\uqgllpo$-module $\widehat V_k$. Note that $\hpi_{-1} = \pi^\mu$. The subspaces $\ker f_k$ and $\im f_k$ are $\uqgllpo$-submodules of $U_k$ and $U_{k - 1}$ respectively. For each $k = 0, 1, \ldots, (l + 1)l/2$, we have
\begin{equation*}
\tr_{\varphi_k} = \tr_{\hpi_k}|_{\ker f_k} + \tr_{\hpi_{k - 1}} |_{\im f_k}.
\end{equation*}
By the definition of an exact sequence, we have 
\begin{equation*}
\im f_k = \ker f_{k - 1}.
\end{equation*}
Hence,
\begin{equation*}
\sum_{k = 0}^{l(l + 1)/2} (-1)^k \tr_{\hpi_k} = \tr_{\hpi_{-1}}|_{\im f_0} - \tr_{\hpi_{l(l + 1)/2}} |_{\ker f_{l(l + 1)/2}}.
\end{equation*}
Finally, having in mind that
\begin{equation*}
\im f_0 = V^\mu, \qquad \ker f_{l(l + 1)/2} = 0,
\end{equation*}
we obtain
\begin{equation*}
\sum_{k = 0}^{l(l + 1)/2} (-1)^k \tr_{\hpi_k} = \tr_{\pi^\mu}.
\end{equation*}
From the definition of $\widehat V_k$ it follows that
\begin{equation*}
\sum_{k = 0}^{(l + 1)l/2} (-1)^k \tr_{\hpi_k} = \sum_{w \in \calW} (-1)^{l(w)} \widetilde \tr_{\pi^{\, w \cdot \mu}}.
\end{equation*}
Thus, we have
\begin{equation*}
\tr_{\pi^\mu} = \sum_{w \in \calW} (-1)^{l(w)} \widetilde \tr_{\pi^{\, w \cdot \mu}}.
\end{equation*}
We determine that
\begin{multline*}
\tr_{\varphi^\mu}(x) = \tr_{\pi^\mu \circ \varepsilon}(x) = \tr_{\pi^\mu}(\varepsilon(x)) = \sum_{w \in \calW} (-1)^{l(w)} \tr_{\widetilde \pi^{\, w \cdot \mu}}(\varepsilon(x)) \\= \sum_{w \in \calW} (-1)^{l(w)} \tr_{\widetilde \pi^{\, w \cdot \mu} \circ \varepsilon}(x) = \sum_{w \in \calW} (-1)^{l(w)} \tr_{\widetilde \varphi^{\, w \cdot \mu}}(x),
\end{multline*}
and obtain the equation
\begin{equation}
\calT^\mu(\zeta) = \sum_{w \in \calW} (-1)^{l(w)} \widetilde \calT^{\, w \cdot \mu}(\zeta) = \sum_{w \in \mathrm S_{l + 1}} \sgn(w) \, \widetilde \calT^{\, w \cdot \mu}(\zeta). \label{twtt}
\end{equation}
Introduce the notations
\begin{equation}
\widetilde \calS^\mu (\zeta) = \widetilde \calT^{\mu - \rho}(\zeta), \qquad \calS^\mu (\zeta) = \calT^{\mu - \rho}(\zeta). \label{tss}
\end{equation}
It follows from (\ref{twtt}) that
\begin{equation}
\calS^\mu(\zeta) = \sum_{w \in \mathrm S_{l + 1}} \sgn(w) \, \widetilde \calS^{\, w(\mu)}(\zeta), \label{swss}
\end{equation}
and we obtain
\begin{equation*}
\calC_l \, \calS^\mu(\zeta) = \det (\calQ_a(q^{2 \mu_b /s} \zeta))_{a, \, b = 1}^{l + 1}.
\end{equation*}
Thus, we come to the following determinant representation
\begin{equation*}
\calC_l \, \calT^\mu(\zeta) = \det (\calQ_a(q^{2 (\mu_b + \rho_b)/s} \zeta))_{a, \, b = 1}^{l + 1}.
\end{equation*}
The simplest consequence of this representation is the equation
\begin{equation}
\calT^{(\mu_1 + \nu, \, \ldots, \, \mu_{l + 1} + \nu)}(\zeta) = \calT^\mu(q^{2 \nu/s} \zeta). \label{tls}
\end{equation}
For the trivial one-dimensional representation $\varphi^0$ one has
\begin{equation*}
\varphi^0_\zeta(e_i) = 0, \qquad \varphi^0_\zeta(q^{\nu h_i}) = 1,
\end{equation*}
and it follows from the structure of the universal $R$-matrix, see, for example, the paper \cite{TolKho92}, that $\calT^0(\zeta) = 1_{\uqlbm}$. Using equation (\ref{tls}), we obtain that
\begin{equation}
\calT^{\nu \omega_{l + 1}}(\zeta) = 1_{\uqlbm} \label{tno}
\end{equation}
for any complex number $\nu$. It is also clear that equations (\ref{twtt}) and (\ref{swss}) imply
\begin{equation*}
\calT^{w \cdot \mu}(\zeta) = \sgn(w) \calT^\mu(\zeta), \qquad \calS^{w(\mu)}(\zeta) = \sgn(w) \calS^\mu(\zeta).
\end{equation*}
In particular,
\begin{equation}
\calT^{\mu_1, \, \ldots, \, \mu_i, \, \mu_{i + 1}, \, \ldots, \, \mu_{l + 1}}(\zeta) = - \calT^{\mu_1, \, \ldots, \, \mu_{i + 1} - 1, \, \mu_i + 1, \, \ldots, \, \mu_{l + 1}}(\zeta) \label{tmt}
\end{equation}
and
\begin{equation}
\calS^{\mu_1, \, \ldots, \, \mu_i, \, \mu_{i + 1}, \, \ldots, \, \mu_{l + 1}}(\zeta) = - \calS^{\mu_1, \, \ldots, \, \mu_{i + 1}, \, \mu_i, \, \ldots, \, \mu_{l + 1}}(\zeta). \label{sms}
\end{equation}

\subsection{\texorpdfstring{$TQ$-relations}{TQ-relations}}

Given $1 \le a \le l + 1$, construct the matrix
\begin{equation*}
\left[ \begin{array}{cccc}
\calQ_1(q^{2 \mu_1/s} \zeta) & \cdots & \calQ_1(q^{2 \mu_{1 + 1}/s} \zeta) & \calQ_1(q^{2 \mu_{l + 2}/s} \zeta) \\
\vdots & \ddots & \vdots & \vdots \\[.5em]
\calQ_{l + 1}(q^{2 \mu_1/s} \zeta) & \cdots & \calQ_{l + 1}(q^{2 \mu_{1 + 1}/s} \zeta) & \calQ_{l + 1}(q^{2 \mu_{l + 2}/s} \zeta) \\[.5em]
\calQ_a(q^{2 \mu_1/s} \zeta) & \cdots & \calQ_a(q^{2 \mu_{1 + 1}/s} \zeta) & \calQ_a(q^{2 \mu_{l + 2}/s} \zeta)
\end{array} \right],
\end{equation*}
where $\mu_b$, $b = 1, \ldots, l + 2$, are arbitrary complex numbers. The determinant of this matrix certainly vanishes. Expanding it over the last row, we obtain the equation
\begin{equation}
\sum_{b = 1}^{l + 2} (-1)^{b - 1} \calS^{(\mu_1, \ldots, \widehat \mu_b, \ldots, \mu_{l + 2})}(\zeta) \calQ_a(q^{2 \mu_b/s} \zeta) = 0, \label{tqr}
\end{equation}
where $\calS^\mu(\zeta)$ is defined by the second equation of (\ref{tss}). We call this equation the master $TQ$-relation. Assuming that
\begin{equation*}
\mu_b = \rho_b + 1 = l/2 - b + 2,
\end{equation*}
we obtain the more familiar equation
\begin{equation*}
\sum_{b = 1}^{l + 2} (-1)^{b - 1} \calT^{\, \omega_{b - 1}}(\zeta) \calQ_a(q^{2 \mu_b/s} \zeta) = 0,
\end{equation*}
where we assume that $\omega_0 = (0, \, \ldots, \, 0)$.

\subsection{\texorpdfstring{$TT$-relations and $T$-system}{TT-relations and T-system}}

It is clear that
\begin{equation*}
\calS^{\mu_b, \, \mu_{l + 3}, \, \ldots, \, \mu_{2 l + 2}}(\zeta) = \sum_{a = 1}^{l + 1} c_a^{\mu_{l + 3}, \, \ldots, \, \mu_{2 l + 2}}(\zeta) \calQ_a(q^{2 \mu_b/s} \zeta)
\end{equation*}
for some scalars $c_a^{\mu_{l + 3}, \, \ldots, \, \mu_{2 l + 2}}(\zeta)$. Multiplying (\ref{tqr}) by $c_a^{\mu_{l + 3}, \, \ldots, \, \mu_{2 l + 2}}(\zeta)$ and summing over $a$, we obtain
\begin{equation}
\sum_{b = 1}^{l + 2} (-1)^b \, \calS^{\mu_1, \, \ldots, \, \widehat{\mu}_b, \, \ldots, \, \mu_{l + 2}}(\zeta) \, \calS^{\mu_b, \, \mu_{l + 3}, \, \ldots, \, \mu_{2 l + 2}}(\zeta) = 0. \label{sss}
\end{equation}
In fact, this equation has the form of the Pl\"ucker relation which describe the image of the embedding of a Grassmannian in a projective space, see, for example, the paper \cite{KleLak72}. We call it the master $TT$-relation.

Given $1 \le a \le l$ and $m \in \bbZ_{>0}$, assume that the $(2 l + 2)$-tuple $(\mu_1, \, \ldots, \, \mu_{2 l + 2})$ has the form 
\begin{equation*}
(m + \rho_0, \, \ldots, m + \rho_a, \, \rho_{a + 1}, \, \ldots, \, \rho_{l + 1}, \, m + \rho_1, \ldots, \, m + \rho_{a - 1}, \, \rho_a, \, \ldots, \, \rho_l).
\end{equation*}
It follows from (\ref{sms}) that
\begin{equation*}
\calS^{\ldots, \, \nu, \, \ldots, \, \nu, \, \ldots}(\zeta) = 0.
\end{equation*}
Therefore, in the case under consideration, only three terms of the sum entering (\ref{sss}) are nonzero, namely, for $b$ equal to $1$, $a + 1$ and $l + 1$. Now using (\ref{sms}) and rewriting (\ref{sss}) in terms of $\calT^\mu(\zeta)$ we come to the equation
\begin{equation*}
\calT^{\, m \, \omega_a}(q^{2/s} \zeta) \calT^{\, m \, \omega_a}(\zeta) = \calT^{\, (m - 1) \, \omega_a}(q^{2/s} \zeta) \calT^{\, (m + 1) \, \omega_a}(\zeta) + \calT^{\, m \, \omega_{a + 1}}(q^{2/s} \zeta) \calT^{\, m \, \omega_{a - 1}}(\zeta).
\end{equation*}
Introducing the notation
\begin{equation*}
\calT_{a, \, m}(\zeta) =  \calT^{\, m \, \omega_a}(q^{(a - m)/s} \zeta),
\end{equation*}
we obtain
\begin{equation}
\calT_{a, \, m}(q^{-1/s} \zeta) \calT_{a, \, m}(q^{1/s} \zeta) = \calT_{a, \, m - 1}(\zeta) \calT_{a, \, m + 1}(\zeta) + \calT_{a - 1, \, m}(\zeta) \calT_{a + 1, \, m}(\zeta). \label{ts}
\end{equation}
This is the well known $T$-system in terms of universal objects, see for the review \cite{KunNakSuz11}.

It is known that $\calT_{a, \, m}(\zeta)$ can be expressed as a polynomial in $\calT_{1, \, 1}(\zeta)$, $\ldots$, $\calT_{1, \, l}(\zeta)$ with various $\zeta$ via the quantum Jacobi-Trudy identities \cite{BazRes90, KunNakSuz11}. The identities in question have the form
\begin{equation}
\calT_{a, \, m}(\zeta) = \det \big( \calT_{a - i + j, \, 1}(q^{(i + j - m - 1)/s} \zeta) \big)_{i, \, j = 1, \ldots, m} \, , \label{tas}
\end{equation}
where $\calT_{a, \, 1}(\zeta) = 0$ unless $0 \le a \le l + 1$, and $\calT_{0, 1}(\zeta) = \calT_{l + 1, \, 1}(\zeta) = 1$. To prove them, remind the Jacobi identity for determinants, see, for example, the book \cite[p.~79]{Hir04}. Let $D$ be the determinant of some square matrix. Denote by $D \Big [{}^i_j \Big]$ the determinant of the same matrix with the $i$th row and the $j$th column removed, and by $D \Big [{}^{i \, m}_{j \, \, n} \Big]$ the determinant of that matrix with the $i$th and $m$th rows and the $j$th and $n$th columns removed. The Jacobi identity states that
\begin{equation}
D \Big [{}^i_i \Big] D \Big [{}^j_j \Big] = D \Big [{}^{i \, \, j}_{i \, \, j} \Big] D + D \Big [{}^i_j \Big] D \Big [{}^j_i \Big]. \label{ji}
\end{equation}
Let us define
\begin{equation*}
\calF_{a, \, m}(\zeta) = \det \big( \calT_{a - i + j, \, 1}(q^{(i + j - m - 1)/s} \zeta) \big)_{i, \, j = 1, \ldots, m} \, ,
\end{equation*}
and write the following Jacobi identity
\begin{multline*}
\calF_{a, \, m + 1}(\zeta) \Big [{}^{m + 1}_{m + 1} \Big] \calF_{a, \, m + 1}(\zeta) \Big [{}^1_1 \Big] \\
= \calF_{a, \, m + 1}(\zeta) \Big [{}^{m + 1 \, \, 1}_{m + 1 \, \, 1} \Big] \calF_{a, \, m + 1}(\zeta) + \calF_{a, \, m + 1}(\zeta) \Big [{}^{\ \ \ 1}_{m + 1} \Big] \calF_{a, \, m + 1}(\zeta) \Big [{}^{m + 1}_{\ \ \ 1} \Big].
\end{multline*}
It is not difficult to demonstrate that this identity coincides with (\ref{ts}) written as an equation for $\calF_{a, \, m}(\zeta)$. Since $\calF_{a, \,1}(\zeta) = \calT_{a, \,1}(\zeta)$ for all $a = 1, \ldots, l$, equation (\ref{tas}) is true.
 
\subsection{\texorpdfstring{$Q$-system and nested Bethe ansatz equations}{Q-system and nested Bethe ansatz equations}}

In this section we generalize the consideration given in the paper \cite{BazFraLukMenSta11} for the integrable system associated with the Yangian $\mathrm Y(\mathfrak{gl}_n)$ to the case of the integrable system associated with the quantum loop algebra $\uqlsllpo$. Let $\bm a = (a_1, \, \ldots, a_p)$, $1 \le p \le l + 1$, be a $p$-tuple of distinct elements of the set $\{1, \ldots, l + 1\}$. Define the generalized universal $Q$-operator $\calQ_{\bm a}(\zeta)$ by the equation
\begin{equation*}
\calQ_{\bm a}(\zeta) = \det \big( \calQ_{a_i}(q^{(p - 2 j + 1)/s} \zeta) \big)_{i, \, j = 1, \ldots, p} \, .
\end{equation*}
We use the symbol '$\varnothing$' for the empty tuple and set
\begin{equation*}
\calQ_\varnothing(\zeta) = 1_{\uqlbm}.
\end{equation*}
by definition. The universal $Q$-operators $\calQ_{\bm a}(\zeta)$ are antisymmetric with respect to the transpositions of the components of $\bm a$. Hence, if $\bm a = (a_1, \, \ldots, a_p)$ and $\bm b = (b_1, \, \ldots, \, b_q)$ are tuples of distinct elements of $\{1, \ldots, l + 1\}$ having no common elements, then
\begin{equation*}
\calQ_{\bm a \cup \bm b}(\zeta) = (-1)^{p q} \calQ_{\bm b \cup \bm a}(\zeta).
\end{equation*}
Here and below we use the symbol `$\cup$' for concatenation of tuples. By definition, we have
\begin{equation*}
\calQ_{1, \, \ldots, \, l + 1}(\zeta) = \calC_l \calT^0(\zeta),
\end{equation*}
and it follows from (\ref{tno}) that
\begin{equation}
\calQ_{1, \, \ldots, \, l + 1}(\zeta) = \calC_l. \label{q1lpo}
\end{equation}

Let $\bm a = (a_1, \, \ldots, \, a_p)$ be a $p$-tuple of distinct elements of the set $\{1, \ldots, l + 1\}$, while $b$ and $c$ are elements of $\{1, \, \ldots, \, l + 1 \, \}$ not included in $\bm a$. One can easily demonstrate that the Jacobi identity
\begin{equation*}
D_{\bm a, \, b, \, c}(\zeta) D_{\bm a, \, b, \, c}(\zeta) \Big[ {}^{1 \ p + 2}_{1 \ p + 2} \Big] = D_{\bm a, \, b, \, c}(\zeta) \Big [{}^{p + 2}_{p + 2} \Big] D_{\bm a, \, b, \, c}(\zeta) \Big[ {}^{1}_{1} \Big] - D_{\bm a, \, b, \, c}(\zeta) \Big[ {}^{\ \ 1}_{p + 2} \Big] D_{\bm a, \, b, \, c}(\zeta) \Big [{}^{p + 2}_{\ \ 1} \Big],
\end{equation*}
where
\begin{equation*}
D_{\bm a, \, b, \, c}(\zeta) = \det \left(  \calQ_{a_i}(q^{(p - 2 j + 1)/s} \zeta)  \right)_{i, \, j = 0, \, 1, \, \ldots, \, p, \, p + 1}
\end{equation*}
with $a_0 = b$ and $a_{p + 1} = c$, implies the equation
\begin{equation*}
\calQ_{b \cup \bm a \cup c}(\zeta) \calQ_{\bm a}(\zeta) = \calQ_{b \cup \bm a}(q^{-1/s} \zeta) \calQ_{\bm a \cup c}(q^{1/s} \zeta) - \calQ_{\bm a \cup c}(q^{-1/s} \zeta) \calQ_{b \cup \bm a}(q^{1/s} \zeta).
\end{equation*}
Note that the case $\bm a = \emptyset$ is allowed if we set
\begin{equation*}
D_{\emptyset, \, b, \, c}(\zeta) \Big[ {}^{1 \ 2}_{1 \ 2} \big] = 1.
\end{equation*}
It is convenient, using the antisymmetry of generalized universal $Q$-operators under the permutations of the indices, to rewrite the above equation in the form
\begin{equation}
\calQ_{\bm a \cup b \cup c}(\zeta) \calQ_{\bm a}(\zeta) = \calQ_{\bm a \cup b}(q^{-1/s} \zeta) \calQ_{\bm a \cup c}(q^{1/s} \zeta) - \calQ_{\bm a \cup  c}(q^{-1/s} \zeta) \calQ_{\bm a \cup b}(q^{1/s} \zeta). \label{cqqq}
\end{equation}

Consider a chain of length $n$ living on the representation space
\begin{equation*}
U = \underbracket[.5pt]{\bbC^{l + 1} \otimes \cdots \otimes \bbC^{l + 1}}_n
\end{equation*}
of the representation
\begin{equation*}
\psi = \underbracket[.5pt]{\varphi^{\omega_1} \otimes_\Delta \cdots \otimes_\Delta \varphi^{\omega_1}}_n,
\end{equation*}
see section \ref{s:lwer} for the explicit form of the representation $\varphi^{\omega_1}$. Let $v_i$, $i = 1, \ldots, l + 1$, be the elements of the standard basis of $\bbC^{l + 1}$. The linear span of the $n$-fold tensor products of the vectors $v_i$ form a basis of $U$. We define the $Q$-operator $\nover{\bm Q}_a(\zeta)$ acting on $U$ by the equation
\begin{equation*}
\rme^{n F_{l + 1}(\zeta^s)} \, \nover{\bm Q}_a(\zeta) = \psi (\calQ_a(\zeta)),
\end{equation*}
see equations (\ref{xfff}), (\ref{rof}) and (\ref{fmz}). For the generalized $Q$-operator $\nover{\bm Q}_{\bm a}(\zeta)$ we have
\begin{equation*}
\rme^{n \sum_{j = 1}^p F_{l + 1}(q^{p - 2j + 1} \zeta^s)} \, \nover{\bm Q}_{\bm a}(\zeta) = \psi(\calQ_{\bm a}(\zeta)).
\end{equation*}
In fact, we have
\begin{equation*}
\nover{\bm Q}_{\bm a}(\zeta) = \det \big(\nover{\bm Q}_{a_i}(q^{(p - 2 j + 1)/s} \zeta) \big)_{i, \, j = 1, \ldots, p} \, .
\end{equation*}
Equation (\ref{cqqq}) implies that
\begin{equation}
\nover{\bm Q}_{\bm a \cup b \cup c}(\zeta) \nover{\bm Q}_{\bm a}(\zeta) = \nover{\bm Q}_{\bm a \cup b}(q^{-1/s} \zeta) \nover{\bm Q}_{\bm a \cup c}(q^{1/s} \zeta) - \nover{\bm Q}_{\bm a \cup c}(q^{-1/s} \zeta) \nover{\bm Q}_{\bm a \cup b}(q^{1/s} \zeta). \label{cbc}
\end{equation}
As follows from the definition (\ref{fmz}) of the function $F_{l + 1}(z)$, it satisfies the equation
\begin{equation*}
\sum_{j = 1}^{l + 1} F_{l + 1} (q^{l - 2 j + 2} \zeta) = - \log (1 - \zeta)
\end{equation*}
and, using (\ref{q1lpo}), we obtain
\begin{equation}
\nover{\bm Q}_{1, \, \ldots, \, l + 1}(\zeta) = (1 - \zeta^s)^n \, \psi(\calC_l). \label{nq1lpo}
\end{equation}
Note also that
\begin{equation}
\nover{\bm Q}_{\emptyset}(\zeta) = 1_U. \label{qnez}
\end{equation}

The operators $\bm Q_{\bm a}(\zeta)$ commute for any $\bm a$ and $\zeta$, see equation (\ref{yyyy}). Although we do not know a rigorous proof of this fact, there are many indications that the operators $\bm Q_{\bm a}(\zeta)$ are diagonalizable, and we assume that this is indeed the case. Therefore, all these operators can be diagonalized simultaneously, see, for example, the book \cite[Theorem 6.8]{HofKun71}.

Let $\bm k = (k_1, \, \ldots, \, k_{l + 1})$ be a tuple of non-negative integers such that $k_1 + \cdots + k_{l + 1} = n$. Denote by $U_{\bm k}$ the linear span of the basis vectors containing $k_i$ vectors $v_i$ for any $i = 1, \ldots, l + 1$. Clearly, it is a subspace of $U$. It is evident that
\begin{equation}
\psi(q^{\nu h_i}) v = q^{\nu(k_i - k_{i + 1})} v \label{pqv}
\end{equation}
for any $v \in U_{\bm k}$. Hence, the subspace $U_{\bm k}$ is invariant with respect the action of the operators  $\psi(q^{\nu h_i})$. One can demonstrate that a vector $v$ belongs to $U_{\bm k}$ if an only if
\begin{equation*}
\psi \big( q^{\nu (n + \sum_{j = i}^l (l - j + 1) \, h_j - \sum_{j = 1}^{i - 1} j \, h_j)} \big) v = q^{\nu k_i} v
\end{equation*}
for any $\nu \in \bbC$ and $i = 1, \ldots, l + 1$. The operators $\nover{\bm Q}_{\bm a}(\zeta)$ commute with the operators $\psi(q^{\nu h_i})$, see equation (\ref{ypa}). Therefore, $U_{\bm k}$ is an invariant subspace for these operators as well.

Restricting ourselves to $U_{\bm k}$, we can consider (\ref{cbc}) as an equation for the eigenvalues of the operators $\nover{\bm Q}_{\bm a}(\zeta)$, which we denote by $\nover{Q}_{\bm a, \, \bm k}(\zeta)$. Thus, we have the equation
\begin{multline}
\nover{Q}_{\bm a \cup b \cup c, \, \bm k}(\zeta)\nover{Q}_{\bm a, \, \bm k}(\zeta)\\ = \nover{Q}_{\bm a \cup b, \, \bm k}(q^{-1/s} \zeta) \nover{Q}_{\bm a \cup c, \, \bm k}(q^{1/s} \zeta) - \nover{Q}_{\bm a \cup c, \, \bm k}(q^{-1/s} \zeta) \nover{Q}_{\bm a \cup b, \, \bm k}(q^{1/s} \zeta). \label{qzqz}
\end{multline}
This equation at $\zeta = (q^{-1} \, \zeta_{\bm a \cup b, \, \bm k, \, m})^{1/s}$ takes the form
\begin{multline*}
\nover{Q}_{\bm a \cup b \cup c, \, \bm k} \big( (q^{-1} \, \zeta_{\bm a \cup b, \, \bm k, \, m})^{1/s} \big) \nover{Q}_{\bm a, \, \bm k} \big( (q^{-1} \, \zeta_{\bm a \cup b, \, \bm k, \, m})^{1/s} \big) \\
= \nover{Q}_{\bm a \cup c, \, \bm k} \big( (\zeta_{\bm a \cup b, \, \bm k, \, m})^{1/s} \big) \nover{Q}_{\bm a \cup b, \, \bm k} \big( (q^{-2} \, \zeta_{\bm a \cup b, \, \bm k, \, m} \big)^{1/s}).
\end{multline*}
while at $\zeta = (q \zeta_{\bm a \cup b, \, \bm k, \, m})^{1/s}$ we have
\begin{multline*}
\nover{Q}_{\bm a \cup b \cup c, \, \bm k} \big( (q \, \zeta_{\bm a \cup b, \, \bm k, \, m})^{1/s} \big) \nover{Q}_{\bm a, \, \bm k} \big( (q \, \zeta_{\bm a \cup b, \, \bm k, \, m})^{1/s} \big) \\*
= {} - \nover{Q}_{\bm a \cup c, \, \bm k} \big( (\zeta_{\bm a \cup b, \, \bm k, \, m})^{1/s} \big) \nover{Q}_{\bm a \cup b, \, \bm k} \big( (q^2 \, \zeta_{\bm a \cup b, \, \bm k, \, m})^{1/s} \big).
\end{multline*}
Taking the ratio of the above two equations, we obtain
\begin{multline}
-1 = \frac{\nover{Q}_{\bm a, \, \bm k} \big( (q^{-1} \, \zeta_{\bm a \cup b, \, \bm k, \, m})^{1/s} \big)}
{\nover{Q}_{\bm a, \, \bm k} \big( (q \, \zeta_{\bm a \cup b, \, \bm k, \, m})^{1/s} \big)} \\*
\times \frac{\nover{Q}_{\bm a \cup b, \, \bm k} \big( (q^2 \, \zeta_{\bm a \cup b, \, \bm k, \, m})^{1/s} \big) \nover{Q}_{\bm a \cup b \cup c, \, \bm k} \big( (q^{-1} \, \zeta_{\bm a \cup b, \, \bm k, \, m})^{1/s} \big)}
{\nover{Q}_{\bm a \cup b, \, \bm k} \big( (q^{-2} \, \zeta_{\bm a \cup b, \, \bm k, \, m})^{1/s} \big) \nover{Q}_{\bm a \cup b \cup c, \, \bm k} \big( (q \, \zeta_{\bm a \cup b, \, \bm k, \, m})^{1/s} \big)}. \label{me}
\end{multline}
Given an $(l + 1)$-tuple $\bm a$ of distinct elements of the set $\{1, \ldots, l + 1\}$,
we define
\begin{equation*}
\bm a_0 = \varnothing, \qquad \bm a_i = (a_1, \, \ldots, \, a_i), \quad i = 1, \ldots, l + 1,
\end{equation*}
and using (\ref{me}), for each $i = 1, \ldots, l$, write the equation
\begin{equation}
-1 = \frac{\nover{Q}_{\bm a_{i - 1}, \, \bm k} \big( (q^{-1} \, \zeta_{\bm a_i, \, \bm k, \, m})^{1/s} \big) \nover{Q}_{\bm a_i, \, \bm k} \big( (q^2 \, \zeta_{\bm a_i, \, \bm k, \, m})^{1/s} \big) \nover{Q}_{\bm a_{i + 1}, \, \bm k} \big( (q^{-1} \, \zeta_{\bm a_i, \, \bm k, \, m})^{1/s} \big)}
{\nover{Q}_{\bm a_{i - 1}, \, \bm k} \big( (q \, \zeta_{\bm a_i, \, \bm k, \, m})^{1/s} \big) \nover{Q}_{\bm a_i, \, \bm k} \big( (q^{-2} \, \zeta_{\bm a_i, \, \bm k, \, m})^{1/s} \big) \nover{Q}_{\bm a_{i + 1}, \, \bm k} \big( (q \, \zeta_{\bm a_i, \, \bm k, \, m})^{1/s} \big)}. \label{be}
\end{equation}
We call $i$ in the above equation the level.

Introduce one more notation
\begin{equation*}
\bm D_a = \psi(\calD_a),
\end{equation*}
where $\calD_a$ is defined by equation (\ref{cda}). It follows from equation (\ref{pqv}) that the eigenvalues of the operators $\bm D_a$ on $U_{\bm k}$ are
\begin{equation}
D_{a, \, \bm k} = \frac{s}{2(l + 1)} \big( \sum_{j = 1}^{a_i - 1} j \, (k_j - k_{j + 1} + t_j) - \sum_{j = a_i}^l (l - j + 1) (k_j - k_{j + 1} + t_j) \big). \label{dak}
\end{equation}
From the explicit form  (\ref{lzeta}) of the $L$-operator $\bm L(\zeta)$, equation (\ref{lazeta}) and relations (\ref{trbp})--(\ref{trtr}), it follows that the eigenvalues of the operators $\nover{\bm Q}_{\bm a}(\zeta)$ on $W_{\bm k}$ have the form
\begin{equation}
\nover{Q}_{\bm a, \bm k}(\zeta) = c_{\bm a, \, \bm k} \prod_{i = 1}^p \zeta^{D_{a_i, \, \bm k}} \prod_{j = 1}^{k_{\bm a}} (\zeta^s - \zeta_{\bm a, \, \bm k, \, j}), \label{qakz}
\end{equation}
where
\begin{equation*}
k_{\bm a} = \sum_{i = 1}^p k_{a_i},
\end{equation*}
and $c_{\bm a, \, \bm k}$ are some complex coefficients. Here we set $k_\varnothing = 0$.

Now, substitute (\ref{qakz}) into (\ref{be}), and consider three different cases. In all cases we need the equation
\begin{equation*}
2(D_{b, \, \bm k} - D_{c, \, \bm k})/s = - k_b + k_c - \sgn(b - c) \sum_{i = b}^{c - 1} t_i, \quad c > b,
\end{equation*}
which follows from the equation (\ref{dak}). Introducing new parameters $\tau_a$ related to $t_a$ by the equation
\begin{equation*}
t_a = \tau_a - \tau_{a + 1},
\end{equation*}
we obtain
\begin{equation}
2(D_{b, \, \bm k} - D_{c, \, \bm k})/s = - k_b + k_c + \tau_b - \tau_c. \label{dbdc}
\end{equation}
For the first level, equation (\ref{qnez}) implies
\begin{equation*}
-1 = \frac{Q_{\bm a_1, \, \bm k} \big( (q^2 \, \zeta_{\bm a_1, \, \bm k, \, m})^{1/s} \big) Q_{\bm a_2, \, \bm k} \big( (q^{-1} \, \zeta_{\bm a_1, \, \bm k, \, m})^{1/s} \big)}{Q_{\bm a_1, \, \bm k} \big( (q^{-2} \, \zeta_{\bm a_1, \, \bm k, \, m})^{1/s} \big) Q_{\bm a_2, \, \bm k} \big( (q \, \zeta_{\bm a_1, \, \bm k, \, m})^{1/s} \big)}.
\end{equation*}
Using (\ref{qakz}) and (\ref{dbdc}), we come to the equation
\begin{equation*}
q^{\tau_{a_2} - \tau_{a_1}} = 
\prod_{\substack{j = 1 \\ j \ne m}}^{k_{\bm a_1}} \frac{q^2 \, \zeta_{\bm a_1, \, \bm k, \, m} - \zeta_{\bm a_1, \, \bm k, \, j}} {\zeta_{\bm a_1, \, \bm k, \, m} - q^2 \zeta_{\bm a_1, \, \bm k, \, j}}
\prod_{j = 1}^{k_{\bm a_2}} \frac{\zeta_{\bm a_1, \, \bm k, \, m} - q \zeta_{\bm a_2, \, \bm k, \, j}} {q \, \zeta_{\bm a_1, \, \bm k, \, m} - \zeta_{\bm a_2, \, \bm k, \, j}}.
\end{equation*}
Similarly, for $i = 2, \ldots, l - 1$ we obtain
\begin{equation*}
q^{\tau_{a_{i + 1}} - \tau_{a_i}} = 
\prod_{j = 1}^{k_{\bm a_{i - 1}}} \frac{\zeta_{\bm a_i, \, \bm k, \, m} - q \zeta_{\bm a_{i - 1}, \, \bm k, \, j}} {q \, \zeta_{\bm a_i, \, \bm k, \, m} - \zeta_{\bm a_{i - 1}, \, \bm k, \, j}}
\prod_{\substack{j = 1 \\ j \ne m}}^{k_{\bm a_i}} \frac{q^2 \, \zeta_{\bm a_i, \, \bm k, \, m} - \zeta_{\bm a_i, \, \bm k, \, j}} {\zeta_{\bm a_i, \, \bm k, \, m} - q^2 \zeta_{\bm a_i, \, \bm k, \, j}}
\prod_{j = 1}^{k_{\bm a_{i + 1}}} \frac{\zeta_{\bm a_i, \, \bm k, \, m} - q \zeta_{\bm a_{i + 1}, \, \bm k, \, j}} {q \, \zeta_{\bm a_i, \, \bm k, \, m} - \zeta_{\bm a_{i + 1}, \, \bm k, \, j}},
\end{equation*}
and, finally for $i = l$, using (\ref{nq1lpo}),
\begin{equation*}
q^{\tau_{a_{l + 1}} - \tau_{a_l}} \Big( \frac{q \, \zeta_{\bm a_l, \, \bm k, \, m} - 1} {\zeta_{\bm a_l, \, \bm k, \, m} - q} \Big)^n = \prod_{j = 1}^{k_{\bm a_{l - 1}}} \frac{\zeta_{\bm a_l, \, \bm k, \, m} - q \zeta_{\bm a_{l - 1}, \, \bm k, \, j}} {q \, \zeta_{\bm a_l, \, \bm k, \, m} - \zeta_{\bm a_{l - 1}, \, \bm k, \, j}} \prod_{\substack{j = 1 \\ j \ne m}}^{k_{\bm a_l}} \frac{q^2 \, \zeta_{\bm a_l, \, \bm k, \, m} - \zeta_{\bm a_l, \, \bm k, \, j}} {\zeta_{\bm a_l, \, \bm k, \, m} - q^2 \zeta_{\bm a_l, \, \bm k, \, j}}.
\end{equation*}
Thus, we come to the nested Bethe ansatz equations. We have $l!$ variants of the equations corresponding to $l!$ paths on the Hasse diagram, see the paper \cite{BazFraLukMenSta11} for the Yangian case.

At present, the families of universal $Q$-operators $\calQ_{\bm a}(\zeta)$, $Q$-operators $\nover{\bm Q}_{\bm k}(\zeta)$, or eigenvalues $\nover{Q}_{\bm a, \, \bm k}(\zeta)$ are called $Q$-systems. Their study was started in the papers \cite{BazLukZam96, BazLukZam97, BazLukZam99, ProStr99, ProStr00}, see also \cite{KazSorZab08, Tsu97, Tsu10, Tsu13a, KazLeuVol16} for more recent results on $T$- and $Q$-systems. Other related systems called $QQ$-systems were recently proposed for effective solving of the nested Bethe ansatz equations \cite{MarVol17, BajGraJacNep2020, Nep20}.

\section{Conclusion}

We have investigated the quantum integrable systems associated with the quantum loop algebras $\uqlsllpo$. Analyzing products of $\ell$--weights of the $q$-oscillator representations of the Borel subalgebras $\uqlbp$ of $\uqlsllpo$ we proved the factorization relations for the transfer operators related to the infinite dimensional evaluation representations of $\uqlsllpo$. This allowed us to construct the determinant representations of the transfer operators related to the finite dimensional representations of $\uqlsllpo$. We derived the universal $TQ$-relations and the universal $TT$-relations, and demonstrated that the known functional relations are their partial cases. The operatorial $Q$-system was constructed and the nested Bethe equations were derived. The results obtained are completely consistent with the results obtained for $l = 1$ and $l = 2$  in the papers \cite{BooGoeKluNirRaz14a, NirRaz16a, BooGoeKluNirRaz14b}.

As a byproduct we have a conjectured formula for the components $\Lambda^{\mu +}_{\bm m, \, i}(\zeta, \, u)$ of the $\ell$-weight $\bm \Lambda_{\bm m}^\mu(\zeta)$ of the evaluation $\uqlsllpo$-module $\widetilde V^\lambda$. It can be obtained from the expression (\ref{ximup}) for $\bm \Xi^{\mu +}_{\bm m, \, i}(\zeta, \, u)$ after the substitution $n_{a, \, i} = m_{a, \, a + i}$. The result is 
\begin{multline*}
\Lambda^{\mu +}_{\bm m, \, i}(\zeta, \, u) = \frac{\prod_{a = 1}^{i - 1} (1 - q^{2 \mu_a - 2 \sum_{j = i}^{l + 1} m_{a j} + i - 2 a + 1} \zeta^s u) }{\prod_{a = 1}^i (1 - q^{2 \mu_a - 2 \sum_{j = i + 1}^{l + 1} m_{a j} + i - 2 a + 1} \zeta^s u)} \\*
\times \frac{\prod_{a = 1}^{i + 1} (1 - q^{2 \mu_a - 2 \sum_{j = i + 2}^{l + 1} m_{a j} + i - 2 a + 3} \zeta^s u)}{\prod_{a = 1}^i (1 - q^{2 \mu_a - 2 \sum_{j = i + 1}^{l + 1} m_{a j} + i - 2 a + 3} \zeta^s u)}.
\end{multline*}
The highest $\ell$-weight of the module $\widetilde V^\lambda$ is rational, see equations(\ref{lzp}) and (\ref{lzm}). In this case the components $\Lambda^{\mu -}_{\bm m, \, i}(\zeta, \, u)$ are uniquely determined by the components $\Lambda^{\mu +}_{\bm m, \, i}(\zeta, \, u)$, see, for example, the paper \cite{MukYou14}. In our case we have
\begin{multline*}
\Lambda^{\mu -}_{\bm m, \, i}(\zeta, \, u^{-1}) = \frac{\prod_{a = 1}^{i - 1} (1 - q^{- 2 \mu_a + 2 \sum_{j = i}^{l + 1} m_{a j} - i + 2 a - 1} \zeta^{-s} u^{-1}) }{\prod_{a = 1}^i (1 - q^{- 2 \mu_a + 2 \sum_{j = i + 1}^{l + 1} m_{a j} - i + 2 a - 1} \zeta^{-s} u^{-1})} \\*
\times \frac{\prod_{a = 1}^{i + 1} (1 - q^{- 2 \mu_a + 2 \sum_{j = i + 2}^{l + 1} m_{a j} - i + 2 a - 3} \zeta^{-s} u^{-1})}{\prod_{a = 1}^i (1 - q^{- 2 \mu_a + 2 \sum_{j = i + 1}^{l + 1} m_{a j} - i + 2 a - 3} \zeta^{-s} u^{-1})}.
\end{multline*}
Recall that the component $\lambda^\mu_{\bm m}$ is given by equation (\ref{psimum}).

It would be interesting to generalize the consideration of the present paper to the supersymmetric case.

\vspace{1em}

\section*{Acknowledgments}

This work was supported in part by the RFBR grant \#~20-51-12005. The author is grateful to H.~ Boos, F.~ G\"ohmann, A.~ Kl\"umper, and Kh.~S.~ Nirov, in collaboration with whom some important results, used in this paper, were previously obtained, for useful discussions.

\appendix

\section{\texorpdfstring{Highest $\ell$-weight of evaluation $\uqlsllpo$-modules}{Highest l-weight of evaluation Uq(L(sll+1)-modules}} \label{a:hwfr}

We start with equation (5.8) of the paper \cite{NirRaz17b}
\begin{align}
& e_i \, v_{\bm m} = \zeta^{s_i} [\mu_i - \mu_{i+1} - \sum_{j = i + 2}^{l + 1} (m_{i j} - m_{i + 1, \, j}) 
- m_{i, \, i + 1} + 1]_q \, [m_{i, \, i + 1}]_q \, v_{{\bm m} - \epsilon_{i, \, i + 1}} \notag \\*
& \hspace{.5em} {} + \zeta^{s_i} q^{\mu_i - \mu_{i + 1} - 2 m_{i, \, i + 1} 
- \sum_{j = i + 2}^{l + 1} (m_{i j} - m_{i + 1, \, j})} 
\sum_{j = 1}^{i - 1} q^{\sum_{k = j + 1}^{i - 1} (m_{k i} - m_{k,\, i + 1})} \,
[m_{j,\, i + 1}]_q \, v_{{\bm m} - \epsilon_{j, \, i + 1}  + \epsilon_{j i}} \notag \\*
& \hspace{11.5em} {} - \zeta^{s_i} \sum_{j = i + 2}^{l + 1} q^{- \mu_i + \mu_{i + 1} - 2 + \sum_{k = j}^{l + 1} (m_{i k} - m_{i + 1, \, k})} \, [m_{i j}]_q \, v_{{\bm m} - \epsilon_{i j} + \epsilon_{i + 1, \, j}}. 
\label{eivm}
\end{align}
Here and below ${\bm m} + k \epsilon_{i j}$ means shifting by $k$ the entry $m_{i j}$ in the $l(l + 1)/2$-tuple ${\bm m}$. Using this relation, we first derive some auxiliary equations.

First find the expression for $e_{\delta - \alpha_i} v_{\bm 0 + k \, \epsilon_{i, \, i + 1}}$. Equations (\ref{edmaa}) and (\ref{edmab}) give
\begin{equation*}
e_{\delta - \alpha_i}  = [e_{\alpha_{i + 1}}, \, [ \, \ldots, \, [e^{}_l, \, [e_{\alpha_{i - 1}}, \, [ \, \ldots, \, [e_{\alpha_1}, \, e_{\delta - \theta}] \, \ldots \, ] \, ] \, ] \, \ldots \, ] \, ],
\end{equation*}
and direct calculations lead to the equation
\begin{equation}
e_{\alpha_j} v_{\bm 0 + k \, \epsilon_{i, \, i + 1}} = \zeta^{s_i} \delta_{j i} [k]_q [\mu_i - \mu_{i + 1} - k + 1]_q  v_{\bm 0 + (k - 1) \epsilon_{i, \, i + 1}}. \label{eajv}
\end{equation}
Hence, we have
\begin{equation*}
e_{\delta - \alpha_i} \, v_{\bm 0 + \epsilon_{i, \, i + 1}} = e_{\alpha_{i + 1}} \ldots \, e^{}_l \, e_{\alpha_{i - 1}} \ldots \, e_{\alpha_1} \, e_{\delta - \theta} \, v_{\bm 0 + \epsilon_{i, \, i + 1}}.
\end{equation*}
It follows from equation (5.7) of the paper \cite{NirRaz17b} that
\begin{equation*}
e_{\delta - \theta} \, v_{\bm 0 + k \, \epsilon_{i, \, i + 1}} = \zeta^{s - s_{1, \, l + 1}} \, q^{\mu_1 + \mu_{l + 1} + k \delta_{i, \, l + 1}} \, v_{\bm 0 +  k \, \epsilon_{i, \, i + 1} + \epsilon_{1, \, l + 1}}.
\end{equation*}
Now, using equation (\ref{eivm}), we obtain
\begin{equation}
e_{\delta - \alpha_i} v_{\bm 0 + k \, \epsilon_{i, \, i + 1}} = \zeta^{s - s_i} (-1)^{i - 1} q^{\mu_i + \mu_{i + 1} - i + 1} \,  \, v_{\bm 0 + (k + 1)\epsilon_{i, \, i + 1}}. \label{edmaiv}
\end{equation}

We also need the corresponding expression for $e'_{n \delta; \, \alpha_i} v_{\bm 0 + k \, \epsilon_{i, \, i + 1}}$. It follows from equation (\ref{cwby5}) that
\begin{equation*}
e'_{\delta; \, \alpha_i} v_{\bm 0 + k \, \epsilon_{i, \, i + 1}} = e_{\alpha_i} \, e_{\delta - \alpha_i} v_{\bm 0 + k \, \epsilon_{i, \, i + 1}} - q^2  e_{\delta - \alpha_i} e_{\alpha_i} v_{\bm 0 + k \, \epsilon_{i, \, i + 1}},
\end{equation*}
and equations (\ref{edmaiv}) and (\ref{eajv}) give
\begin{multline}
e'_{\delta; \, \alpha_i} v_{\bm 0 + k \, \epsilon_{i, \, i + 1}} = \zeta^s (-1)^{i - 1} q^{\mu_i + \mu_{i + 1} - i + 1} \\
\times \big( [k + 1]_q [\mu_i - \mu_{i + 1} - k]_q - q^2 \, [k]_q [\mu_i - \mu_{i + 1} - k + 1]_q \big) v_{\bm 0 + k \, \epsilon_{i, \, i + 1}}. \label{epdav}
\end{multline}

To obtain the last necessary expression, we start with equation (\ref{cwby1}) which gives
\begin{equation*}
e_{\alpha_i + n \delta} \, v_{\bm 0 + \epsilon_{i, \, i + 1}} = [2]_q^{-1} \big( e_{\alpha_i + (n - 1)\delta} \, e'_{\delta; \, \alpha_i}  \, v_{\bm 0 + \epsilon_{i, \, i + 1}} - e'_{\delta; \, \alpha_i} e_{\alpha_i + (n - 1)\delta}  \, v_{\bm 0 + \epsilon_{i, \, i + 1}} \big).
\end{equation*}
One can demonstrate that this equation is satisfied by the substitution
\begin{equation}
e_{\alpha_i + n \delta} \, v_{\bm 0 + k \, \epsilon_{i, \, i + 1}} = \zeta^{s_i + n s} (-1)^{i n} q^{n (2 \mu_i -2 k - i + 3)} [k]_q [\mu_i - \mu_{i+ 1} - k + 1]_q  v_{\bm 0 + (k - 1) \epsilon_{i, \, i + 1}}  \label{eaindve}
\end{equation}
if we take into account equation (\ref{epdav}).

Finally, using equation (\ref{cwby5}), we write
\begin{equation*}
e'_{n \delta; \, \alpha_i} v_{\bm 0} = e\strut_{\alpha_i + (n - 1)\delta} \, e\strut_{\delta - \alpha_i} v_{\bm 0} - q^2 e\strut_{\delta - \alpha_i}  \, e\strut_{\alpha_i + (n - 1)\delta} v_{\bm 0}
\end{equation*}
and, taking into account equations (\ref{eaindve}) and (\ref{edmaiv}), get
\begin{equation*}
e'_{n \delta; \, \alpha_i} v_{\bm 0} = \zeta^{n s} \kappa_q^{-1} (-1)^{i n - 1} q^{2(n - 1) \mu_i - (i - 1)n} (q^{2 \mu_i} - q^{2 \mu_{i + 1}}) \, v_{\bm 0}.
\end{equation*}
Choosing $o_i = (-1)^i$ and substituting the above expression into the first equation of (\ref{phiiu}), we get
\begin{equation*}
\phi^+_i(u) v_{\bm 0} = (1 - (q - q^{-1}) \sum_{n = 1}^\infty ((-1)^{i - 1} u)^n e'_{n \delta; \, \alpha_i}) v_{\bm 0} = \frac{1 - q^{2 \mu_{i + 1} - i + 1} \zeta^s u} {1 - q^{2 \mu_i - i + 1} \zeta^s u} v_{\bm 0}.
\end{equation*}
Thus, the vector $v_{\bm 0}$ is the highest $\ell$-weight vector of $\ell$-weight
\begin{equation*}
\bm \Lambda^\mu_{\bm 0} = (\lambda^\mu_{\bm 0}, \, \bm \Lambda_{\bm 0}^{\mu +}, \, \bm \Lambda_{\bm 0}^{\mu -}),
\end{equation*}
 where
\begin{equation*}
\lambda^\mu_{\bm 0} = \sum_{i = 1}^l (\mu_i - \mu_{i + 1}) \omega_i,
\end{equation*}
and for the components of $\bm \Lambda_{\bm 0}^{\mu +}(u)$ we have the expressions
\begin{equation}
\Lambda_{\bm 0, \, i}^{\mu +}(\zeta, u) = \frac{1 - q^{2 \mu_{i + 1} - i + 1} \zeta^s u} {1 - q^{2 \mu_i - i + 1} \zeta^s u}.  \label{lzp}
\end{equation}
In the same way we obtain
\begin{equation}
\Lambda_{\bm 0, \, i}^{\mu -}(\zeta, u) = \frac{1 - q^{- 2 \mu_{i + 1} + i - 1} \zeta^{-s} u^{-1}} {1 - q^{- 2 \mu_i + i - 1} \zeta^{-s} u^{-1}}.  \label{lzm}
\end{equation}

\newcommand{\noopsort}[1]{}
\providecommand{\bysame}{\leavevmode\hbox to3em{\hrulefill}\thinspace}
\providecommand{\href}[2]{#2}
\providecommand{\curlanguage}[1]{%
 \expandafter\ifx\csname #1\endcsname\relax
 \else\csname #1\endcsname\fi}

\end{document}